\documentclass[pdflatex, sn-mathphys-ay]{sn-jnl}

\usepackage{graphicx}%
\usepackage{multirow}%
\usepackage{amsmath,amssymb,amsfonts}%
\usepackage{amsthm}%
\usepackage{mathrsfs}%
\usepackage[title]{appendix}%
\usepackage{xcolor}%
\usepackage{textcomp}%
\usepackage{manyfoot}%
\usepackage{booktabs}%
\usepackage{algorithm}%
\usepackage{algorithmicx}%
\usepackage{algpseudocode}%
\usepackage{listings}%
\usepackage{mathtools}

\usepackage[subrefformat=parens]{subcaption}
\usepackage{bm}
\newcommand{\rme}{\mathrm{e}\,}
\newcommand{\rmi}{\mathrm{i}\,}
\newcommand{\td}{\mathrm{d}}
\newcommand{\pd}{\partial}
\newcommand{\nab}{\nabla}
\newcommand{\rot}{\nabla \times}
\newcommand{\eps}{\epsilon}
\newcommand{\veps}{\varepsilon}
\newcommand{\vphi}{\varphi}
\newcommand{\bv}{\boldsymbol}
\newcommand{\vq}{\bv{q}}
\newcommand{\vp}{\bv{p}}
\newcommand{\vu}{\bv{u}}
\newcommand{\vv}{\bv{v}}
\newcommand{\vx}{\bv{x}}
\newcommand{\vz}{\bv{z}}
\newcommand{\vL}{\bv{L}}
\newcommand{\vB}{\bv{B}}
\newcommand{\vecrho}{\bv{\rho}}
\newcommand{\pr}{\prime}
\newcommand{\hrvz}{\hat{\bv{z}}}

\theoremstyle{thmstyleone}%
\theoremstyle{thmstyletwo}%

\theoremstyle{thmstylethree}%

\begin{document}

\title[Simulated annealing]{Simulated annealing of reduced magnetohydrodynamic systems}


\author*[1]{\fnm{M.} \sur{Furukawa}}\email{furukawa@tottori-u.ac.jp}

\author[2]{\fnm{P. J.} \sur{Morrison}}\email{morrison@physics.utexas.edu}

\affil*[1]{\orgdiv{Faculty of Engineering},
\orgname{Tottori University}, 
\orgaddress{
\street{Minami 4-101, Koyama-cho}, 
\city{Tottori-shi}, \postcode{680-8552}, 
\country{Japan}}} 

\affil[2]{\orgdiv{Department of Physics and Institute for Fusion Studies},
\orgname{University of Texas at Austin}, 
\orgaddress{
\city{Texas}, \postcode{78712}, 
\country{USA}}}


\abstract{
Theory of simulated annealing (SA), a method for equilibrium and 
stability analyses for Hamiltonian systems, is reviewed.  The SA 
explained in this review is based on a double bracket formulation that derives from Hamiltonian structure.  
In addition to  general theoretical aspects, the 
explicit formulation as well as numerical applications are presented.  Both finite and infinite degree-of-freedom systems are treated, in particular,   the  heavy top, a toy model mimicking low-beta reduced magnetohydrodynamics (MHD) and low- and high-beta reduced MHD.   Numerical results successfully demonstrate the usefulness of SA for
equilibrium and stability analyses. At the same time, the  results raise some future issues that are
discussed in the paper.
}




\maketitle

\tableofcontents 

\section{Introduction}
\label{sec:introduction}

Simulated annealing (SA) is a type of relaxation method for Hamiltonian
systems based on an artificial dynamics that uses  the
Hamiltonian structure.  In usual  Hamiltonian dynamics, the energy
(Hamiltonian) is conserved because of the antisymmetry of the
Poisson bracket, while the artificial dynamics of SA is constructed in such a way 
that the time evolution changes the energy (Hamiltonian) monotonically.   It does this by acting twice with the Poisson bracket and, consequently, SA relaxes to  a  stationary state of the energy as time progresses.   

If the Hamiltonian system is noncanonical, the Poisson bracket possesses
a null space and the null space leads  to Casimir invariants that are
conserved during the time evolution for any Hamiltonian.
Because the artificial dynamics of SA is constructed by acting twice with  the Poisson
bracket, the Casimir invariants are preserved during the time evolution. 
Because  SA extremizes the energy on a constant Casimir leaf, which is a
subspace of the phase space of the system defined by the level sets of the 
Casimir invariants, it in effect finds a solution of the energy-Casimir variational principle,  a variational principle that made its way into the  plasma and fluid literature in the early work of \citet{KO-1958} and \citet{ Arnold-1965-1}  \citep[see e.g.] [for review]   {Morrison-1998}.   The equilibria obtained by SA of noncanonical Hamiltonian systems can
have a variety of structure because of the possible variety of  Casimir invariants.

The ideal fluid and MHD were  shown to be  noncanonical Hamiltonian systems  by \citet{Morrison-1980} \citep[see also][] {Morrison-1982, Morrison-1998}.   Therefore, SA can be used for equilibrium calculations of such systems.   Reduced MHD systems are also Hamiltonian systems,  as was shown by   \citet{MH84}; these will be treated
in this paper explicitly. 

Originally, equilibrium calculations by such artificial dynamics were 
developed for two-dimensional vortical motion of neutral 
fluids in \citet{Vallis-1989, Carnevale-1990} and placed in a general
Hamiltonian systems setting in  \citet{Shepherd-1990}.  
However, the method of these references is limited and is now known to only work for a small class of  equilibria.  To correct for this the method  was generalized by   \cite{Flierl-Morrison-2011}, where  the term ``simulated annealing'' was introduced, and where it was shown to work for a variety of equilibria.
They developed a double bracket that is constructed from 
the Poisson bracket and a definite symmetric kernel.  
The Dirac SA (DSA) dynamics was also introduced, 
that utilizes a Dirac bracket instead of the Poisson bracket
in the construction of the double bracket.
They presented numerically a variety of non-trivial equilibria of
two-dimensional neutral fluids and two-layer quasigeostrophic flows.

The first application of SA to MHD systems  \citep{Chikasue-2015-PoP} was on   low-beta reduced
MHD \citep{Strauss-1976} in a two-dimensional rectangular domain with doubly 
periodic boundary conditions.  
Numerical results with several ratios of kinetic energy to the magnetic
energy were presented.  It was shown that upon relaxation to 
stationary states, fine structure remained  when the kinetic energy is comparable to or greater
than the magnetic energy.  It was also pointed out that 
the relaxation path, i.e., which of kinetic or magnetic energies
decreases earlier, can affect the resultant stationary state.
This subtlety arises because the low-beta reduced MHD has multiple
fields to be relaxed.   As explained in the discussion section of the present paper, 
 each Casimir invariant should be adjusted to have a desired value prior to the time evolution of SA, since the value
does not change during the time evolution.  A method for the adjustment was developed in \cite{Chikasue-2015-JFM}.  

Next,  SA  was applied to   low-beta reduced MHD in a cylindrical plasma  in \cite{Furukawa-2017}. 
By performing SA with an initial condition that is 
a sum  of a cylindrically symmetric equilibrium and a
small-amplitude helical perturbation accompanying magnetic islands, 
an equilibrium with magnetic islands was obtained.  In further work, toroidal equilibria were calculated by SA in 
\cite{Furukawa-2018} by using the high-beta reduced MHD model 
\citep{Strauss-1977}.  An example described therein was that of an  axisymmetric tokamak equilibria with a large aspect ratio and a circular cross section.  The Shafranov shift was shown to increase
as  beta was increased, although the magnitude of the shift did not
fully agree  with the analytic theory based on the
large-aspect-ratio expansion.  This was because the toroidicity
completely disappears in  high-beta reduced MHD, while it remains in
the analytic theory.  Some equilibria with poloidal rotation were also calculated by SA, and
examined based on a mapping between such equilibria with poloidal rotation
and static equilibria.  Toroidally-averaged stellarator equilibria were also calculated.

Simulated annealing can be used not only for equilibrium calculations
but also for stability analyses \citep{Furukawa-2022}.  
We know that equilibria obtained by SA that decreases the total energy
of the system are stable at least linearly since they locate at energy
minima.  However, equilibria that are obtained by other methods,
such as solving the Grad-Shafranov equation 
\citep{Lust-1957, Grad-1958, Shafranov-1958},
are not necessarily stable.  
Suppose we know such an equilibrium, and we  perform SA starting from
an initial condition that is a sum  of the known equilibrium and a
small-amplitude perturbation.
If SA recovers the original equilibrium, it is linearly stable.
However, if the perturbation grows during the  time evolution of SA, the
equilibrium is not at an energy minimum. 

In the numerical demonstration of the stability analyses, 
it was shown that the perturbation grows in a short time if the
equilibrium is unstable.  On the other hand, SA required a long time for
recovering the original equilibrium if it is stable.  Therefore,
accelerated 
relaxation is indispensable for SA to be practically useful.
In \citet{Furukawa-2022}, a method for accelerating the relaxation was
developed by introducing time dependence in the symmetric kernel of the
double bracket. 

Another kind of SA based on a metriplectic bracket introduced in \citet{pjm84,pjm86} \citep[see][for recent work and historical summary]{pjmU24} has also been studied extensively in  \citet{Bressan-2018, Bressan-2023}.  
Metriplectic dynamics is  a 
combination of the Hamiltonian dynamics and dissipative dynamics.
The dissipative mechanism is realized by a metric bracket.
The metriplectic dynamics was shown to successfully obtain 
equilibria of two-dimesional Euler flow, axisymmetric toroidal
equilibria that are a solution to the Grad--Shafranov equation,
and force-free MHD equilibria.
Metriplectic dynamics is also explained in 
\cite{Morrison-2017}; this paper covers wider topics on geometric
aspects of plasma physics and numerical algorithms for them.

The rest of present paper is organized as follows.
In Sec.~\ref{sec:HamiltonianSystem},
  Hamiltonian theory is reviewed for systems of both finite and
infinite degrees of freedom.
It starts from general theory, then proceeds to some examples
such as a free rigid body and the  heavy top.
A toy model mimicking aspects of low-beta reduced MHD is also introduced.
For systems with infinite degrees of freedom, 
two-dimensional Euler flow, low-beta reduced MHD in both a two-dimensional
rectangular domain and in a cylindrical geometry, and high-beta reduced
MHD are considered.
Then, the  theory of SA is explained in Sec.~\ref{sec:SA}.
It reviews the double bracket formulation of SA for systems with both
finite and infinite degrees of freedom.  SA by  metriplectic brackets is also described briefly.
Section~\ref{sec:SAFinDoF} is devoted to  some numerical examples of
SA for the  heavy top and a toy model mimicking low-beta reduced MHD.
Analyses of equilibrium and stability are also presented.
Sections \ref{sec:linearStability} to \ref{sec:acceleration} cover numerical studies of SA for low-beta and high-beta reduced MHD.  Section~\ref{sec:linearStability} is on   linear stability 
analyses using SA, while Sec.~\ref{sec:toroidal} is on the equilibrium 
calculations in toroidal geometry.
Section~\ref{sec:helical} shows that helically-deformed equilibria can
be obtained by SA of low-beta reduced MHD in cylindrical geometry.
Section~\ref{sec:superAlfvenic} describes our  numerical studies  of flowing
equilibria in two-dimensional rectangular domain.  
An equilibrium with magnetic islands is  introduced in 
Sec.~\ref{sec:island}.
Two methods for accelerated relaxation are described  in 
Sec.~\ref{sec:acceleration}.
Section~\ref{sec:discussion} contains  discussion on several issues that remain to
be solved.   Finally, our  summary and conclusions are given in 
Sec.~\ref{sec:summaryConclusion}.

\section{Hamiltonian systems}
\label{sec:HamiltonianSystem}

In Sec.~\ref{sec:HamiltonianSystem}, 
the theory of Hamiltonian systems of finite and infinite degrees of
freedom is reviewed.
Section~\ref{subsec:HamiltonianFinDoF} is on systems with finite degrees
of freedom.  Starting from a canonical case, a noncanonical case is
briefly introduced.  Explicit examples are the  free rigid body, the  heavy
top \citep[see][]{sudarshan} and a toy model mimicking low-beta reduced MHD.
Section \ref{subsec:HamiltonianInfDoF} describes Hamiltonian theory of
infinite-dimensional systems such as two-dimensional Euler flow and  low-
and high-beta reduced MHD.

\subsection{System with finite degrees of freedom}
\label{subsec:HamiltonianFinDoF}

\subsubsection{General theory}
\label{subsubsec:HamiltonianFinDoFGeneral}

A canonical Hamiltonian system is governed by 
Hamilton's equations
\begin{equation}
 \dot{q}^{i}=
 \frac{\pd H(\vq, \vp)}{\pd p_{i}}
\qquad \mathrm{and}\qquad
 \dot{p}_{i} =
 -\frac{\pd H(\vq, \vp)}{\pd q^{i}},
\label{eq:HamiltonEq-pi}
\end{equation}
with $i = 1, 2, \cdots, N$, where 
$\vq = ( q^{1}, q^{2}, \cdots, q^{N} )^{\mathsf{T}}$
and
$\vp = ( p_{1}, p_{2}, \cdots, p_{N} )^{\mathsf{T}}$
are canonical coordinates and canonical momenta of a system with $N$
degrees of freedom, respectively, 
$H(\vq, \vp)$ is a Hamiltonian,
and a dot $\dot{\,\,}$ denotes time derivative.
Defining a Poisson bracket as
\begin{equation}
 [ f , g  ]
\coloneqq 
 \frac{\pd f}{\pd q^{i}} \frac{\pd g}{\pd p_{i}} 
- \frac{\pd f}{\pd p_{i}} \frac{\pd g}{\pd q^{i}},
\label{eq:canonicalPoissonBracket}
\end{equation}
where $f(\vq, \vp)$ and $g(\vq, \vp)$ are arbitrary functions,
the canonical equations are written as
\begin{align}
 \dot{q}^{i}
&=
 [ q^{i} , H ].
\label{eq:HamiltonEqPoissonBracket-qi}
\\
 \dot{p}_{i}
&=
 [ p_{i} , H ],
\label{eq:HamiltonEqPoissonBracket-pi}
\end{align}
These equations are rewritten by introducing
phase space coordinates 
$\vz := ( z^{1}, z^{2}, \cdots, z^{2N} )$ 
with
$z^{i} = q^{i}$ for $i = 1, 2, \cdots , N$
and 
$z^{i} = p_{i - N}$ for $i = N+1, N+2, \cdots , 2N$
as
\begin{equation}
 \dot{z}^{i}
=
 [ z^{i} , H ].
\label{eq:HamiltonEqPoissonBracket-zi}
\end{equation}
Further, by introducing a canonical Poisson tensor as
\begin{equation}
 J_{\mathrm{c}} 
\coloneqq
 \begin{pmatrix}
  0_{N} & I_{N}
\\
 -I_{N} & 0_{N}
 \end{pmatrix}
\label{eq:canonicalPoissonTensor}
\end{equation}
where $0_{N}$ and $I_{N}$ are $N \times N$ zero and unit matrices,
respectively, 
the Poisson bracket is expressed as 
\begin{equation}
 [ f , g ]
=
 \frac{\pd f}{\pd z^{i}}
 J_{\mathrm{c}}^{ij}
 \frac{\pd g}{\pd z^{j}},
\label{eq:canonicalPoissonBracket-z}
\end{equation}
and the canonical equations are rewritten as
\begin{equation}
 \dot{z}^{i} 
=
 J_{\mathrm{c}}^{ij} \frac{\pd H(\vz)}{\pd z^{j}}.
\label{eq:HamiltonEqCanonicalPoissonTensor-zi}
\end{equation}

By changing the phase space variables from $\vz$ to $\bar{\vz}$ as 
$\bar{z}^{i} = \bar{z}^{i}(\vz)$,
  Hamilton's equations (\ref{eq:HamiltonEqCanonicalPoissonTensor-zi})
become
\begin{equation}
 \dot{\bar{z}}^{i}
=
 J^{ij}(\bar{\vz}) \frac{\pd \bar{H}(\bar{\vz})}{\pd \bar{z}^{j}},
\label{eq:HamiltonEqPoissonTensor-zi}
\end{equation}
where the Hamiltonian is transformed to $\bar{H}(\bar{\vz})$,
and the Poisson tensor is transformed to 
\begin{equation}
J ( \bar{\vz} )
=
 ( J^{ij}( \bar{\vz} ) )
=
 \left(
   \frac{\pd \bar{z}^{i}}{\pd z^{k}}
   J_{\mathrm{c}}^{k \ell}
   \frac{\pd \bar{z}^{j}}{\pd z^{\ell}}
 \right).
\label{eq:PoissonTensor}
\end{equation}
The Poisson tensor (\ref{eq:PoissonTensor}) is antisymmetric 
by definition, but it does not have 
canonical form when $\bar{\vz}$ are noncanonical coordinates. 
Equation~(\ref{eq:HamiltonEqPoissonTensor-zi}) can also be written as
\begin{equation}
 \dot{\bar{z}}^{i}
=
 [ \bar{z}^{i} , \bar{H} ],
\label{eq:HamiltonEqPoissonBracket-zbari}
\end{equation}
where the Poisson bracket is given by
\begin{equation}
 [ f, g ]
=
 \frac{\pd f}{\pd \bar{z}^{i}}
 J^{ij}
 \frac{\pd g}{\pd \bar{z}^{j}}.
\label{eq:PoissonBracket-zbar}
\end{equation}

Let us now consider a dynamical system that need not be generated by a transformation such as that above. This system is governed by
\begin{align}
\label{eq:evolutionEqGeneralNonCanonicalPoissonTensor}
 \dot{u}^{i}
=& 
 J^{ij}(\vu) \frac{\pd H(\vu)}{\pd u^{j}}
= 
 [ u^{i}, H(\vu) ],
\\
 [ f, g ]
\coloneqq& \,
 \frac{\pd f}{\pd u^{i}} J^{ij} \frac{\pd g}{\pd u^{j}}, 
\label{eq:PoissonBracketGeneralNonCanonical}
\end{align}
where $\vu \coloneqq ( u^{1}, u^{2}, \cdots, u^{M} )^{\mathsf{T}}$
is a vector of noncanonical variables of  an  $M$-dimensional phase space,
$H(\vu)$ is a Hamiltonian, 
$J(\vu) := ( J^{ij}(\vu) )$ is an antisymmetric 
Poisson tensor,
and $[ f , g ]$ is the Poisson bracket for arbitrary functions 
$f(\vu)$ and $g(\vu)$.
The dimension $M$ of the phase space can be odd.
If $\mathrm{rank}\, J = 2 N < M$, 
the Poisson tensor $J$ has a $(M - 2N)$-dimensional null space.
The eigenvectors of the zero eigenvalues determine directions in which
the system cannot evolve.  
Surfaces perpendicular to the eigenvectors define Casimir invariants.
The Casimir invariants $C_{k}(\vu)$ ($k = 1, 2, \cdots M - 2N$) satisfy
\begin{equation}
 J^{ij} \frac{\pd C_{k}}{\pd u^{j}} = 0.
\label{eq:CasimirGeneralNonCanonical}
\end{equation}
The gradient of $C_{k}$ points in  the direction that the system is prohibited
from  evolving.

The dynamics are not affected even if we plug in an energy-Casimir
function
\begin{equation}
 F(\vu) \coloneqq H(\vu) + \lambda_{k} C_{k}(\vu)
\label{eq:energyCasimirGeneralNonCanonical}
\end{equation}
into the evolution Eq.~(\ref{eq:evolutionEqGeneralNonCanonicalPoissonTensor}).
Here $\lambda_{k}$ are Lagrange multipliers.  
The evolution equation reads
\begin{equation}
 \dot{u}^{i}
=
 J^{ij}(\vu) \frac{\pd F(\vu)}{\pd u^{j}}.
\label{eq:evolutionEqGeneralNonCanonicalEnergyCasimir}
\end{equation}
Equilibria of this system are given by
\begin{equation}
\frac{\pd F(\vu)}{\pd u^{j}} = 0.
\label{eq:equilibriumGeneralNonCanonicalEnergyCasimir}
\end{equation}
For an equilibrium $\vu_{\mathrm{e}}$ given by 
Eq.~(\ref{eq:equilibriumGeneralNonCanonicalEnergyCasimir}), the 
linearized equations are given by 
\begin{equation}
\delta   \dot{u}^{i}
=
 J^{ij} (\vu_{\mathrm{e}})
 \frac{\pd^{2} F}{\pd u^{j} \pd u^{k}} (\vu_{\mathrm{e}})
 \delta u^{k},
\label{eq:linEvoEqGeneralNonCanonicalEnergyCasimir}
\end{equation}
where $\delta u^{i}$ is a perturbation away from equilibrium.
By assuming the time dependence of the perturbation is 
$\delta u^{i} = \delta \tilde{u}^{i} \rme^{-\rmi \omega t}$
with $\delta \tilde{u}^{i}$ being a constant, 
  linear stability can be analyzed by 
solving the following eigenvalue problem: 
\begin{equation}
 -\rmi \omega \delta \tilde{u}^{i}
=
 J^{ij} (\vu_{\mathrm{e}})
 \frac{\pd^{2} F}{\pd u^{j} \pd u^{k}} (\vu_{\mathrm{e}})
 \delta \tilde{u}^{k}.
\label{eq:EigEqGeneralNonCanonicalEnergyCasimir}
\end{equation}

Lastly in the present  
Sec.~\ref{subsubsec:HamiltonianFinDoFGeneral},
we define the  energy of a  linearized mode as
\begin{equation}
 \delta^{2} H
\coloneqq
 \frac{1}{2} 
 \frac{\pd^{2} F}{\pd u^{j} \pd u^{k}} (\vu_{\mathrm{e}})
 \delta u^{j} \delta u^{k},
\label{eq:linModeEnergyGeneralNonCanonicalEnergyCasimir}
\end{equation}
where $\delta u^{j}$ is an eigenvector 
of the eigenvalue problem~(\ref{eq:EigEqGeneralNonCanonicalEnergyCasimir}).
The time derivative of $\delta^{2} H$ is easily seen to be zero, 
\begin{align}
 \frac{\td \delta^{2} H}{\td t}
&=
 \frac{1}{2} 
 \frac{\pd^{2} F}{\pd u^{j} \pd u^{k}} (\vu_{\mathrm{e}})
 ( \delta \dot{u}^{j} \delta u^{k} + \delta u^{j} \delta \dot{u}^{k} )
=
 \frac{\pd^{2} F}{\pd u^{j} \pd u^{k}} (\vu_{\mathrm{e}})
 \delta u^{j} \delta  \dot{u}^{k}
\nonumber
\\
&=
 \frac{\pd^{2} F}{\pd u^{j} \pd u^{k}} (\vu_{\mathrm{e}})
 \delta u^{j} 
 J^{k \ell} (\vu_{\mathrm{e}})
 \frac{\pd^{2} F}{\pd u^{\ell} \pd u^{i}} (\vu_{\mathrm{e}})
 \delta u^{i}
\nonumber
\\
&=
 J^{k \ell} (\vu_{\mathrm{e}})
 \left(
   \frac{\pd^{2} F}{\pd u^{k} \pd u^{j}} (\vu_{\mathrm{e}})
   \delta u^{j} 
 \right) 
 \left(
   \frac{\pd^{2} F}{\pd u^{\ell} \pd u^{i}} (\vu_{\mathrm{e}})
   \delta u^{i}
 \right)= 0,
\label{eq:conservationLinModeEnergyGeneralNonCanonicalEnergyCasimir}
\end{align}
where the symmetry of the Hessian matrix 
$\left( \pd^{2} F / ( \pd u^{k} \pd u^{j} ) \right)$ 
and the antisymmetry of $J$
were used.
Therefore, as a measure of the mode energy for systems with finite
degrees of freedom,
we adopt 
\begin{equation}
 \tilde{H}
 \coloneqq
 \frac{\delta^{2} H }
      { \frac{1}{2} | \delta \vu |^{2} }.
\label{eq:mesureLinModeEnergyGeneralNonCanonicalEnergyCasimir}
\end{equation}

\subsubsection{Free rigid body}
\label{subsubsec:HamiltonianFinDoFFRB}

We sometimes find a set of variables that forms a closed
subset with a proper Poisson bracket in the $2N$-dimensional phase
space.  This is called reduction.
An example is rotational dynamics of the  free rigid body.  
If we choose angular momenta as the dynamical variables, we obtain a 
three-dimensional phase space, where 
the Hamiltonian, the Poisson bracket, and the evolution equations are
given by 
\begin{align}
 H ( \vL ) 
\coloneqq &\, 
 \frac{1}{2} \sum_{i = 1}^{3} \frac{ L_{i}^{2} }{ I_{i} },
\label{eq:frbHamiltonian}
\\
 [ f( \vL ) , g( \vL )  ]
\coloneqq &\,
 -\eps_{ijk} L_{k} \frac{\pd f}{\pd L_{i}} \frac{\pd g}{\pd L_{j}},
\label{eq:frbPoissonBracket}
\\
 \dot{L}_{i}
= &\,
 [ L_{i} , H ],
\label{eq:frbEvolutionEqPoissonBracket}
\end{align}
respectively.
Here, $\vL \coloneqq  ( L_{1}, L_{2}, L_{3} )^{\mathsf{T}}$ is the angular momenta,
$I_{i} \, (i = 1, 2, 3)$ are the principal moments of inertia in a frame
fixed to the 
rigid body, $f$ and $g$ are arbitrary functions of $\vL$, 
and $\eps_{ijk}$ is the Levi-Civita symbol.
The Poisson tensor is given by $J = ( J_{ij} )$ with $
J_{ij} = [ L_{i} , L_{j} ] = -\eps_{ijk} L_{k}$, or
\begin{equation}
 J
=
 \begin{pmatrix}
  0 & -L_{3} & L_{2} \\
  L_{3} & 0 & -L_{1} \\
  -L_{2} & L_{1} & 0
 \end{pmatrix}.
\label{eq:frbPoissonTensor}
\end{equation}
Using the Poisson tensor, the evolution equations can be written as
\begin{equation}
 \dot{L}_{i} = J_{ij} \frac{\pd H}{\pd L_{j}}.
\label{eq:frbEvolutionEqPoissonTensor}
\end{equation}
Since the phase space is odd dimensional, the determinant of the Poisson
tensor is zero and consequently there must be a Casimir invariant.   The system cannot evolve in the direction of the null
space of the Poisson tensor.  This degeneracy of the Poisson tensor defines
a Casimir invariant, which is $| \vL |$ in the present case.
Therefore, $C( |\vL| )$ is conserved by  the dynamics where $C$ is an
arbitrary function.

\subsubsection{Heavy top}
\label{subsubsec:htHamiltonianFinDoF}

Another example  of noncanonical dynamics is that of the  heavy top.  
A unit vector in the direction opposite to the gravitational
acceleration is taken to be $\vecrho = ( \rho_{1}, \rho_{2}, \rho_{3} )$,
where the components are taken in a frame fixed to the top.
Then the phase space variables are 
$\vu = ( u_{1}, u_{2}, \cdots, u_{6} )^{\mathsf{T}} 
= ( L_{1}, L_{2}, L_{3}, \rho_{1}, \rho_{2}, \rho_{3} )^{\mathsf{T}}$.
The Hamiltonian, the Poisson bracket, the Poisson tensor, and the
evolution equations are given by
\begin{align}
 H ( \vu ) 
\coloneqq &\, 
 \frac{1}{2} \sum_{i = 1}^{3} \frac{ L_{i}^{2} }{ I_{i} }
 + G \rho_{3},
\label{eq:htHamiltonian}
\\
 [ f( \vu ) , g( \vu )  ]
\coloneqq  &\,
 -\eps_{ijk} L_{k} \frac{\pd f}{\pd L_{i}} \frac{\pd g}{\pd L_{j}}
 - \eps_{ijk} \rho_{k} 
   \left(
      \frac{\pd f}{\pd L_{i}} \frac{\pd g}{\pd \rho_{j}}
    - \frac{\pd g}{\pd L_{i}} \frac{\pd f}{\pd \rho_{j}}
   \right),
\label{eq:htPoissonBracket}
\\
 J
&=
 ( [ u_{i} , u_{j} ] )
\nonumber
\\
&=
 \begin{pmatrix}
   0 & -L_{3} & L_{2} & 0 & -\rho_{3} & \rho_{2}
   \\
   L_{3} & 0 & -L_{1} & \rho_{3} & 0 & -\rho_{1} 
   \\
   -L_{2} & L_{1} & 0 & -\rho_{2} & \rho_{1} & 0
   \\
   0 & -\rho_{3} & \rho_{2} & 0 & 0 & 0
   \\
   \rho_{3} & 0 & -\rho_{1} & 0 & 0 & 0
   \\
   -\rho_{2} & \rho_{1} & 0 & 0 & 0 & 0
 \end{pmatrix},
\label{eq:htPoissonTensor}
\\
 \dot{u}_{i}
= &\,
 [ u_{i} , H ]
\label{eq:htEvolutionEqPoissonBracket}
= 
 J_{ij} \frac{\pd H}{\pd u_{j}},
\end{align}
respectively.  A measure of the effect of  gravity is expressed by the parameter 
$G = mg\ell$, 
where $m$ is the mass, $g$ is the magnitude of the gravitational
acceleration, and $\ell$ is the the distance of the center of mass of
from the fixed point of the top.

\begin{figure}[h]
 \centering
 \includegraphics[width = 0.7\textwidth]{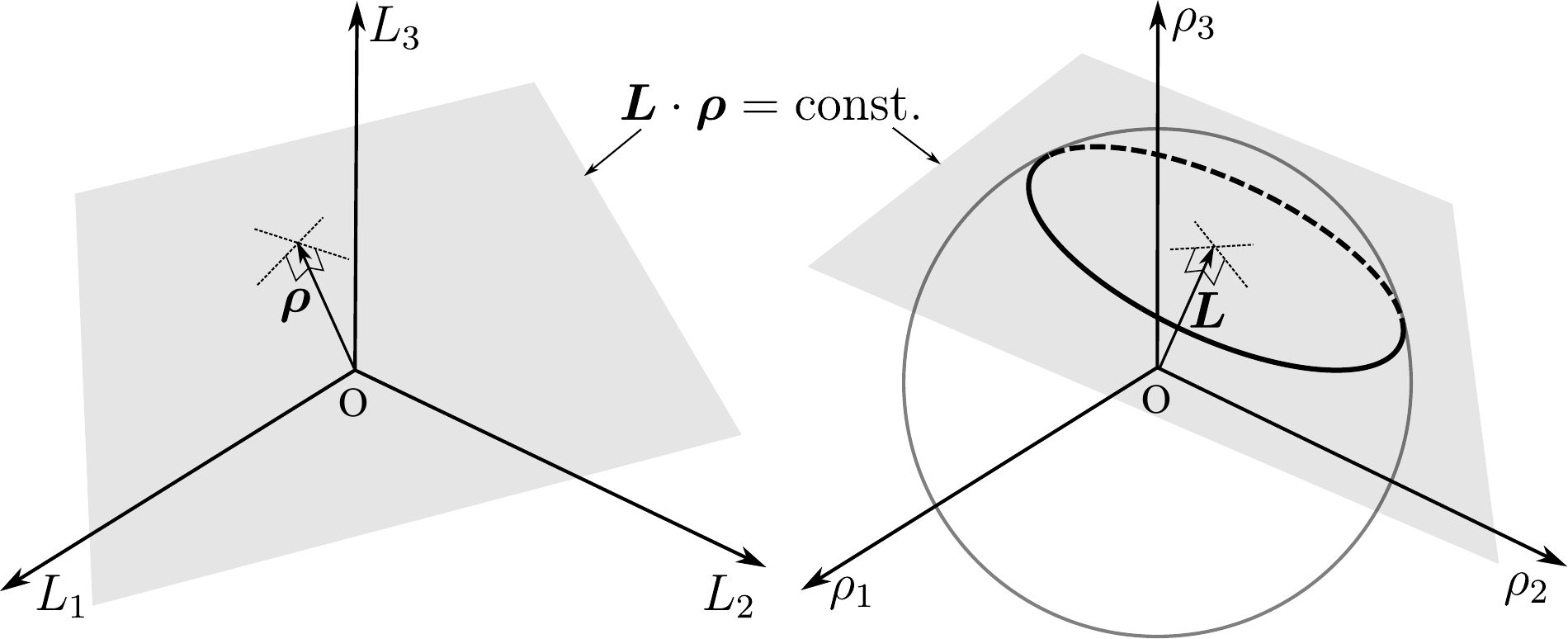}
 \bigskip
 \caption{A phase space depiction of the dynamics of the  heavy top.
The dynamics is restricted to  a  four-dimensional subspace in the
 six-dimensional phase space as shown by the gray plane $\vL \cdot \vecrho = \mathrm{const.}$ in the
 $\vL$-space and the spherical surface 
$| \vecrho | = \mathrm{const.}$ in the $\vecrho$-space.
The $\vecrho$ vector can change only its direction,
but the $\vL$ vector can change both its direction and magnitude.}
 \label{fig:htPhaseSpace}
\end{figure}

The Casimir invariants are given by 
\begin{align}
 C_{1}
\coloneqq &\,
 C_{1}( | \vecrho |^{2} / 2 ),
\label{eq:htCasimir1}
\\
 C_{2}
\coloneqq  &\,
 C_{2}( \vL \cdot \vecrho ).
\label{eq:htCasimir2}
\end{align}
The phase space is depicted in  Fig.~\ref{fig:htPhaseSpace}, which shows 
the dynamics restricted, because of the constancy of $C_{1,2}$,  to a  four-dimensional subspace in the
six-dimensional phase space.
Moreover, the  system follows a trajectory that conserves the energy in the
four-dimensional subspace. Two dimensions are drawn as the plane 
$\vL \cdot \vecrho = \mathrm{const.}$ in the $\vL$-space.
The other two dimensions are the surface of the sphere 
$| \vecrho | = \mathrm{const.}$ drawn in the $\vecrho$-space.
In the $\vL$-space, the direction of $\vecrho$ changes, while 
$| \vecrho |$ does not change.  Therefore, the distance of the 
plane $\vL \cdot \vecrho = \mathrm{const.}$ from the origin does not
change.  Similarly, $\vL$ drawn in the $\vecrho$-space can change both  
in  direction and the magnitude.  The intersection of the plane 
$\vL \cdot \vecrho = \mathrm{const.}$ and the sphere 
$| \vecrho | = \mathrm{const.}$ changes in time.
However, since the distance of the plane $\vL \cdot \vecrho =
\mathrm{const.}$ from the origin in the $\vecrho$-space is smaller than
or equals to $| \vecrho |$ according to 
\begin{equation}
 \frac{\vL \cdot \vecrho}{| \vL |}
\leq 
 \frac{ | \vL |  | \vecrho |}{| \vL |}
= | \vecrho |,
\end{equation}
the intersection always exists.

For later use, equilibria and stability of the heavy top 
are briefly summarized in the remainder  of  this 
Sec.~\ref{subsubsec:htHamiltonianFinDoF}.
To this end, let us define an energy-Casimir function $F$ as
\begin{equation}
 F := H + \lambda_{1} C_{1} + \lambda_{2} C_{2} ,
\label{eq:htEnergyCasimirFunction}
\end{equation}
where the Hamiltonian $H$ is given by
Eq.~(\ref{eq:htHamiltonian}), the 
two Casimir invariants $C_{i}$ ($i = 1, 2$)   by 
Eqs.~(\ref{eq:htCasimir1}) and (\ref{eq:htCasimir2}),
and the $\lambda_{i}$ are the Lagrange multipliers.

The first partial derivatives of $F$ are  given by
\begin{align}
 \frac{\pd F}{\pd L_{i}}
&=
 \frac{L_{i}}{I_{i}} 
 + \lambda_{2} C_{2}^{\pr}( \vL \cdot \vecrho ) \rho_{i},
\label{eq:htEnergyCasimirFirstDerivative1}
\\
 \frac{\pd F}{\pd \rho_{i}}
&=
 G \delta_{i3} 
 + \lambda_{1} C_{1}^{\pr}( | \vecrho |^{2} / 2 ) \rho_{i}
 + \lambda_{2} C_{2}^{\pr}( \vL \cdot \vecrho ) L_{i},
\label{eq:htEnergyCasimirFirstDerivative2}
\end{align}
where the 
$L_{i} / I_{i}$ term is not summed over  $i$, 
and $\delta_{i3}$ is used only for the index of $\rho_{i}$.
The prime denotes the derivative with respect to the argument, which
will not be written explicitly hereafter.
Equilibria are given by setting the first derivatives of $F$ to   zero.
Since the   parameter $G$ only appears when $i = 3$, equilibria
may be classified into two categories.
One is equilibria with $\rho_{3} \neq 0$,
and the other is equilibria with $\rho_{1} \neq 0$ or 
$\rho_{2} \neq 0$.
Let us explain them one by one.

When $\rho_{3} \neq 0$, 
Eqs.~(\ref{eq:htEnergyCasimirFirstDerivative1})
and (\ref{eq:htEnergyCasimirFirstDerivative2})
with $i = 3$ can be solved to obtain
\begin{align}
 \lambda_{1}
&=
 -\frac{1}{C_{1}^{\pr} \rho_{3}}
 \left(
   G - \frac{L_{3}^{2}}{I_{3} \rho_{3}}
 \right),
\label{eq:htEnergyCasimirRho3Lambda1}
\\
 \lambda_{2}
&=
 -\frac{L_{3}}{C_{2}^{\pr} I_{3} \rho_{3}}.
\label{eq:htEnergyCasimirRho3Lambda2}
\end{align}
Here, we assumed that 
$C_{1}^{\pr} \neq 0$ 
and 
$C_{2}^{\pr} \neq 0$.
From 
Eqs.~(\ref{eq:htEnergyCasimirFirstDerivative1}), 
(\ref{eq:htEnergyCasimirFirstDerivative2}), 
and $\lambda_{i}$ in 
Eqs.~(\ref{eq:htEnergyCasimirRho3Lambda1})
and (\ref{eq:htEnergyCasimirRho3Lambda2}),
we obtain
\begin{equation}
 \left(
  \begin{array}{cc}
   \frac{1}{I_{1}}
   & 
   -\frac{L_{3}}{I_{3} \rho_{3}}
   \\
   - \frac{L_{3}}{I_{3}}
   &
   - G + \frac{L_{3}}{I_{3} \rho_{3}}
  \end{array}
 \right)
 \left(
  \begin{array}{c}
   L_{1}
   \\ 
   \rho_{1}
  \end{array}
 \right)
=
 \left(
  \begin{array}{c}
   0
   \\
   0
  \end{array}
 \right)
\end{equation}
for $i = 1$.
The determinant of the $2 \times 2$ matrix on the left-hand side is
\begin{equation}
 -\frac{G}{I_{1}}
 + \frac{L_{3}^{2}}{I_{3} \rho_{3}}
   \left(
    \frac{1}{I_{2}} - \frac{1}{I_{3}}
   \right),
\end{equation}
which  is generally not zero.
Therefore we obtain $L_{1} = \rho_{1} = 0$.
Similarly, we obtain $L_{2} = \rho_{2} = 0$ 
from $\pd F / \pd L_{2} = 0$ and $\pd F / \pd \rho_{2} = 0$.

When $\rho_{1} \neq 0$, on the other hand,
$\lambda_{i}$ are obtained from 
Eqs.~(\ref{eq:htEnergyCasimirFirstDerivative1})
and (\ref{eq:htEnergyCasimirFirstDerivative2})
with $i = 1$ as
\begin{equation}
 \lambda_{1} =
 \frac{L_{1}^{2}}{C_{1}^{\pr} I_{1} \rho_{1}^{2}},
\label{eq:htEnergyCasimirRho1Lambda1}
\qquad 
 \lambda_{2}
=
 -\frac{L_{1}}{C_{2}^{\pr} I_{1} \rho_{1}}.
\end{equation}
Then 
Eqs.~(\ref{eq:htEnergyCasimirFirstDerivative1})
and (\ref{eq:htEnergyCasimirFirstDerivative2})
with $i = 2$ 
yield
\begin{equation}
 \left(
  \begin{array}{cc}
   \frac{1}{I_{2}}
   & 
   -\frac{L_{1}}{I_{1} \rho_{1}}
   \\
   - \frac{L_{1}}{I_{1}}
   &
    \frac{L_{1}}{I_{1} \rho_{1}}
  \end{array}
 \right)
 \left(
  \begin{array}{c}
   L_{2}
   \\ 
   \rho_{2}
  \end{array}
 \right)
=
 \left(
  \begin{array}{c}
   0
   \\
   0
  \end{array}
 \right).
\end{equation}
The determinant of the $2 \times 2$ matrix is
\begin{equation}
  \frac{L_{1}^{2}}{I_{1} \rho_{1}}
  \left(
   \frac{1}{I_{2}} - \frac{1}{I_{1}}
  \right),
\end{equation}
which  is generally not zero.
Therefore, we obtain $L_{2} = \rho_{2} = 0$.
From 
Eqs.~(\ref{eq:htEnergyCasimirFirstDerivative1})
and (\ref{eq:htEnergyCasimirFirstDerivative2})
with $i = 3$, 
we obtain 
\begin{equation}
 \left(
  \begin{array}{cc}
   \frac{1}{I_{3}}
   & 
   -\frac{L_{1}}{I_{1} \rho_{1}}
   \\
   - \frac{L_{1}}{I_{1}}
   &
   \frac{L_{1}}{I_{1} \rho_{1}}
  \end{array}
 \right)
 \left(
  \begin{array}{c}
   L_{3}
   \\ 
   \rho_{3}
  \end{array}
 \right)
=
 \left(
  \begin{array}{c}
   0
   \\
   -G
  \end{array}
 \right).
\end{equation}
Except for cases where the determinant of this $2 \times 2$ matrix
vanishes, 
we obtain
\begin{equation}
 \left(
  \begin{array}{c}
   L_{3}
   \\ 
   \rho_{3}
  \end{array}
 \right)
=
 \frac{-G \rho_{1}}
      {  \frac{L_{1}^{2}}{I_{1}}
         \left(
           \frac{1}{I_{3}} - \frac{1}{I_{1}}
         \right)
      }
 \left(
  \begin{array}{c}
   \frac{L_{1}}{I_{1}}
   \\ 
   \frac{\rho_{1}}{I_{3}}
  \end{array}
 \right).
\end{equation}
Therefore $L_{3} \neq 0$ and $\rho_{3} \neq 0$ for 
the equilibrium with $\rho_{1} \neq 0$.

Similarly, when $\rho_{2} \neq 0$, we obtain
\begin{align}
 \lambda_{1}
&=
 \frac{L_{2}^{2}}{C_{1}^{\pr} I_{2} \rho_{2}^{2}},
\qquad 
 \lambda_{2}
=
 -\frac{L_{2}}{C_{2}^{\pr} I_{2} \rho_{2}},
\label{eq:htEnergyCasimirRho2Lambda2}
\\
 L_{1}
&=
 0,
\qquad
 \rho_{1}
=
 0,
\\
 \left(
  \begin{array}{c}
   L_{3}
   \\ 
   \rho_{3}
  \end{array}
 \right)
&=
 \frac{-G \rho_{2}} 
      {  \frac{L_{2}^{2}}{I_{2}}
         \left(
           \frac{1}{I_{3}} - \frac{1}{I_{2}}
         \right)
      }
 \left(
  \begin{array}{c}
   \frac{L_{2}}{I_{2}}
   \\ 
   \frac{\rho_{2}}{I_{3}}
  \end{array}
 \right).
\end{align}

Linear stability of these equilibria can be examined 
by solving 
Eq.~(\ref{eq:linEvoEqGeneralNonCanonicalEnergyCasimir})
or
Eq.~(\ref{eq:EigEqGeneralNonCanonicalEnergyCasimir}).

\subsubsection{A toy model mimicking low-beta reduced MHD}
\label{subsubsec:HamiltonianFinDoFMimicLBRMHD}

We  propose a toy model that tries to mimic features of low-beta reduced MHD (see 
 Sec.~\ref{subsec:HamiltonianInfDoFRectLBRMHD}). The toy model is based on the heavy top 
presented in Sec.~\ref{subsubsec:htHamiltonianFinDoF}, but with a new Hamiltonian  taken to be
\begin{equation}
  H(\vu)
=
 \frac{1}{2} \left(
    \frac{L_{1}^{2}}{I_{1}}
  + \frac{L_{2}^{2}}{I_{2}}
  + \frac{L_{3}^{2}}{I_{3}}
             \right)
 + \frac{1}{2} \left( 
       M_{1} \rho_{1}^{2}
     + M_{2} \rho_{2}^{2}
     + M_{3} \rho_{3}^{2}
     \right).
\label{eq:mimicLBRMHDHamiltonian}
\end{equation}
As presented in 
Sec.~\ref{subsec:HamiltonianInfDoFRectLBRMHD},
the Hamiltonian of low-beta reduced MHD in two dimensions 
is composed of kinetic and magnetic energy terms.
In the Hamiltonian (\ref{eq:mimicLBRMHDHamiltonian}), 
the terms quadratic in $L_{i}$,  being kinetic in origin, mimic the corresponding kinetic energy  of reduced MHD, while the terms quadratic in $\rho_{i}$ mimic  magnetic energy.


Because the Poisson bracket is assumed to be the same, the  Casimir invariants are the same as those of the  original heavy top, i.e., those of \eqref{eq:htPoissonBracket}.  With these ingredients, the energy-Casimir function $F$ is thus given by 
\begin{equation}
 F\coloneqq 
H + \lambda_{1} C_{1} + \lambda_{2} C_{2}, 
\label{eq:mimicLBRMHDEnergyCasimir}
\end{equation}
the evolution equation for this system is given by
\begin{equation}
 \dot{u}^{i}
=
 J^{ij} \frac{\pd F}{\pd u^{j}},
\label{eq:mimicLBRMHDEvolutionEq}
\end{equation}
where 
$\lambda_{1}$ and $\lambda_{2}$ are Lagrange multipliers, 
$\vu := ( L_{1}, L_{2}, L_{3}, \rho_{1}, \rho_{2}, \rho_{3} )^{\mathsf{T}}$,
and the Poisson tensor is given by 
Eq.~(\ref{eq:htPoissonTensor}).

Equilibria are obtained by setting to zero  the gradient of the energy-Casimir
function (\ref{eq:mimicLBRMHDEnergyCasimir}),   
\begin{align}
 \frac{\pd F}{\pd L_{i}}
&=
  \frac{L_{i}}{I_{i}}
+ \lambda_{2} C_{2}^{\pr}(\vL \cdot \vecrho) \rho_{i} =0,
\label{eq:mimicLBRMHDdFdL}
\\
 \frac{\pd F}{\pd \rho_{i}}
&=
 M_{i} \rho_{i}
+ \lambda_{1} C_{1}^{\pr}( | \vecrho |^{2} / 2 ) \rho_{i}
+ \lambda_{2} C_{2}^{\pr}(\vL \cdot \vecrho) L_{i}=0.
\label{eq:mimicLBRMHDdFdrho}
\end{align}
Note that $L_{i} / I_{i}$ term and $M_{i} \rho_{i}$ term are not summed over  $i$.
The prime denotes derivative with respect to the argument, which
will not be written explicitly hereafter.

Equations~(\ref{eq:mimicLBRMHDdFdL}) and (\ref{eq:mimicLBRMHDdFdrho})
for any of $i = 1$, $2$, or $3$ 
are written in a matrix form as
\begin{equation}
 \begin{pmatrix}
  \frac{1}{I_{i}} & \lambda_{2} C_{2}^{\pr}
\\
  \lambda_{2} C_{2}^{\pr} & M_{i} + \lambda_{1} C_{1}^{\pr}
 \end{pmatrix}
 \begin{pmatrix}
  L_{i}
  \\
  \rho_{i}
 \end{pmatrix}
=
 \begin{pmatrix}
  0
  \\
  0
 \end{pmatrix}.
\label{eq:mimicLBRMHDdFdLdrhodL-i}
\end{equation}
This equation has non-zero solution $(L_{i}, \rho_{i})^{\mathsf{T}}$
when the determinant of the $2 \times 2$ matrix on the left-hand side
vanishes, which leads to
\begin{equation}
 \lambda_{2}^{2}
=
 \frac{ M_{i} + \lambda_{1} C_{1}^{\pr} }{ I_{i} (C_{2}^{\pr})^{2} }.
\label{eq:mimicLBRMHDEquilibriumSolvableCondition}
\end{equation}
For some $j \neq i$, Eqs.~(\ref{eq:mimicLBRMHDdFdL}) and
(\ref{eq:mimicLBRMHDdFdrho}) are 
\begin{equation}
 \begin{pmatrix}
  \frac{1}{I_{j}} & \lambda_{2} C_{2}^{\pr}
\\
  \lambda_{2} C_{2}^{\pr} & M_{j} + \lambda_{1} C_{1}^{\pr}
 \end{pmatrix}
 \begin{pmatrix}
  L_{j}
  \\
  \rho_{j}
 \end{pmatrix}
=
 \begin{pmatrix}
  0
  \\
  0
 \end{pmatrix}.
\label{eq:mimicLBRMHDdFdLdrhodL-j}
\end{equation}
If $I_{j} \neq I_{i}$ and/or $M_{j} \neq M_{i}$,
the determinant of the $2 \times 2$ matrix in 
Eq.~(\ref{eq:mimicLBRMHDdFdLdrhodL-j}) does not vanish.
Therefore, we obtain $( L_{j}, \rho_{j} )^{\mathsf{T}} = \bv{0}$.

Now, we have four unknowns $L_{i}$, $\rho_{i}$, $\lambda_{1}$ and
$\lambda_{2}$.
First, we give values of $C_{1}$ and $C_{2}$, and then solve 
Eqs.~(\ref{eq:htCasimir1}) and (\ref{eq:htCasimir2}) 
for $L_{i}$ and $\rho_{i}$.
Then, we solve Eqs.~(\ref{eq:mimicLBRMHDdFdLdrhodL-i}) for
$\lambda_{1}$ and $\lambda_{2}$ to obtain
\begin{equation}
 \lambda_{1}
=
 - \frac{1}{C_{1}^{\pr} } 
 \left( M_{i} - \frac{ L_{i}^{2} }{ I_{i} \rho_{i}^{2} } \right) \qquad \mathrm{and}\qquad
 \lambda_{2}
=
 -\frac{ L_{i} }{ C_{2}^{\pr} I_{i} \rho_{i} } \,.
\label{eq:mimicLBRMHDLam1}
\end{equation}
Here, we assumed $C_{1}^{\pr} \neq 0$ and $C_{2}^{\pr} \neq 0$.
These $\lambda_{1}$ and $\lambda_{2}$ satisfy 
Eq.~(\ref{eq:mimicLBRMHDEquilibriumSolvableCondition}). 
Note that $\rho_{i}$ must not be zero.

Linear stability of these equilibria can be analyzed by  studying 
Eq.~(\ref{eq:linEvoEqGeneralNonCanonicalEnergyCasimir}) with the present $F$.
The $6 \times 6$ Hessian matrix 
$\pd^{2} F / ( \pd u^{j} \pd u^{k} )$
is explicitly obtained upon differentiating   Eqs.~(\ref{eq:mimicLBRMHDdFdL}) and 
(\ref{eq:mimicLBRMHDdFdrho}),  yielding 
\begin{align}
  \frac{\pd^{2} F}{\pd L_{i} \pd L_{j}}
&=
  \frac{1}{I_{i}} \delta_{ij}
+ \lambda_{2} C_{2}^{\pr\pr} \rho_{i} \rho_{j},
\label{eq:d2FdLdL}
\\
  \frac{\pd^{2} F}{\pd L_{i} \pd \rho_{j}}
&=
  \lambda_{2} \left(
    C_{2}^{\pr} \delta_{ij}
  + C_{2}^{\pr\pr} \rho_{i} L_{j}
              \right),
\label{eq:d2FdLdrho}
\\
 \frac{\pd^{2} F}{\pd \rho_{i} \pd \rho_{j}}
&=
 M_{i} \delta_{ij}
+ \lambda_{1} \left(
    C_{1}^{\pr} \delta_{ij}
  + C_{1}^{\pr\pr} \rho_{i} \rho_{j}
              \right)
+ \lambda_{2} C_{2}^{\pr\pr} L_{i} L_{j}.
\label{eq:d2Fdrhodrho}
\end{align}
Again, $\delta_{ij} / I_{i}$ in Eq.~(\ref{eq:d2FdLdL})
and $M_{i} \delta_{ij}$ in Eq.~(\ref{eq:d2Fdrhodrho}) are not summed over 
$i$. 

\subsection{System with Infinite degrees of freedom}
\label{subsec:HamiltonianInfDoF}

\subsubsection{Two-dimensional Euler flow}
\label{subsec:HamiltonianInfDoF2DEuler}

One of the simplest examples of a noncanonical Hamiltonian system with 
infinite dimensions is two-dimensional Euler fluid flow \citep[see][]{Morrison-1982}.
Suppose the two-dimensional velocity field $\vv(x,y,t)$ is given by 
$\vv = \hrvz \times \nab_{\perp} \vphi$,
where $\vphi(x,y,t)$ is the stream function,
$\nab_{\perp}$ is the gradient operator in the $x$--$y$ plane,  
and $\hrvz$ is the unit vector perpendicular to the $x$--$y$ plane.
The vorticity in the $z$ direction is given by
$U \coloneqq \hrvz \cdot \rot \vv = \bigtriangleup_{\perp} \vphi$,
where $\bigtriangleup_{\perp}$ is the Laplacian in the $x$--$y$ plane.
The governing equation of $U$ is 
\begin{equation}
 \frac{\pd U}{\pd t} 
= 
 [ U , \vphi ],
\label{eq:2DEulerVorticityEq}
\end{equation}
where the `inner' Poisson bracket or Jacobian is defined by
\begin{equation}
 [ f , g ]
\coloneqq 
 \hrvz \cdot \nab f \times \nab g
=  
 \frac{\pd f}{\pd x} \frac{\pd g}{\pd y}
-\frac{\pd f}{\pd y} \frac{\pd g}{\pd x}.
\label{eq:2DEulerPoissonBracket}
\end{equation}
A Hamiltonian and a Lie--Poisson bracket for functionals 
are defined
as
\begin{align}
 H [U]
\coloneqq &\,
 \frac{1}{2} \int_{\mathcal{D}} \td^{2} x  \,
 | \nab_{\perp} ( \bigtriangleup_{\perp}^{-1} U ) |^{2},
\label{eq:2DEulerHamiltonian}
\\
 \{ F, G \}
\coloneqq &\,
 \int_{\mathcal{D}} \td^{2} x  \,
 U \left[ \frac{\delta F}{\delta U} , \frac{\delta G}{\delta U}  \right],
\label{eq:2DEulerFunctionalPoissonBracket}
\end{align}
respectively, where $F[U]$ and $G[U]$ are arbitrary functionals of $U$,
and $\mathcal{D}$ is a two-dimensional domain in the $x$--$y$ plane.
Functional derivatives such as  $\delta F / \delta U$ are defined through 
a variation of $F$ as
\begin{align}
 \delta F
&=
 \lim_{\veps \rightarrow 0}
 \frac{1}{\veps}
 \int_{\mathcal{D}} \td^{2} x  \,
 \left( 
   F[ U + \veps \delta U ] - F[U]
 \right)
\nonumber\\
&=
 \left. \frac{\td}{\td \veps} F[ U + \veps \delta U ] 
 \right\vert_{\veps = 0}
=: 
 \int_{\mathcal{D}} \td^{2} x  \,
 \delta U \frac{\delta F}{\delta U}.
\end{align}
By using the Poisson bracket for functionals, 
the vorticity equation (\ref{eq:2DEulerVorticityEq}) can be written as
\begin{equation}
 \frac{\pd U}{\pd t} 
= \{ U , H \}.
\label{eq:2DEulerEvolutionEqFunctionalPoissonBracket}
\end{equation}
Understanding  Eq.~\eqref{eq:2DEulerEvolutionEqFunctionalPoissonBracket} 
may need some care, so details are  explained in  Appendix~\ref{sec:detail2DEulerEvolutionEqFunctionalPoissonBracket}.

The antisymmetric Poisson operator $\mathcal{J}$ associated with the Poisson bracket of \eqref{eq:2DEulerFunctionalPoissonBracket} is
\begin{equation}
 \mathcal{J}\, \coloneqq [ \circ , U ],
\label{eq:2DEulerFunctionalPoissonTensor}
\end{equation}
in terms of which the Poisson bracket can be expressed as 
\begin{equation}
 \{ F , G \} 
=
 \int_{\mathcal{D}} \td^{2} x  \,
 \frac{\delta F}{\delta U}
 \mathcal{J}
 \frac{\delta G}{\delta U}.
\label{eq:2DEulerFunctionalPoissonBracket-2}
\end{equation}
Note that $\mathcal{J}$ takes the argument $\circ$ from its right.
Then the evolution equation
(\ref{eq:2DEulerEvolutionEqFunctionalPoissonBracket})
can also be written as
\begin{equation}
 \frac{\pd U}{\pd t} 
= 
 \mathcal{J} \frac{\delta H}{\delta U}.
\label{eq:2DEulerEvolutionEqPoissonOperator} 
\end{equation}

There exists an infinite number of Casimir invariants  for  this system, viz. 
\begin{equation}
 C 
\coloneqq
 \int_{\mathcal{D}} \td^{2} x  \,
 f(U)\,,
\label{eq:2DEulerCasimir} 
\end{equation}
where $f(U)$ is an arbitrary function. It can easily be shown that $\{F,C\}=0$ for all functionals $F$.

\subsubsection{Low-beta reduced MHD in a  two-dimensional rectangular domain}
\label{subsec:HamiltonianInfDoFRectLBRMHD}

Another noncanonical Hamiltonian system is that of the  low-beta reduced MHD system of  \citet{Strauss-1976} whose Hamiltonian structure was given by \citet{MH84}.  This system  describes two-dimensional dynamics in the plane perpendicular to
a strong ambient magnetic field. 
The velocity and magnetic fields are expressed as
\begin{align}
 \vv
&=
 \hrvz \times \nab_{\perp} \vphi,
\label{eq:RectLBRMHDVelocityField}
\\
 \vB
&=
 \hrvz + \hrvz \times \nab_{\perp} \psi,
\label{eq:RectLBRMHDMagneticField}
\end{align}
where the magnetic field is normalized by the strong magnetic field in
the $z$ direction, and the velocity field is by the Alfv\'en velocity.
If we assume translational symmetry in the $z$ direction, 
the governing equations of the low-beta reduced MHD are given by
\begin{align}
 \frac{\pd U}{\pd t}
&=
 [ U , \vphi ] + [ \psi , J ],
\label{eq:rectLBRMHDVorticityEq}
\\
 \frac{\pd \psi}{\pd t}
&=
 [ \psi , \vphi ],
\label{eq:rectLBRMHDOhmLaw}
\end{align}
where
$U \coloneqq \hrvz \cdot \rot \vv = \bigtriangleup_{\perp} \vphi$ 
is the same definition used for  two-dimensional Euler flow, 
$J \coloneqq- \hrvz \cdot \rot \vB = \bigtriangleup_{\perp} \psi$,
and the Poisson bracket  $[ \, , \,]$ is  the same as Eq.~(\ref{eq:2DEulerPoissonBracket}).
The noncanonical variables are 
$\vu = ( u^{1}, u^{2} )^{\mathsf{T}} = ( U , \psi )^{\mathsf{T}}$.
The Hamiltonian, the Lie-Poisson bracket, and the evolution equations
are, respectively,  given by 
\begin{align}
 H [\vu] 
\coloneqq&\,
 \frac{1}{2} \int_{\mathcal{D}} \td^{2} x  \,
 \left(
 | \nab_{\perp} ( \bigtriangleup_{\perp}^{-1} U ) |^{2} 
 + | \nab_{\perp} \psi |^{2} 
 \right),
\label{eq:rectLBRMHDHamiltonian}
\\
 \{ F, G \}
\coloneqq &\,
 \int_{\mathcal{D}} \td^{2} x  \,
 \left(
 U \left[ \frac{\delta F}{\delta U} , \frac{\delta G}{\delta U}  \right]
+\psi \left(
       \left[ \frac{\delta F}{\delta U} , \frac{\delta G}{\delta \psi}  \right]
      +\left[ \frac{\delta F}{\delta \psi} , \frac{\delta G}{\delta U}  \right]
      \right)
 \right),
\label{eq:rectLBRMHDFunctionalPoissonBracket}
\\
 \frac{\pd u^{i}}{\pd t} 
= &\,
 \{ u^{i} , H \},
\label{eq:rectLBRMHDEvolutionEqFunctionalPoissonBracket}
\end{align}
where $F[\vu]$ and $G[\vu]$ are arbitrary functionals of $\vu$.
For   low-beta reduced MHD,  the  antisymmetric Poisson operator $\mathcal{J} = ( \mathcal{J}_{ij} )$ 
can be defined 
as
\begin{equation}
 \mathcal{J}  
\coloneqq
 \begin{pmatrix}
  [ \circ , U ] & [ \circ , \psi ]
  \\
  [ \circ , \psi ] & 0
 \end{pmatrix}, 
\end{equation}
and the Poisson bracket reads
\begin{equation}
 \{ F , G \} 
=
 \int_{\mathcal{D}} \td^{2} x  \,
 \frac{\delta F}{\delta u^{i}}
 \mathcal{J}^{ij}
 \frac{\delta G}{\delta u^{j}}.
\end{equation}
Note, as before,   $\mathcal{J}$ takes the arguments $\circ$ from its right.
Then the evolution equation 
(\ref{eq:rectLBRMHDEvolutionEqFunctionalPoissonBracket})
can also be written as
\begin{equation}
 \frac{\pd u^{i}}{\pd t} 
= 
 \mathcal{J}^{ij} \frac{\delta H}{\delta u^{j}}.
\label{eq:rectLBRMHDEvolutionEqPoissonOperator} 
\end{equation}

The Casimir invariants are given by
\begin{equation}
 C_{1}[\vu]
\coloneqq 
 \int_{\mathcal{D}} \td^{2} x  \,
   f(\psi),
\qquad \mathrm{and}\qquad 
 C_{2}[\vu]
\coloneqq 
 \int_{\mathcal{D}} \td^{2} x  \,
   U g(\psi),
\label{eq:rectLBRMHDCasimir-2}
\end{equation}
where $f(\psi)$ and $g(\psi)$ are arbitrary functions.

\subsubsection{Low-beta reduced MHD in cylindrical geometry}
\label{subsec:HamiltonianInfDoFCylLBRMHD}

In  cylindrical geometry
under periodic boundary condition in the axial direction, 
Eqs.~(\ref{eq:rectLBRMHDVorticityEq}) and
(\ref{eq:rectLBRMHDOhmLaw}) become
\begin{align}
 \frac{\pd U}{\pd t}
&=
 [ U , \vphi ] + [ \psi , J ] - \veps \frac{\pd J}{\pd \zeta},
\label{eq:cylLBRMHDVorticityEq}
\\
 \frac{\pd \psi}{\pd t}
&=
 [ \psi , \vphi ] - \veps \frac{\pd \vphi}{\pd \zeta},
\label{eq:cylLBRMHDOhmLaw}
\end{align}
where $\veps := a / R_{0}$ is the inverse aspect ratio with the length
of the cylinder and the minor radius being $2 \pi R_{0}$ and $a$,
respectively.  The toroidal angle is $\zeta \coloneqq  z / R_{0}$.
Using the cylindrical coordinates $(r, \theta, z)$, 
the Poisson bracket (\ref{eq:2DEulerPoissonBracket}) becomes
\begin{equation}
 [ f, g ]
=
 \frac{1}{r}
 \left(
   \frac{\pd f}{\pd r} \frac{\pd g}{\pd \theta}
  -\frac{\pd f}{\pd \theta} \frac{\pd g}{\pd r}
 \right).
\label{eq:cylLBRMHDPoissonBracket}
\end{equation}

The Hamiltonian is the same as  that of Eq.~(\ref{eq:rectLBRMHDHamiltonian}).
The Poisson bracket   for arbitrary functionals $F[\vu]$ and $G[\vu]$
and the Poisson tensor are given, respectively,  by
\begin{align}
 \{ F, G \}
:= &\,
 \int_{\mathcal{D}} \td^{3} x \,
 \left(
 U \left[ \frac{\delta F}{\delta U} , \frac{\delta G}{\delta U}  \right]
+\psi \left(
       \left[ \frac{\delta F}{\delta U} , \frac{\delta G}{\delta \psi}  \right]
      +\left[ \frac{\delta F}{\delta \psi} , \frac{\delta G}{\delta U}  \right]
      \right)
 \right.
\nonumber
\\
&\hspace*{2.5cm}
 + \left.
 \veps \left(
    \frac{\delta F}{\delta U} 
    \frac{\pd}{\pd \zeta} \frac{\delta G}{\delta \psi}
  - \frac{\delta G}{\delta U} 
    \frac{\pd}{\pd \zeta} \frac{\delta F}{\delta \psi} 
        \right)
 \right),
\label{eq:cylLBRMHDFunctionalPoissonBracket}
\\
 \mathcal{J}   
:= &\,
 \begin{pmatrix}
  [ \circ , U ] & [ \circ , \psi ] + \veps \frac{\pd}{\pd \zeta}
  \\
  [ \circ , \psi ]  + \veps \frac{\pd}{\pd \zeta}  & 0
 \end{pmatrix}.
\label{eq:cylLBRMHDFunctionalPoissonTensor}
\end{align}
Note that again $\mathcal{J}$ takes the arguments $\circ$ from its right.

Using the Poisson bracket (\ref{eq:cylLBRMHDFunctionalPoissonBracket})
and the Poisson tensor (\ref{eq:cylLBRMHDFunctionalPoissonTensor}),
the evolution equations (\ref{eq:cylLBRMHDVorticityEq})
and (\ref{eq:cylLBRMHDOhmLaw}) can be  rewritten as
\begin{equation}
 \frac{\pd u^{i}}{\pd t}
=
 \{ u^{i} , H \}=
 \mathcal{J}^{ij} \frac{\delta H}{\delta u^{j}}\,, 
\label{eq:cylLBRMHDEvolutionEqFunctionalPoissonBracket}
\end{equation}
and the Casimir invariants are 
\begin{equation}
 C_{\mathrm{v}}[\vu]
\coloneqq 
 \int_{\mathcal{D}} \td^{3} x  \,  U \qquad \mathrm{and} \qquad
 C_{\mathrm{m}}[\vu]
\coloneqq  
 \int_{\mathcal{D}} \td^{3} x  \,  \psi \,.
\label{eq:cylLBRMHDCasimir-2}
\end{equation}

If we focus on single helicity dynamics that includes 
only a family of Fourier modes  with  mode numbers 
$\ell (m, n)$ where is an $\ell$ integer and  $m$ and $n$ are specified poloidal and toroidal
mode numbers, respectively, 
the $\zeta$-derivative terms can be absorbed in the bracket
terms.
By adopting a helical flux
\begin{equation}
 \psi_{\mathrm{h}}
:=
 \psi + \frac{\veps n}{2 m} r^{2}
\end{equation}
as a state variable as 
$\vu = ( u^{1}, u^{2} )^{\mathsf{T}} = ( U, \psi_{\mathrm{h}} )^{\mathsf{T}}$,
the Hamiltonian, the Lie-Poisson bracket, the Poisson tensor, and the 
evolution equations  become, respectively, 
\begin{align}
 H [\vu] 
\coloneqq &\,
 \frac{1}{2} \int_{\mathcal{D}} \td^{3} x  \,
 \left(
 \left\vert \nab_{\perp} ( \bigtriangleup_{\perp}^{-1} U ) \right\vert^{2} 
 + \left\vert 
    \nab_{\perp} \left( \psi_{\mathrm{h}} - \frac{\veps n}{2 m} r^{2}
                 \right)
  \right\vert^{2} 
 \right),
\label{eq:cylLBRMHDHelFlxHamiltonian}
\\
 \{ F, G \}
\coloneqq &\,
 \int_{\mathcal{D}} \td^{3} x \,
 \left(
 U \left[ \frac{\delta F}{\delta U} , \frac{\delta G}{\delta U}  \right]
+\psi_{\mathrm{h}} \left(
       \left[ \frac{\delta F}{\delta U} , \frac{\delta G}{\delta \psi_{\mathrm{h}}}  \right]
      +\left[ \frac{\delta F}{\delta \psi_{\mathrm{h}}} , \frac{\delta G}{\delta U}  \right]
                   \right)
 \right),
\label{eq:cylLBRMHDHelFlxFunctionalPoissonBracket}
\\
 \mathcal{J} 
\coloneqq &\,
 \begin{pmatrix}
  [ \circ , U ] & [ \circ , \psi_{\mathrm{h}} ]
  \\
  [ \circ , \psi_{\mathrm{h}} ] & 0
 \end{pmatrix},
\label{eq:cylLBRMHDHelFlxFunctionalPoissonTensor}
\\
 \frac{\pd u^{i}}{\pd t}
=&
 \{ u^{i} , H \}
\label{eq:cylLBRMHDHelFlxEvolutionEqFunctionalPoissonBracket}
=
 \mathcal{J}^{ij} \frac{\delta H}{\delta u^{j}}.
\end{align}
Again, note that $\mathcal{J}$ takes the arguments $\circ$ from its right.

For this case, the  Casimir invariants are given by 
\begin{equation}
  C_{1}[\vu]
\coloneqq 
 \int_{\mathcal{D}} \td^{3} x  \,
   f(\psi_{\mathrm{h}}) 
\label{eq:cylLBRMHDHelFlxCasimir-1}
\qquad \mathrm{and}\qquad 
 C_{2}[\vu]
\coloneqq
 \int_{\mathcal{D}} \td^{3} x  \,
   U g(\psi_{\mathrm{h}}),
\end{equation}
where $f$ and $g$ are arbitrary functions.

\subsubsection{High-beta reduced MHD in toroidal geometry}
\label{subsec:HamiltonianInfDoFTorHBRMHD}

Lastly, the evolution equations for  high-beta reduced MHD \citep{Strauss-1977}
are given by
\begin{align}
 \frac{\pd U}{\pd t}
&=
 [ U , \vphi ] + [ \psi , J ] - \veps \frac{\pd J}{\pd \zeta}
 + [ P, h ],
\label{eq:HBRMHDVorticityEq}
\\
 \frac{\pd \psi}{\pd t}
&=
 [ \psi , \vphi ] - \veps \frac{\pd \vphi}{\pd \zeta},
\label{eq:HBRMHDOhmLaw}
\\
 \frac{\pd P}{\pd t}
&=
 [ P, \vphi ],
\label{eq:HBRMHDPressureEq}
\end{align}
where $U$, $\psi$, $\vphi$ and $J$ are the same as those of   low-beta
reduced MHD in cylindrical geometry, 
$P$ is the normalized pressure, 
and $h := \veps r \cos \theta$ expresses the toroidicity.
The pressure is normalized by the typical magnetic pressure, the  brackets $[\,,\,]$ are the same as those of Eq.~(\ref{eq:2DEulerPoissonBracket}), and the  state vector is 
$\vu = ( u^{1} , u^{2} , u^{3} )^{\mathsf{T}} := ( U , \psi , P )^{\mathsf{T}}$.
The Hamiltonian, the Poisson bracket for functionals, 
the Poisson tensor, and the evolution equations \citep{MH84} are given, respectively, by
\begin{align}
  H [\vu] 
\coloneqq &\,
 \int_{\mathcal{D}} \td^{3} x  \,
 \left(
   \frac{1}{2} 
   | \nab_{\perp} ( \bigtriangleup_{\perp}^{-1} U ) |^{2} 
 + \frac{1}{2}  
   | \nab_{\perp} \psi |^{2} 
 - h P
 \right),
\label{eq:torHBRMHDHamiltonian}
\\
 \{ F, G \}
\coloneqq &\,
 \int_{\mathcal{D}} \td^{3} x \,
 \left(
 U \left[ \frac{\delta F}{\delta U} , \frac{\delta G}{\delta U}  \right]
+\psi \left(
       \left[ \frac{\delta F}{\delta U} , \frac{\delta G}{\delta \psi}  \right]
      +\left[ \frac{\delta F}{\delta \psi} , \frac{\delta G}{\delta U}  \right]
      \right)
 \right.
\label{eq:torHBRMHDFunctionalPoissonBracket}
\\
&\hspace*{1cm}
 + \left.
 P \left(
       \left[ \frac{\delta F}{\delta U} , \frac{\delta G}{\delta P}  \right]
      +\left[ \frac{\delta F}{\delta P} , \frac{\delta G}{\delta U}  \right]
    \right)
+ \veps \left(
    \frac{\delta F}{\delta U} 
    \frac{\pd}{\pd \zeta} \frac{\delta G}{\delta \psi}
  - \frac{\delta G}{\delta U} 
    \frac{\pd}{\pd \zeta} \frac{\delta F}{\delta \psi} 
        \right)
 \right),
\nonumber
\\
 \mathcal{J} 
\coloneqq &\,
 \begin{pmatrix}
  [ \circ , U ] & [ \circ , \psi ] + \veps \frac{\pd}{\pd \zeta} 
  & [ \circ, P ]
  \\
  [ \circ , \psi ]  + \veps \frac{\pd}{\pd \zeta}  & 0 & 0
  \\
  [ \circ, P ]  & 0 & 0
 \end{pmatrix},
\label{eq:torHBRMHDFunctionalPoissonTensor}
\\
  \frac{\pd u^{i}}{\pd t}
=&
 \{ u^{i} , H \} =
 \mathcal{J}^{ij} \frac{\delta H}{\delta u^{j}}.
\label{eq:torHBRMHDEvolutionEqFunctionalPoissonBracket}
\end{align}
Again, note that $\mathcal{J}$ takes the arguments $\circ$ from its right.

The Casimir invariants are 
\begin{equation}
 C_{\mathrm{v}}[\vu]
\coloneqq
 \int_{\mathcal{D}} \td^{3} x  \,  U \,,
\quad
 C_{\mathrm{m}}[\vu]
\coloneqq
 \int_{\mathcal{D}} \td^{3} x  \,  \psi \,\quad 
\mathrm{and}\quad
 C_{\mathrm{p}}[\vu]
\coloneqq 
 \int_{\mathcal{D}} \td^{3} x  \,  f(P) ,
\label{eq:torHBRMHDCasimir-3}
\end{equation}
where $f$ is an arbitrary function.

\section{Simulated annealing}
\label{sec:SA}

Let us now turn to the theory of SA, which is explained in Section~\ref{sec:SA}. This theory will be used for the computation of equilibrium states    in Secs.~\ref{sec:SAFinDoF}--\ref{sec:acceleration}. 
In Sec.~\ref{subsec:SAdoubleBracket},  the double bracket of SA is
presented and its properties are discussed, both for finite and infinite-dimensional systems.  Then, in Sec.~\ref{subsec:SAmetriplecticBracket} we briefly introduce 
another kind of SA by means of a  metriplectic bracket.

\subsection{Simulated annealing by double bracket}
\label{subsec:SAdoubleBracket}

\subsubsection{Finite degrees of freedom}
\label{subsubsec:SAdoubleBracketFinDoF}

In Sec.~\ref{subsec:HamiltonianFinDoF},
it was explained that 
Hamiltonian systems are governed by equations of  the following type:
\begin{equation}
 \dot{u}^{i} = J^{ij} \frac{\pd H(\vu)}{\pd u^{j}},
 \label{eq:evolutionEqHamiltonianSystemFinDoF}
\end{equation}
where $\vu = ( u^{1}, \cdots , u^{M} )^{\mathsf{T}}$ are  the phase space
variables, 
$H(\vu)$ is the Hamiltonian, 
and  $J = ( J^{ij} )$ is the Poisson tensor.
The antisymmetry of $J$ guarantees the energy is conserved, as is easliy shown, 
\begin{align}
 \frac{\td H(\vu)}{\td t}
&=
 \frac{\pd H}{\pd u^{i}} \dot{u}^{i}
=
 \frac{\pd H}{\pd u^{i}}
 J^{ij} \frac{\pd H}{\pd u^{j}}
\nonumber
\\
&=
- \frac{\pd H}{\pd u^{i}}
 J^{ij} \frac{\pd H}{\pd u^{j}}
= 0.
\end{align}
The Casimir invariants are also conserved during the time evolution, but  this is because of the null space of $J$. 

Consider an artificial dynamics governed by equations of the form 
\begin{align}
 \dot{u}^{i} 
= 
 J^{ij} K_{jk} J^{k \ell} \frac{\pd H(\vu)}{\pd u^{\ell}}
 \label{eq:evolutionEqSAFinDoF-1}
=
 [ u^{i} , u^{j} ] K_{jk} [ u^{k} , H ],
\end{align}
where $K_{jk}$ is a matrix with a definite sign.
We assume here  that this matrix  is positive definite.
The time evolution of $H(\vu)$ according to 
Eq.~(\ref{eq:evolutionEqSAFinDoF-1})
is then
\begin{align}
 \frac{\td H(\vu)}{\td t}
&=
 \frac{\pd H}{\pd u^{i}} \dot{u}^{i}
=
 \frac{\pd H}{\pd u^{i}}
 J^{ij} K_{jk} J^{k \ell} \frac{\pd H}{\pd u^{\ell}}
\nonumber
\\  
&=
\left(
- J^{ji}  \frac{\pd H}{\pd u^{i}}
\right)
K_{jk} J^{k \ell} \frac{\pd H}{\pd u^{\ell}}
 \leq 0.
\end{align}
Therefore, the energy of the system monotonically decreases, and 
the system approaches a minimum energy state until  $\td H / \td t = 0$ 
or $J^{ij} ( \pd H / \pd u^{j} ) = 0$, 
which corresponds to 
an equilibrium of the original system 
(\ref{eq:evolutionEqHamiltonianSystemFinDoF}).
If we take $K$ as negative definite, the energy monotonically increases
to approach an energy maximum.

Of note, is that SA dynamics preserves Casimir invariants of the original system.
In fact, it is easily shown that $\td C_{k} / \td t = 0$ because 
of $J^{ij} ( \pd C_{k} / \pd u^{j} ) \equiv 0$.

To obtain a wider class of equilibria it is necessary to constrain the dynamics.  To do this \citet{Flierl-Morrison-2011} used Dirac constraint theory, by constructing  a Dirac bracket,  which imposes  
additional constraints $C_{\ell}$ that differ  from the original Casimir
invariants. As part of the construction, each  $C_{\ell}$ must possess a counterpart, i.e., the set of $C_{\ell}$s must be evenly  splint  into such pairs where $[C_{\ell}, C_{\ell'}]\neq 0$.  If this split is not possible, then other paired constraints  for any $C_{\ell}$ can be manufactured according to
\begin{align}
 C_{\ell + 1}
\coloneqq &\,
 [ C_{\ell}, H ]\,.
\end{align}
If $C_{\ell}$ does not change during the course of the  time evolution, then 
$C_{\ell + 1}$ must be always zero.
By using this pair of constraints, a Dirac bracket can be constructed as 
\begin{equation}
 [ f, g ]_{\mathrm{D}}
=
 [ f, g ]
- \begin{pmatrix}
    [ f, C_{\ell} ]
    \\
    [ f, C_{\ell + 1} ]
  \end{pmatrix}^{\mathsf{T}}
  \begin{pmatrix}
   [ C_{\ell} , C_{\ell} ]  & [ C_{\ell}, C_{\ell + 1} ]
   \\
   [ C_{\ell + 1}, C_{\ell} ] & [ C_{\ell + 1}, C_{\ell + 1} ]
  \end{pmatrix}^{-1}
 \begin{pmatrix}
    [ C_{\ell}, g ]
    \\
    [ C_{\ell + 1}, g ]
  \end{pmatrix}.
\label{eq:DiracBracketFinDoFExample}
\end{equation}
Note that this definition is valid when the inverse matrix 
on the right hand side exists, or when $[C_{\ell}, C_{\ell + 1}] \neq 0$.  

The number of additional constraints can be increased in a similar
manner.  Suppose we have $( M - 2N )$ Casimir invariants originally, and
we impose $L$ constraints additionally.  
Then $C_{M-2N+1}$ and $C_{M-2N+2} := [ C_{M-2N+1}, H ]$ is the first
pair,
$C_{M-2N+3}$ and $C_{M-2N+4} := [ C_{M-2N+3}, H ]$ 
is the second pair, and the last pair is 
$C_{M-2N+2L-1}$ and $C_{M-2N+2L} := [ C_{M-2N+2L-1}, H ]$.
By defining a matrix 
$\mathcal{C} = ( \mathcal{C}^{ij} ) := ( [ C_{i}, C_{j} ] )^{-1}$,
the Dirac bracket is given by
\begin{equation}
 [ f, g ]_{\mathrm{D}}
:=
 [ f, g ] 
- [ f, C_{i} ] \mathcal{C}^{ij} [ C_{j}, g ],
\label{eq:DiracBracketFinDoF}
\end{equation}
where $i$ and $j$ take on  integer values from $M-2N+1$ to $M-2N+2L$.
Here $\mathcal{C}$ must exist for this Dirac bracket to be valid.

In terms of a Dirac bracket of the form of \eqref{eq:DiracBracketFinDoF}, the evolution equation for  DSA is given by 
\begin{equation}
 \dot{u}^{i} 
=
 [ u^{i} , u^{j} ]_{\mathrm{D}} K_{jk} [ u^{k} , H ]_{\mathrm{D}}.
 \label{eq:evolutionEqDSAFinDoF}
\end{equation}

\subsubsection{Infinite degrees of freedom}
\label{subsubsec:SAdoubleBracketInfDoF}

The governing equations of 
systems with infinite degrees of freedom have the following form
\begin{equation}
 \frac{\pd u^{i}}{\pd t} 
= \{ u^{i} , H  \}
 \label{eq:evolutionEqHamiltonianSystemPoissonBracketInfDoF}
= \mathcal{J}^{ij} \frac{\delta H}{\delta u^{j}}, 
\end{equation}
where, as described in Sec.~\ref{subsec:HamiltonianInfDoF},  $\mathcal{J}^{ij}$ is now an operator.  On the basis of this form,   \citet{Flierl-Morrison-2011}  defined an  artificial dynamics generated by a double bracket according to  
\begin{align}
 \frac{\pd u^{i}}{\pd t}
&= (( u^{i} , H )),
 \label{eq:evolutionEqSAInfDoF}
\\
(( F, G ))
&= 
 \int_{\mathcal{D}} \td^{N} x^{\pr} \,
 \int_{\mathcal{D}} \td^{N} x^{\pr\pr} \,
  \{ F , u^{i}(\vx^{\pr}) \} 
  \mathcal{K}_{ij}( \vx^{\pr} , \vx^{\pr\pr} )
  \{ u^{j}(\vx^{\pr\pr}) , G \} ,
 \label{eq:doubleBracketInfDoF} 
\end{align}
where $N$ is the spatial dimension
and $\mathcal{K} = ( \mathcal{K}_{ij} )$ is a symmetric kernel with a definite
sign.

Double bracket SA dynamics for infinite degree-of-freedom systems 
can be understood as a replacement of the advection fields
for the dynamical variables $u^{i}$.
This will be shown explicitly case-by-case in 
Secs.~\ref{sec:linearStability} and \ref{sec:toroidal}.

According to the dynamics generated by  Eq.~(\ref{eq:evolutionEqSAInfDoF}),
time evolution of any arbitrary functional $F[\vu]$ is governed by 
\begin{equation}
 \frac{\td F[\vu]}{\td t}
= 
 (( F, H )).
 \label{eq:evolutionEqFunctional} 
\end{equation}
Thus time derivative of the Hamiltonian becomes
\begin{align}
 \frac{\td H[\vu]}{\td t}
&= 
 (( H, H ))
= 
 \int_{\mathcal{D}} \td^{N} x^{\pr} \,
 \int_{\mathcal{D}} \td^{N} x^{\pr\pr} \,
  \{ H , u^{i} \}
  \mathcal{K}_{ij}  
  \{ u^{j} , H \} 
\nonumber
\\
&= 
- \int_{\mathcal{D}} \td^{N} x^{\pr} \,
  \int_{\mathcal{D}} \td^{N} x^{\pr\pr} \,
  \{ u^{i} , H \}
  \mathcal{K}_{ij}  
  \{ u^{j} , H \} 
\leq 0, 
 \label{eq:evolutionEqHamiltonianInfDoF} 
\end{align}
for a positive definite symmetric kernel $\mathcal{K}$.
Therefore, $H$ decreases monotonically and approaches a minimum value
where $\{ u^{i} , H \} = 0$, which   is a stationary state of the
original system (\ref{eq:evolutionEqHamiltonianSystemPoissonBracketInfDoF}).

On the other hand,
the time evolution of a Casimir invariant $C[\vu]$ is given by
\begin{align}
 \frac{\td C[\vu]}{\td t}
&= 
 (( C, H ))
\nonumber
\\
&= 
 \int_{\mathcal{D}} \td^{N} x^{\pr} \,
 \int_{\mathcal{D}} \td^{N} x^{\pr\pr} \,
  \{ C , u^{i} \}
  \mathcal{K}_{ij}  
  \{ u^{j} , H \} 
 \equiv 0.
 \label{eq:evolutionEqCasimirInfDoF}
\end{align}
Therefore, all Casimir invariants of the original system are preserved in
the SA dynamics.


As with finite-dimensional systems,  \citet{Flierl-Morrison-2011} made SA more useful by  using a Dirac bracket, giving   DSA  akin to that in finite dimensions, 
\begin{equation}
 \{ F , G \}_{\mathrm{D}}
:=
 \{ F , G \} 
 - \{ F , C_{i} \} \mathcal{C}^{ij} \{ C_{j}, G \},
\label{eq:DiracBracket}
\end{equation}
where 
$\mathcal{C} = ( \mathcal{C}^{ij} ) \coloneqq ( \{ C_{i}, C_{j} \} )^{-1}$
is an even dimensional matrix, 
where  $C_{i}$ are additional constraints to be incorporated.
Then,   DSA is defined as
\begin{align}
 \frac{\pd u^{i}}{\pd t}
&= (( u^{i} , H ))_{\mathrm{D}},
 \label{eq:evolutionEqDSAInfDoF}
\\
(( F, G ))_{\mathrm{D}}
&= 
 \int_{\mathcal{D}} \td^{N} x^{\pr} \,
 \int_{\mathcal{D}} \td^{N} x^{\pr\pr} \,
  \{ F , u^{i}( \vx^{\pr} ) \}_{\mathrm{D}}
  \mathcal{K}_{ij}( \vx^{\pr} , \vx^{\pr\pr} )
  \{ u^{j}( \vx^{\pr\pr} ) , G \}_{\mathrm{D}}.
 \label{eq:DiracDoubleBracketInfDoF} 
\end{align}

\subsection{Simulated annealing by metriplectic brackets}
\label{subsec:SAmetriplecticBracket}

The SA changes the energy (Hamiltonian) of the system monotonically,
while the Casimir invariants are preserved. 
On the other hand, the metriplectic dynamics changes the entropy monotonically, while the energy
is conserved.  See \citet{pjm84,pjm86} for original papers and see  \citet{pjmU24} for a summary and recent results.

For finite-dimensional  systems, 
let us define a symmetric bracket according to 
\begin{equation}
 ( f , h )
\coloneqq
 \frac{\pd f}{\pd u^{i}} G^{ij} \frac{\pd h}{\pd u^{j}},
\label{eq:symmetricBracketFinDoF}
\end{equation}
where 
$f$ and $h$ are arbitrary functions, 
and $G = (G^{ij} )$ is a symmetric metric-like matrix that ensures  
$( f, h ) = ( h, f )$.
One more important feature imposed on a metriplectic bracket is 
\begin{equation}
 ( f, H ) = 0
\label{eq:symmetricBracketHamiltonianFinDoF}
\end{equation}
for any $f$.  Such a choice can be realized by a projection, for example. 
Then, metriplectic dynamics is generated by a free energy like quantity, 
$F \coloneqq  H  - \mathcal{T} S$, where $\mathcal{T}$ is a global constant temperature and $S$ is an entropy, according to  
\begin{equation}
  \dot{u}^{i}
= 
 [ u^{i} , F ] + ( u^{i} , F ).
\label{eq:metriplecticEvolutionEqFinDoF}
\end{equation}
Here,  for convenience we have scaled away $\mathcal{T}$.
The entropy $S$ is selected from the set of  Casimir
invariant of the Poisson bracket; i.e., Casimirs are candidate entropies that determine ones choice for `thermal equilibrium.' 
Then the entropy evolves as
\begin{equation}
 \frac{\td S}{\td t}
=
 [ S , H ] + [ S , S ] + ( S , H ) + ( S , S )
\geq 0
\label{eq:metriplecticEvolutionEqEntropyFinDoF}
\end{equation} 
for a positive semi-definite $(G^{ij} )$.
On the other hand, 
the Hamiltonian is conserved as 
\begin{equation}
 \frac{\td H}{\td t}
=
  [ H , H ] + [ H , S ] + ( H , H )  + ( H , S )
= 0.
\label{eq:metriplecticEvolutionEqHamiltonianFinDoF}
\end{equation}

For infinite-dimensional systems, 
a symmetric bracket is defined similarly as
\begin{equation}
( F, G )
:= 
 \int_{\mathcal{D}} \td^{N} x^{\pr} \,
 \int_{\mathcal{D}} \td^{N} x^{\pr\pr} \,
  \frac{\delta F}{\delta u^{i}( \vx^{\pr} )}
  \mathcal{G}^{ij}( \vx^{\pr} , \vx^{\pr\pr} )
  \frac{\delta G}{\delta u^{j}( \vx^{\pr\pr} )}, 
 \label{eq:symmetricBracketInfDoF} 
\end{equation}
where $\mathcal{G} := ( \mathcal{G}^{ij} )$ is a symmetric kernel, 
and $F[\vu]$ and $G[\vu]$ are arbitrary functionals of $\vu$.
The kernel is chosen to satisfy $( H, F ) \equiv 0$ for any $F$ 
and $( S, S) \geq 0$.
The evolution equations of the metriplectic dynamics are given by
\begin{equation}
 \frac{\pd u^{i}}{\pd t}
= \{ u^{i} , F \} + ( u^{i} , F ),
 \label{eq:metriplecticEvolutionEqInfDoF}
\end{equation}
where $F[\vu] := H[\vu] + S[\vu]$.
This dynamics increases the entropy functional $S$ monotonically, 
while conserving  $H$.

\section{Simulated annealing of system with finite degrees of freedom}
\label{sec:SAFinDoF}

Section~\ref{sec:SAFinDoF} presents 
some analyses of equilibria and stability of 
Hamiltonian systems with finite degrees of freedom.
Numerical results of SA are also shown.
Section~\ref{subsec:htSA} treats the  heavy top,
while Sec.~\ref{subsec:SAmimicLBRMHD} presents results on a toy model
designed  to mimic reduced MHD.

\subsection{Heavy top}
\label{subsec:htSA}

Some numerical results of SA for the  heavy top will be shown in
Sec.~\ref{subsec:htSA}.  Our first example consists of a   stable equilibrium with two positive energy modes, while a second example  is for an unstable equilibrium with a positive energy mode and a saddle.  These cases have the same equilibrium  point, but  different values of the gravity parameter.  A third example consists of  a stable equilibrium with a positive and a negative energy mode.  Recall, negative energy modes are stable oscillations with negative energy \citep[see e.g.,][]{Morrison-1998}. Linear spectral stability analyses are described for these cases, along with SA results.  
Our final example employs DSA.
In all cases, the principal moments of inertia were chosen to be
$I_{1} = 1$, $I_{2} = 2$, and $I_{3} = 3$.

\begin{figure}[h]
  \begin{minipage}[t]{0.45\textwidth}
    \centering
    \includegraphics[width=\textwidth]{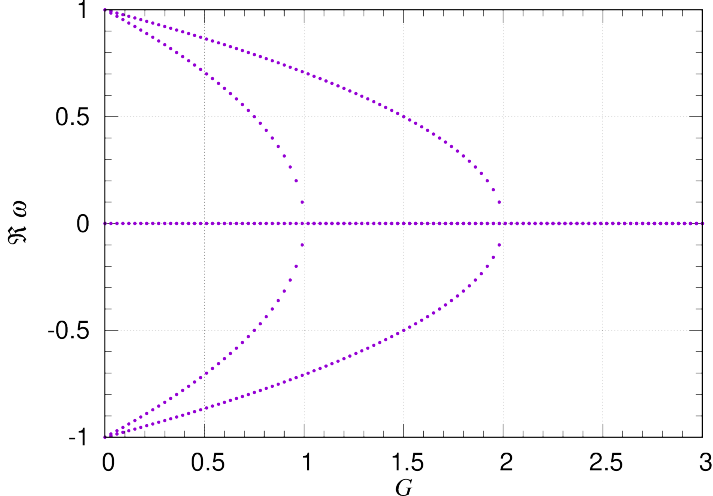}
    \subcaption{$\Re \, \omega$.}
    \label{subfig:htLinearStability-G-omgr-ieq3}
  \end{minipage}
  \begin{minipage}[t]{0.45\textwidth}
    \centering
    \includegraphics[width=\textwidth]{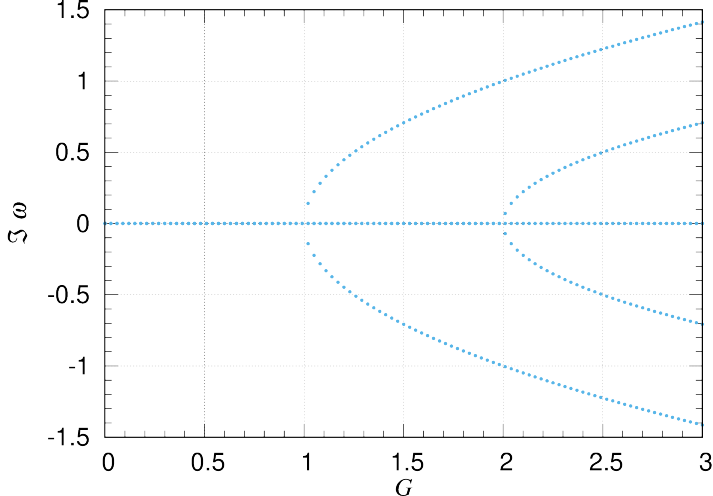}
    \subcaption{$\Im \, \omega$.}
    \label{subfig:htLinearStability-G-omgi-ieq3}
  \end{minipage}
  \caption{Real and imaginary parts of $\omega$ as determined by 
Eq.~(\ref{eq:EigEqGeneralNonCanonicalEnergyCasimir})
for the heavy top with $I_{1} = 1$, $I_{2} = 2$, and $I_{3} = 3$.
The equilibrium point is 
$\rho_{3} = 1$, $L_{3} = 3$, and $L_{1} = L_{2} = \rho_{1} = \rho_{2} = 0$.
The horizontal axis is the gravity parameter $G$.}
  \label{fig:htLinearStability-G-omg-ieq3}
\end{figure}

The first and the second examples are for an equilibrium with 
$\rho_{3} = 1$ and $L_{3} = 3$, and  $L_{1} = L_{2} = \rho_{1} = \rho_{2} = 0$.  
As shown in Sec.~\ref{subsubsec:htHamiltonianFinDoF}, this is an equilibrium point. 
Figure~\ref{fig:htLinearStability-G-omg-ieq3}
shows the real and the imaginary parts of $\omega$, as determined    by  
  Eq.~(\ref{eq:EigEqGeneralNonCanonicalEnergyCasimir}), 
for the heavy top as functions of the gravity parameter $G$.
Note that two of the six eigenvalues are zero, which is expected because of the existence of  two Casimirs \citep[see][]{pjmE86}, and these two are  not plotted.   The equilibrium is linearly stable when $G \leq 1$, and is unstable when
$G > 1$.  The  bifurcations at   $G=1$ and $G=2$ are steady state  bifurcations, i.e., they happen at $\omega=0$ and the two neutrally stable modes yield one  purely growing unstable mode with a corresponding purely damped mode. 

\begin{figure}[h]
    \centering
    \includegraphics[width=0.5\textwidth]{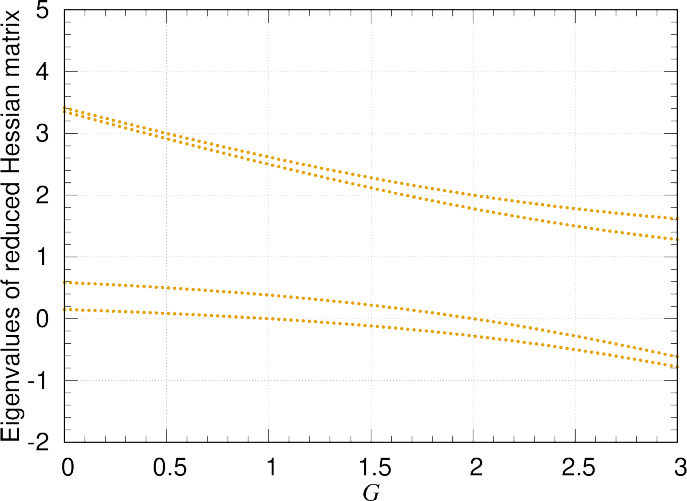}
  \caption{Eigenvalues of reduced Hessian matrix $A_{\mathrm{rHM}}$
for the heavy top with $I_{1} = 1$, $I_{2} = 2$, and $I_{3} = 3$.
The equilibrium was given by 
$\rho_{3} = 1$, $L_{3} = 3$, and $L_{1} = L_{2} = \rho_{1} = \rho_{2} = 0$.
The horizontal axis is the gravity parameter $G$.}
  \label{fig:ht-G-evHssr-ieq3}
\end{figure}

In Fig.~\ref{fig:ht-G-evHssr-ieq3}, 
eigenvalues of the Hessian matrix $( \pd^{2} F / ( \pd u^{i} \pd u^{j} ) )$
are plotted as functions of $G$.  
Here, the original Hessian matrix is a $6 \times 6$ matrix.
However, it includes two directions that are not allowed for
the system to evolve because of  the two Casimir invariants.  Therefore, two
dimensions were removed by using the linearized equations 
\begin{equation}
 \rho_{i} \delta \rho_{i}
=
 0 \qquad\mathrm{and}\qquad
 L_{i} \delta \rho_{i}
 + \rho_{i} \delta L_{i}  =
 0.
\label{eq:htCasimir2Linearized}
\end{equation}
For the equilibrium under consideration, 
we obtain $\delta L_{3} = 0$ and $\delta \rho_{3} = 0$.
Therefore, by using a four dimensional vector of perturbations 
$\delta \vu_{\mathrm{r}} := 
( \delta L_{1}, \delta L_{2}, \delta \rho_{1}, \delta \rho_{2})$,
the second variation of the energy-Casimir function $F$ can be written
as 
\begin{equation}
 \delta^{2} F 
= 
 \delta u_{\mathrm{r}}^{i} A_{\mathrm{rHM} ij} \delta u_{\mathrm{r}}^{j},
\end{equation}
where the subscript ``r'' and ``HM'' stands for ``reduced''
and ``Hessian Matrix'', respectively.
Note that the eigenvalues of $A_{\mathrm{rHM} ij}$ are always real.

When $G < 1$, there exist four positive eigenvalues,
which means that the system is an energy minimum on the Casimir leaf.
When $1 < G < 2$, there exist three positive and one negative
eigenvalues, meaning that a saddle exists, i.e., there is one neutral (stable) degree of freedom, and one unstable mode with its damped counterpart,  in this range of $G$.   In fact, 
we observe that $\Im \, \omega > 0$ in 
Fig.~\ref{fig:htLinearStability-G-omg-ieq3}\subref{subfig:htLinearStability-G-omgi-ieq3} 
showing linear
instability.  
When $G > 2$, there exists two positive and two negative eigenvalues.
In this case, another saddle appeared and we have two purely growing modes with their damped counterparts.   
This can be confirmed also in 
Fig.~\ref{fig:htLinearStability-G-omg-ieq3}\subref{subfig:htLinearStability-G-omgi-ieq3} 
where two $\Im \, \omega > 0$ and two  $\Im \, \omega < 0$ eigenvalues exist for $G > 2$.

\begin{figure}[h]
    \centering
    \includegraphics[width=0.5\textwidth]{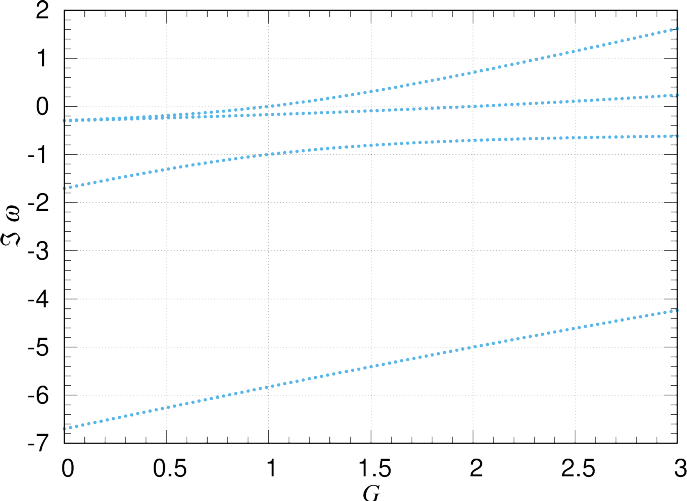}
  \caption{Imaginary parts of eigenvalues of linearized SA equation 
for the heavy top with $I_{1} = 1$, $I_{2} = 2$, and $I_{3} = 3$ and equilibrium with 
$\rho_{3} = 1$, $L_{3} = 3$, and $L_{1} = L_{2} = \rho_{1} = \rho_{2} = 0$.
The horizontal axis is the gravity parameter $G$.}
  \label{fig:htlinearSAp-G-omgi-4D-ieq3}
\end{figure}

Note that similar information can be obtained from the linearized
equations of SA.  
If we assume time dependence of the perturbation as 
$\rme^{-\rmi \omega t}$, we obtain an eigenvalue problem from the linearized
equations of SA. 
The imaginary part of $\omega$ corresponds to the eigenvalue of 
the reduced Hessian matrix $A_{\mathrm{rHM}}$.
If the equilibrium under consideration is at an energy minimum,
all $\Im \, \omega$ should be negative.  On the other hand,
if the equilibrium is not at an energy minimum, there should be at least
one positive $\Im \, \omega$.  
Figure~\ref{fig:htlinearSAp-G-omgi-4D-ieq3} shows $\Im \, \omega$
of the linearized SA equation.

\begin{figure}[h]
    \centering
    \includegraphics[width=0.5\textwidth]{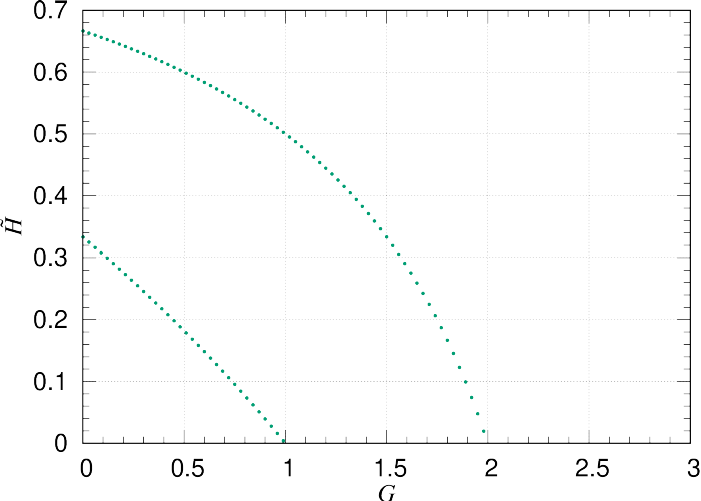}
  \caption{Mode energy $\tilde{H}$ 
for the heavy top with $I_{1} = 1$, $I_{2} = 2$, and $I_{3} = 3$.
The equilibrium was given by 
$\rho_{3} = 1$, $L_{3} = 3$, and $L_{1} = L_{2} = \rho_{1} = \rho_{2} = 0$.
The horizontal axis is the gravity parameter $G$.}
  \label{fig:ht-G-mdener-ieq3}
\end{figure}

Moreover, the  mode energy $\tilde{H}$ was calculated 
according to 
Eq.~(\ref{eq:mesureLinModeEnergyGeneralNonCanonicalEnergyCasimir})
by using the eigenmodes corresponding to the eigenvalue problem of the
original dynamics 
(\ref{eq:EigEqGeneralNonCanonicalEnergyCasimir}).
Figure~\ref{fig:ht-G-mdener-ieq3} shows $\tilde{H}$ as functions
of $G$.  As is to  be expected, 
two pairs of oscillatory modes have positive energies
in $0 \leq G < 1$, and a pair has a positive energy 
in $1 \leq G < 2$.
The pair of modes with $\Im \, \omega \neq 0$ has $\tilde{H} = 0$.
No negative energy mode exists in this equilibrium.

\begin{figure}[h]
  \begin{minipage}[t]{0.45\textwidth}
    \centering
    \includegraphics[width=\textwidth]{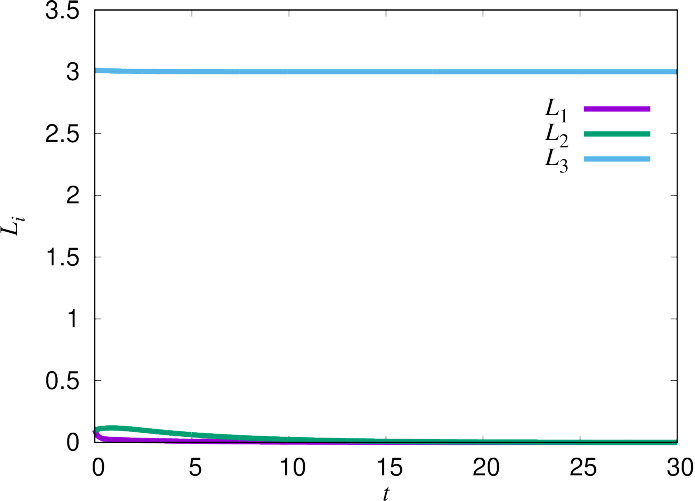}
    \subcaption{$\vL$.}
    \label{subfig:htSA-t-L-ieq3-G0_5}
  \end{minipage}
  \begin{minipage}[t]{0.45\textwidth}
    \centering
    \includegraphics[width=\textwidth]{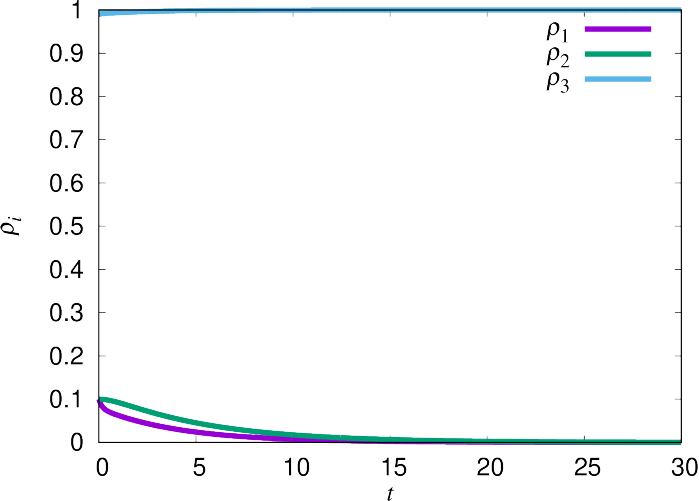}
    \subcaption{$\vecrho$.}
    \label{subfig:htSA-t-rho-ieq3-G0_5}
  \end{minipage}
  \begin{minipage}[t]{0.45\textwidth}
    \centering
    \includegraphics[width=\textwidth]{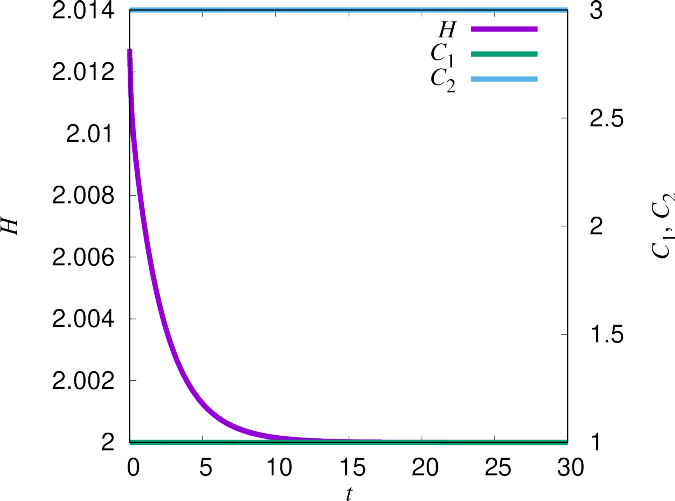}
    \subcaption{$H$, $C_{1}$ and $C_{2}$.}
    \label{subfig:htSA-t-H-C-ieq3-G0_5}
  \end{minipage}
  \caption{Time evolution of SA 
for the heavy top with $I_{1} = 1$, $I_{2} = 2$, and $I_{3} = 3$.
The equilibrium was given by $\rho_{3} = 1$ and $L_{3} = 3$.
The gravity parameter was $G = 0.5$. 
Since the equilibrium is stable, the original equilibrium was recovered
 by SA.}
  \label{fig:htSA-t-L-rho-H-C-ieq3-G0_5}
\end{figure}

Now, let us show SA results.
The numerical results shown here  use  the unit matrix as the symmetric
kernel $K$. 
First, the time evolution for $G = 0.5$ is shown in 
Fig.~\ref{fig:htSA-t-L-rho-H-C-ieq3-G0_5}.
The initial perturbation was given so that the perturbed state has the
same values of $C_{1}$ and $C_{2}$.
Explicitly, $L_{1} = 0.1$, $L_{2} = 0.1$, $L_{3} =3.010$, 
$\rho_{1} =0.1$ $\rho_{2} =0.1$, and $\rho_{3} = 0.990$.
As the time proceeds, the energy $H$ decreased
as seen in 
Fig.~\ref{fig:htSA-t-L-rho-H-C-ieq3-G0_5}\subref{subfig:htSA-t-H-C-ieq3-G0_5},
and the system approaches the equilibrium.
During the time evolution  $C_{1} = | \vecrho |^{2}$ and 
$C_{2} = \vL \cdot \vecrho$  were conserved.
This result was to be expected since the equilibrium has two
positive energy modes for $G = 0.5$.

\begin{figure}[h]
  \begin{minipage}[t]{0.45\textwidth}
    \centering
    \includegraphics[width=\textwidth]{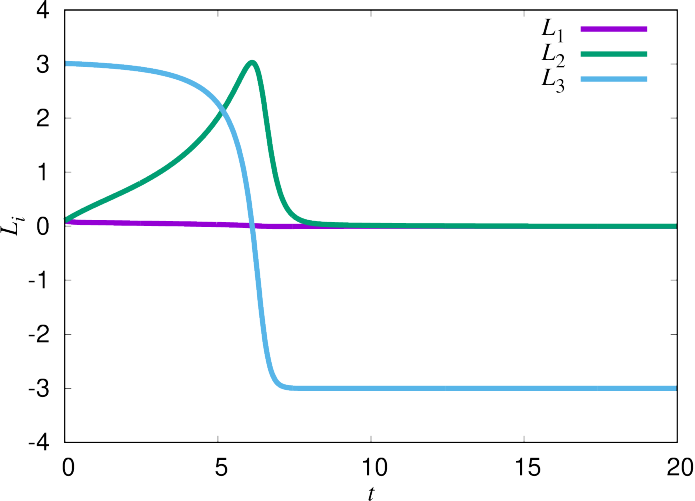}
    \subcaption{$\vL$.}
    \label{subfig:htSA-t-L-ieq3-G1_5}
  \end{minipage}
  \begin{minipage}[t]{0.45\textwidth}
    \centering
    \includegraphics[width=\textwidth]{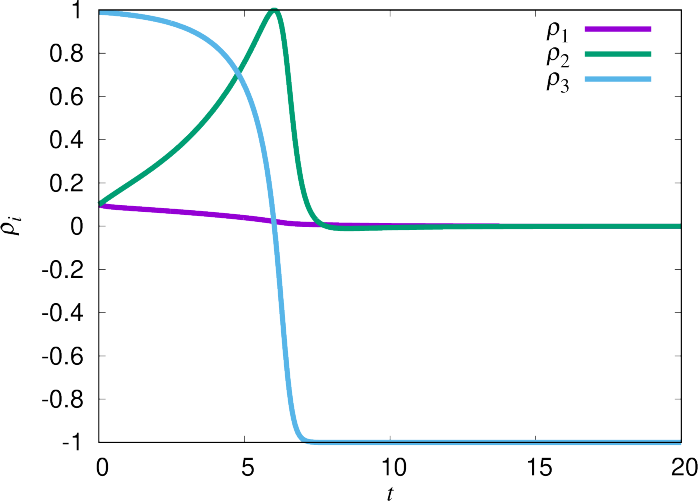}
    \subcaption{$\vecrho$.}
    \label{subfig:htSA-t-rho-ieq3-G1_5}
  \end{minipage}
  \begin{minipage}[t]{0.45\textwidth}
    \centering
    \includegraphics[width=\textwidth]{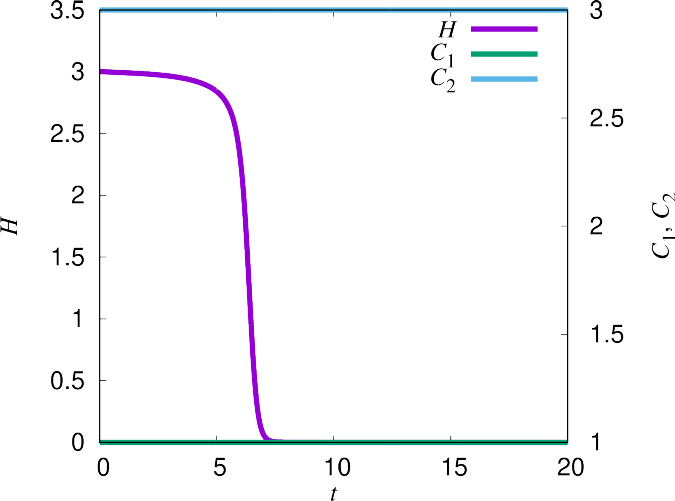}
    \subcaption{$H$, $C_{1}$ and $C_{2}$.}
    \label{subfig:htSA-t-H-C-ieq3-G1_5}
  \end{minipage}
  \caption{Time evolution of SA 
for the heavy top with $I_{1} = 1$, $I_{2} = 2$, and $I_{3} = 3$.
The equilibrium was given by $\rho_{3} = 1$ and $L_{3} = 3$.
The gravity parameter was $G = 1.5$.
Since the equilibrium is unstable, another equilibrium was reached by SA.}
  \label{fig:htSA-t-L-rho-H-C-ieq3-G1_5}
\end{figure}

Next, the time evolution for $G = 1.5$ is shown in 
Fig.~\ref{fig:htSA-t-L-rho-H-C-ieq3-G1_5}.
The initial perturbation was given 
similarly as in the case of 
Fig.~\ref{fig:htSA-t-L-rho-H-C-ieq3-G0_5} 
so that the perturbed state has the
same values of $C_{1}$ and $C_{2}$.
As the time proceeds, the energy $H$ decreased
as seen in 
Fig.~\ref{fig:htSA-t-L-rho-H-C-ieq3-G1_5}\subref{subfig:htSA-t-H-C-ieq3-G1_5},
and $C_{1} = | \vecrho |^{2}$ and $C_{2} = \vL \cdot \vecrho$ were
conserved. 
In this case, another equilibrium 
$L_{3} = -3$ and $\rho_{3} = -1$,
$L_{1} = L_{2} = \rho_{1} = \rho_{2} = 0$ was reached by SA.
This is because the original equilibrium is unstable for $G = 1.5$.
Similar time evolution of SA was obtained for $G=2.5$ 
since the equilibrium is unstable.

\begin{figure}[h]
  \begin{minipage}[t]{0.45\textwidth}
    \centering
    \includegraphics[width=\textwidth]{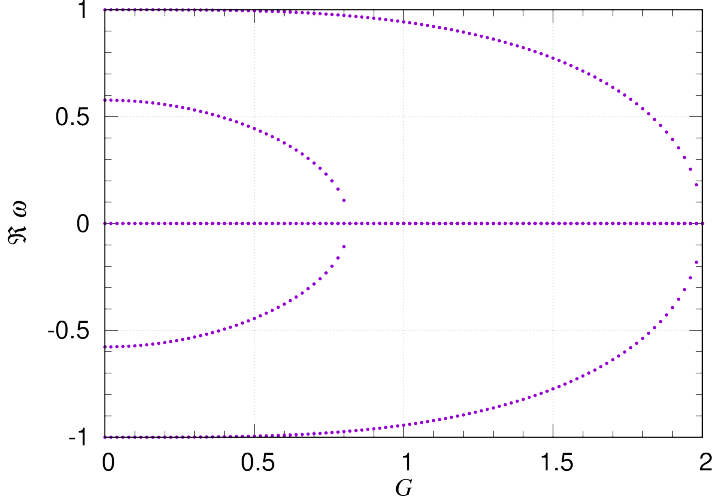}
    \subcaption{$\Re \, \omega$ of eigenvalue problem of the original
   dynamics.}
    \label{subfig:htLinearStability-G-omgr-ieq1}
  \end{minipage}
  \begin{minipage}[t]{0.45\textwidth}
    \centering
    \includegraphics[width=\textwidth]{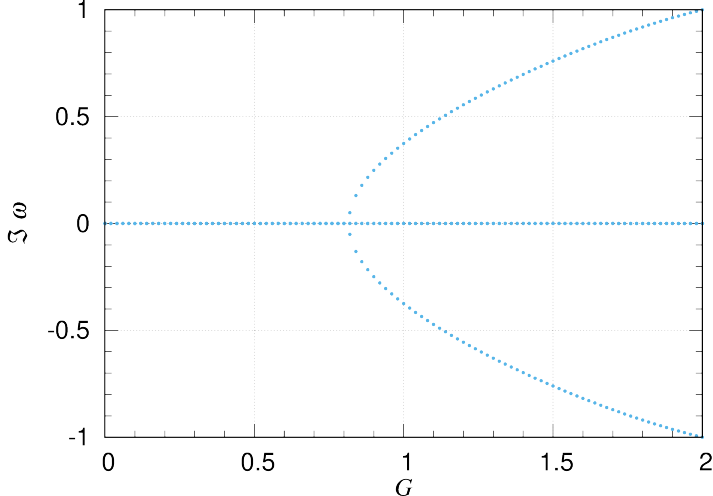}
    \subcaption{$\Im \, \omega$ of eigenvalue problem of the original
   dynamics.}
    \label{subfig:htLinearStability-G-omgi-ieq1}
  \end{minipage}
  \begin{minipage}[t]{0.45\textwidth}
    \centering
    \includegraphics[width=\textwidth]{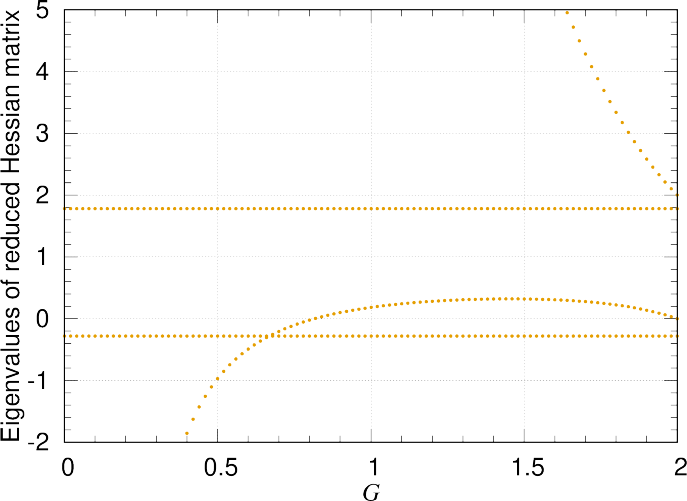}
    \subcaption{Eigenvalues of reduced Hessian matrix $A_{\mathrm{rHM}}$.}
    \label{subfig:ht-G-evHssr-ieq1}
  \end{minipage}
  \begin{minipage}[t]{0.45\textwidth}
    \centering
    \includegraphics[width=\textwidth]{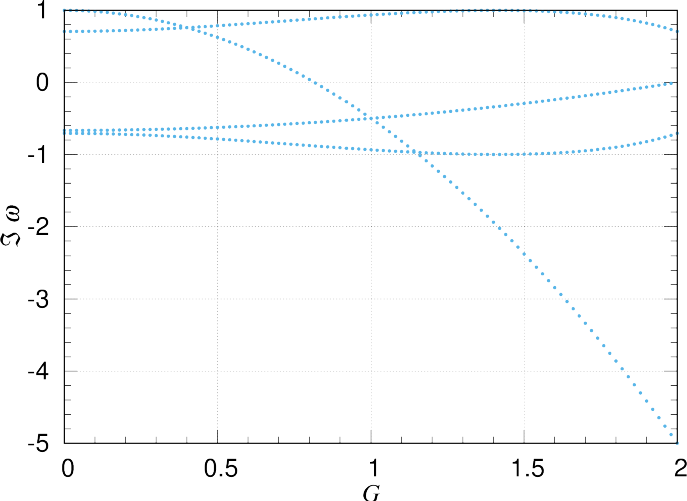}
    \subcaption{Imaginary parts of eigenvalues of linearized SA equation.}
    \label{subfig:htlinearSAp-G-omgi-4D-ieq1}
  \end{minipage}
  \begin{minipage}[t]{0.45\textwidth}
    \centering
    \includegraphics[width=\textwidth]{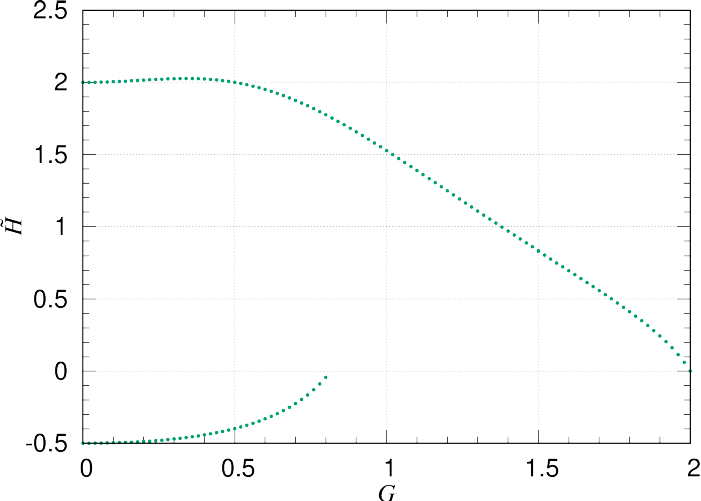}
    \subcaption{Mode energy $\tilde{H}$.}
    \label{subfig:ht-G-mdener-ieq1}
  \end{minipage}
  \caption{Real and imaginary parts of $\omega$ of 
eigenvalue problem for the linearized equations of the original dynamics
 Eq.~(\ref{eq:EigEqGeneralNonCanonicalEnergyCasimir}),
eigenvalues of the reduced Hessian matrix $A_{\mathrm{rHM}}$,
imaginary parts of eigenvalues of linearized SA equation, 
and the mode energy $\tilde{H}$
were shown 
for the heavy top with $I_{1} = 1$, $I_{2} = 2$, and $I_{3} = 3$.
The equilibrium was given by 
$L_{1} = 0.968$, $L_{2} = 0$, $L_{3} = 0.75$,
$\rho_{1} = 0.968$, $\rho_{2} = 0$, $\rho_{3} = 0.25$.
The horizontal axis is the gravity parameter $G$.}
  \label{fig:htLinearStability-G-omg-ieq1}
\end{figure}

Another numerical example is for an equilibrium with a pair of negative
energy modes.
The principal moments of inertia were chosen to be
$I_{1} = 1$, $I_{2} = 2$, and $I_{3} = 3$, which were same as in the
previous cases.
The equilibrium was chosen to have 
$L_{1} = 0.968$, $L_{2} = 0$, $L_{3} = 0.75$,
$\rho_{1} = 0.968$, $\rho_{2} = 0$, $\rho_{3} = 0.25$.
Figures~\ref{fig:htLinearStability-G-omg-ieq1}\subref{subfig:htLinearStability-G-omgr-ieq1}
and 
\ref{fig:htLinearStability-G-omg-ieq1}\subref{subfig:htLinearStability-G-omgi-ieq1}
shows the real and the imaginary parts of $\omega$ 
for the linearized equations of the original dynamics 
Eq.~(\ref{eq:EigEqGeneralNonCanonicalEnergyCasimir}),
respectively.
Note that this equilibrium exists only for $0 \leq G \leq 2$,
and becomes linearly unstable for $G \gtrsim 0.8$.

Figure~\ref{fig:htLinearStability-G-omg-ieq1}\subref{subfig:ht-G-evHssr-ieq1} shows eigenvalues of the
reduced Hessian matrix $A_{\mathrm{rHM}}$.
When $0 \leq G \lesssim 0.8$, two positive and two negative eigenvalues
exist.   Given only  the information shown in 
Fig.~\ref{fig:htLinearStability-G-omg-ieq1}\subref{subfig:ht-G-evHssr-ieq1},
the situation cannot be entirely identified:  either there is 
(i) a pair of positive energy modes and a pair of negative energy modes
or
(ii) there are two saddles.  
For $G \gtrsim 0.8$, we can identify that 
there exists a saddle and a pair of positive energy modes.

Figure~\ref{fig:htLinearStability-G-omg-ieq1}\subref{subfig:htlinearSAp-G-omgi-4D-ieq1}
shows eigenvalues of the linearized SA equation, where the kernel $K$ was
chosen to be the unit matrix.  Then the energy of the system
monotonically decreases as time proceeds.
There exist two positive and two negative eigenvalues 
for $0 \leq G \lesssim 0.8$, while one positive and three negative eigenvalues 
for $ 0.8 \lesssim G \leq 2$.
The existence of the positive eigenvalues of the linearized SA equation
indicates that the dynamics in the direction corresponding to these
eigenvectors is  unstable.  This is true even for $0 \leq G \lesssim 0.8$.
However, the original dynamics shows linear stability for 
$0 \leq G \lesssim 0.8$.
This indicates an existence of a pair of negative energy modes.
Figure~\ref{fig:htLinearStability-G-omg-ieq1}\subref{subfig:ht-G-mdener-ieq1} shows the mode energy
$\tilde{H}$.  Whence, it is  clear that there exists negative energy modes
for $0 \leq G \lesssim 0.8$.  Thus we see that SA can be used to identify negative energy modes. 

\begin{figure}[h]
  \begin{minipage}[t]{0.45\textwidth}
    \centering
    \includegraphics[width=\textwidth]{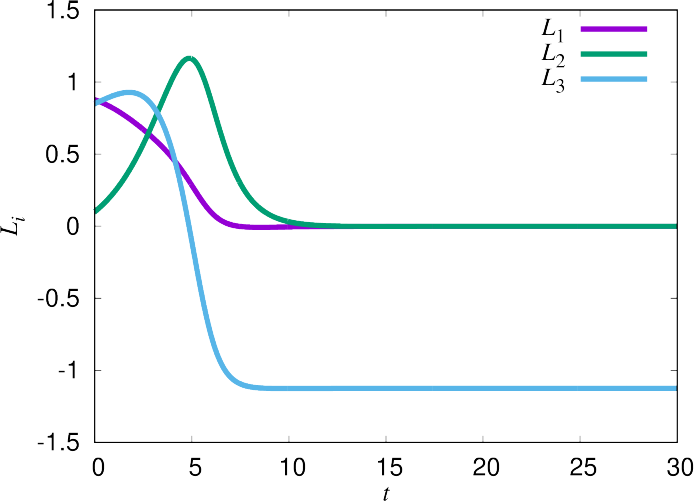}
    \subcaption{$\vL$.}
    \label{subfig:htSA-t-L-ieq1-G0_5}
  \end{minipage}
  \begin{minipage}[t]{0.45\textwidth}
    \centering
    \includegraphics[width=\textwidth]{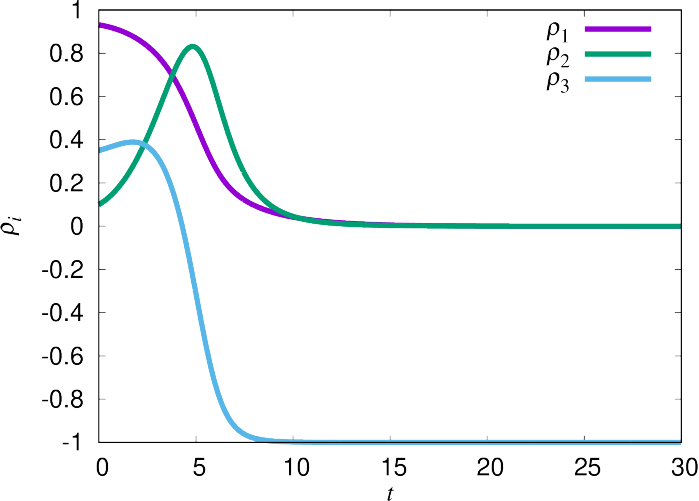}
    \subcaption{$\vecrho$.}
    \label{subfig:htSA-t-rho-ieq1-G0_5}
  \end{minipage}
  \begin{minipage}[t]{0.45\textwidth}
    \centering
    \includegraphics[width=\textwidth]{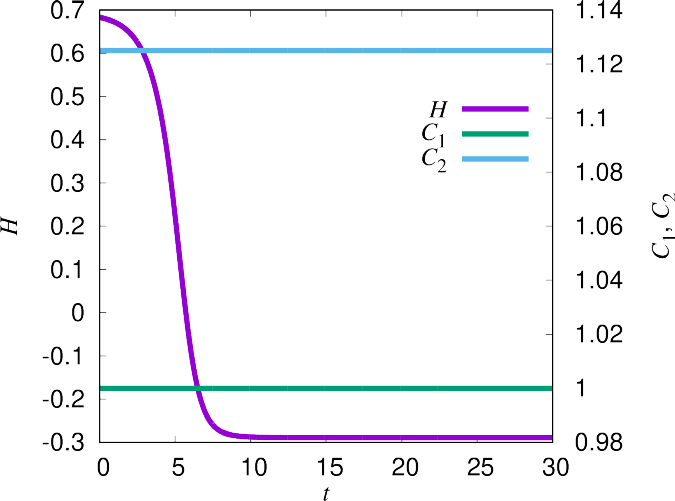}
    \subcaption{$H$, $C_{1}$ and $C_{2}$.}
    \label{subfig:htSA-t-H-C-ieq1-G0_5}
  \end{minipage}
  \caption{Time evolution of SA 
for the heavy top with $I_{1} = 1$, $I_{2} = 2$, and $I_{3} = 3$.
The equilibrium is 
$L_{1} = 0.968$, $L_{2} = 0$, $L_{3} = 0.75$,
$\rho_{1} = 0.968$, $\rho_{2} = 0$, $\rho_{3} = 0.25$.
The gravity parameter is $G = 0.5$. 
Since the equilibrium is stable, the original equilibrium was recovered
 by SA.}
  \label{fig:htSA-t-L-rho-H-C-ieq1-G0_5}
\end{figure}

Now, time evolution of SA is shown in 
Fig.~\ref{fig:htSA-t-L-rho-H-C-ieq1-G0_5}.
The gravity parameter was chosen to be $G = 0.5$,
where the negative energy modes exist.
The initial condition was 
$L_{1} = 0.878$, $L_{2} = 0.1$, $L_{3} =0.85$,
$\rho_{1} =0.93$, $\rho_{2} = 0.1$, and $\rho_{3} = 0.35$,
which has the same values for the  Casimir invariants as those for  the equilibrium.

The energy of the system monotonically decreases as time proceeds,
while the Casimir invariants,  $C_{1} = | \vecrho |^{2}$ and 
$C_{2} = \vL \cdot \vecrho$ in this case, were conserved 
as shown in 
Fig.~\ref{fig:htSA-t-L-rho-H-C-ieq1-G0_5}\subref{subfig:htSA-t-H-C-ieq1-G0_5}.
We also observe 
in Figs.~\ref{fig:htSA-t-L-rho-H-C-ieq1-G0_5}\subref{subfig:htSA-t-L-ieq1-G0_5} 
and \ref{fig:htSA-t-L-rho-H-C-ieq1-G0_5}\subref{subfig:htSA-t-rho-ieq1-G0_5}
that the system did not recover the original equilibrium and reached
another equilibrium.
The existence of the negative energy modes explains this behavior.

The last case of this subsection is a  DSA result.
Let us introduce a new constant 
$C_{3} := \rho_{3}$.
Then, the Dirac bracket is constructed according to 
Eq.~(\ref{eq:DiracBracketFinDoF}).
The counterpart of  $C_{3}$ is given by 
\begin{equation}
 C_{4}
\coloneqq
 [ C_{3}, H ]
=
 \frac{L_{2} \rho_{1}}{I_{2}}
-\frac{L_{1} \rho_{2}}{I_{1}}.
\label{eq:htDiracBracketC3st-2}
\end{equation}
If $C_{3}$ is kept unchanged during the time evolution, 
$C_{4}$ must be always zero since $\dot{C}_{3} = [ C_{3} , H ] \equiv 0$.
The Dirac bracket is properly defined when 
either of $\rho_{1}$ or $\rho_{2}$ is not zero
since 
\begin{equation}
 [ C_{3} , C_{4} ]
=
  \frac{\rho_{1}^{2}}{I_{2}}
+ \frac{\rho_{2}^{2}}{I_{1}}.
\label{eq:htPoissonBracketC3C3st}
\end{equation}
In other words, this formulation breaks down when $\rho_{1} = \rho_{2} =
0$.  

The initial condition for the DSA run is chosen to be 
$L_{1} = 0.878$, $L_{2} = 0.1$, $L_{3} =0.85$,
$\rho_{1} =0.93$, $\rho_{2} = 0.1$, and $\rho_{3} = 0.35$,
which is a perturbed state of an equilibrium with 
$L_{1} = 0.968$, $L_{2} = 0$, $L_{3} = 0.75$,
$\rho_{1} = 0.968$, $\rho_{2} = 0$, $\rho_{3} = 0.25$.
This initial condition is the same as the one for the case of 
Fig.~\ref{fig:htSA-t-L-rho-H-C-ieq1-G0_5}.
The gravity parameter $G=0.5$ was also chosen to be the same as that 
for 
Fig.~\ref{fig:htSA-t-L-rho-H-C-ieq1-G0_5}.

The kernel $K$ for the double bracket was again chosen to be the unit matrix
so that the energy of the system monotonically decreases by  DSA.
If we use the ordinary Poisson bracket for constructing the double
bracket, the SA lead to an equilibrium with $\rho_{3} = -1$ 
that is different from the original
equilibrium without a perturbation as shown in 
Fig.~\ref{fig:htSA-t-L-rho-H-C-ieq1-G0_5}.

\begin{figure}[h]
  \begin{minipage}[t]{0.45\textwidth}
    \centering
    \includegraphics[width=\textwidth]{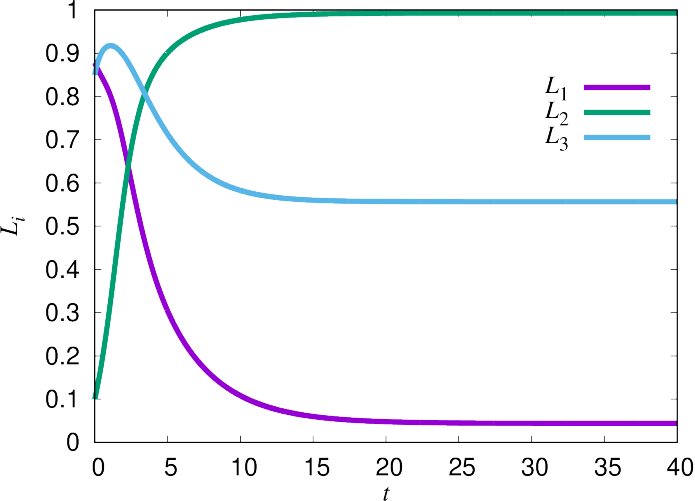}
    \subcaption{$\vL$.}
    \label{subfig:htDSA-t-L-ieq1-G0_5}
  \end{minipage}
  \begin{minipage}[t]{0.45\textwidth}
    \centering
    \includegraphics[width=\textwidth]{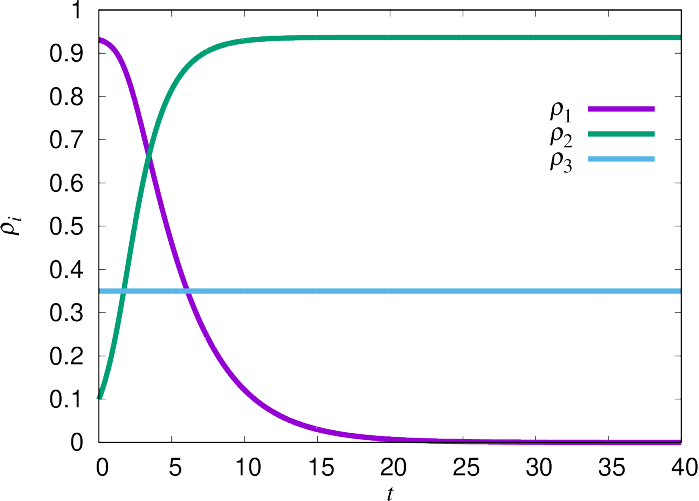}
    \subcaption{$\vecrho$.}
    \label{subfig:htDSA-t-rho-ieq1-G0_5}
  \end{minipage}
  \begin{minipage}[t]{0.45\textwidth}
    \centering
    \includegraphics[width=\textwidth]{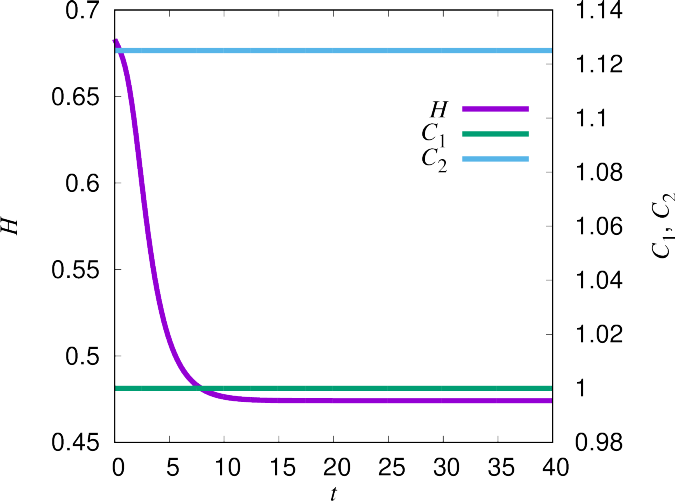}
    \subcaption{$H$, $C_{1}$ and $C_{2}$.}
    \label{subfig:htDSA-t-H-C-ieq1-G0_5}
  \end{minipage}
  \caption{Time evolution of DSA 
for the heavy top with $I_{1} = 1$, $I_{2} = 2$, and $I_{3} = 3$.
The initial condition 
$L_{1} = 0.878$, $L_{2} = 0.1$, $L_{3} =0.85$,
$\rho_{1} =0.93$, $\rho_{2} = 0.1$, $\rho_{3} = 0.35$,
as well as the gravity parameter 
$G=0.5$ were the same as for 
Fig.~\ref{fig:htSA-t-L-rho-H-C-ieq1-G0_5}.
Because of the Dirac constraint, $\rho_{3}$ was kept unchanged from the
  initial condition during DSA.
}
  \label{fig:htDSA-t-L-rho-H-C-ieq1-G0_5}
\end{figure}

Figure~\ref{fig:htDSA-t-L-rho-H-C-ieq1-G0_5} shows 
time evolution of $\vL$, $\vecrho$, $H$, $C_{1} = | \vecrho |^{2}$ 
and $C_{2} = \vL \cdot \vecrho$.
As observed in 
Fig.~\ref{fig:htDSA-t-L-rho-H-C-ieq1-G0_5}\subref{subfig:htDSA-t-rho-ieq1-G0_5},
$C_{3} = \rho_{3}$ was successfully conserved, it remaining  at  its  initial value.

Note that the final state is not an equilibrium originally.
It is a stationary state where the top is somehow supported at a tilted
angle.  Without such a support, the top will flip over to get 
$\rho_{3} = -1$ as in the case of 
Fig.~\ref{fig:htSA-t-L-rho-H-C-ieq1-G0_5}.

The number of Dirac constraints can be increased further.
We have confirmed that $\rho_{1}$ in addition to $\rho_{3}$ can be fixed
at the initial value by adding $C_{5} = \rho_{1}$
and $C_{6} = [ C_{5} , H ]$.
In this case, $\rho_{2}$ is also fixed at the initial condition since 
$C_{1} = | \vecrho |^{2} = 1$ is conserved.
On the other hand, $\vL$ can change in time while keeping 
$C_{2} = \vL \cdot \vecrho$.

\subsection{A toy model mimicking low-beta reduced MHD}
\label{subsec:SAmimicLBRMHD}

Equilibrium and stability analyses similar to the heavy top presented in 
Sec.~\ref{subsec:htSA}  can be performed   for the  toy model of Sec.~\ref{subsubsec:HamiltonianFinDoFMimicLBRMHD} that mimics an aspect of  low-beta reduced MHD.   In the present subsection,  numerical results examining effects of addition of 
Hamiltonian dynamics to SA dynamics are presented.  This toy model was created to answer whether the addition of the
Hamiltonian dynamics to SA dynamics can accelerate relaxation to an
equilibrium.   We have tried  some numerical tests, and the results show  that the
relaxation was not affected significantly.   On the other hand,  the relaxation can be either accelerated or decelerated by the addition
of the Hamiltonian dynamics to SA for  low-beta reduced MHD
as shown in Sec.~\ref{subsec:additionHamiltonianDynamics}.
Therefore, we need to further investigate what determines the fastest
path to the equilibrium both analytically and numerically.  Examination of toy models like the present one, however, may shed light on this important issue. 

Here, we solve
\begin{equation}
 \dot{u}^{i} = \tilde{f}^{i} + c f^{i},
\label{eq:mimicLBMHDEvolutionEqMixHamiltonian}
\end{equation}
where 
\begin{equation}
 f^{i}
\coloneqq
 [ u^{i} , H ] 
\qquad \mathrm{and}\qquad
 \tilde{f}^{i}
\coloneqq
 [ u^{i}, u^{j} ] K_{jk} [ u^{k} , H ].
\label{eq:mimicLBMHDVectorFieldSA}
\end{equation}
Note that the kernel $K$ in Eq.~(\ref{eq:mimicLBMHDVectorFieldSA})
is taken to be the unit matrix.  The parameters are  chosen to be 
$I_{1} = I_{2} = I_{3} = 1$, 
$M_{1} = M_{2} = 2$, and $M_{3} = 1$, while the  equilibrium considered  is 
$L_{1} = L_{2} = 0$, $L_{3} = 1/2$, 
$\rho_{1} = \rho_{2} = 0$, and $\rho_{3} = 1$.
The Casimir invariants are chosen to be 
$C_{1} = | \vecrho |^{2} = 1$ 
and 
$C_{2} = \vL \cdot \vecrho = 1/2$.
This equilibrium is linearly stable with positive energy modes only.

\begin{figure}[h]
 \begin{minipage}[t]{0.32\textwidth}
  \centering
  \includegraphics[width=\textwidth]{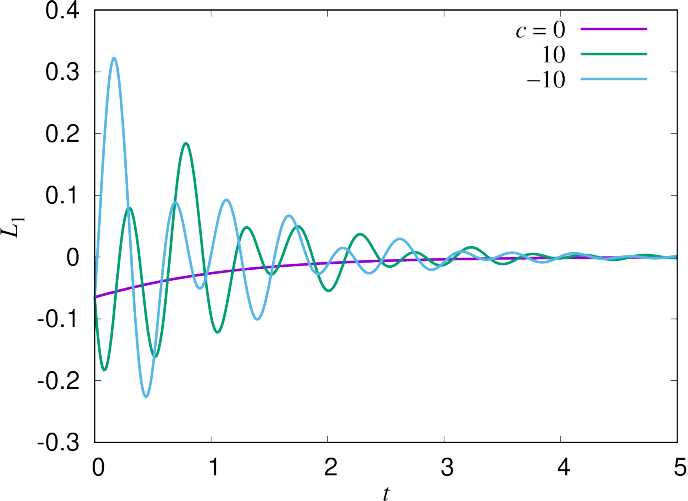}
  \subcaption{$L_{1}$.}
  \label{subfig:mimicLBRMHD-t-L1-c-case1}
 \end{minipage}
\hfill
 \begin{minipage}[t]{0.32\textwidth}
  \centering
  \includegraphics[width=\textwidth]{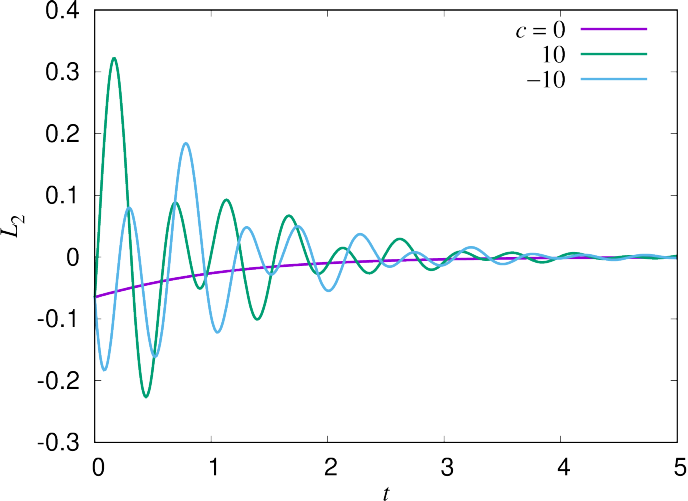}
  \subcaption{$L_{2}$.}
  \label{subfig:mimicLBRMHD-t-L2-c-case1}
 \end{minipage}
\hfill
 \begin{minipage}[t]{0.32\textwidth}
  \centering
  \includegraphics[width=\textwidth]{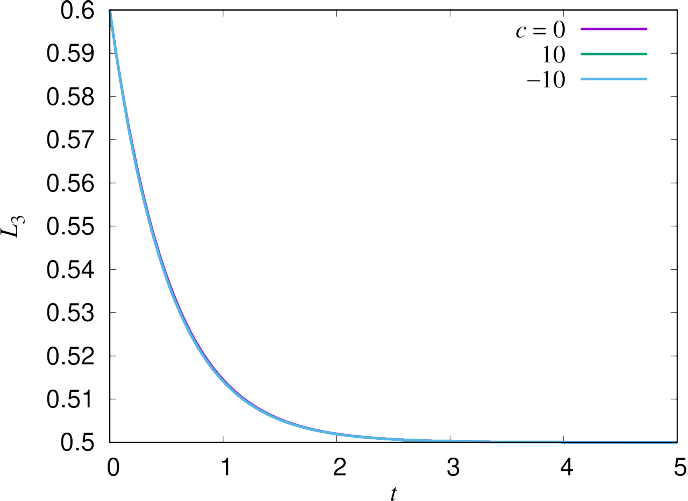}
  \subcaption{$L_{3}$.}
  \label{subfig:mimicLBRMHD-t-L3-c-case1}
 \end{minipage}
\\
 \begin{minipage}[t]{0.32\textwidth}
  \centering
  \includegraphics[width=\textwidth]{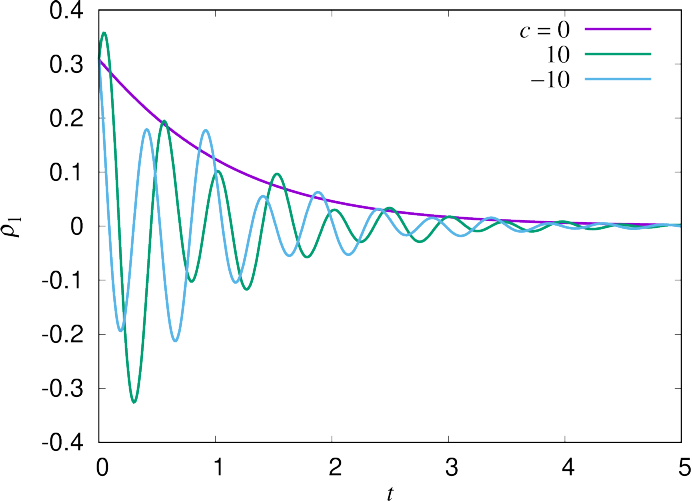}
  \subcaption{$\rho_{1}$.}
  \label{subfig:mimicLBRMHD-t-rho1-c-case1}
 \end{minipage}
\hfill
 \begin{minipage}[t]{0.32\textwidth}
  \centering
  \includegraphics[width=\textwidth]{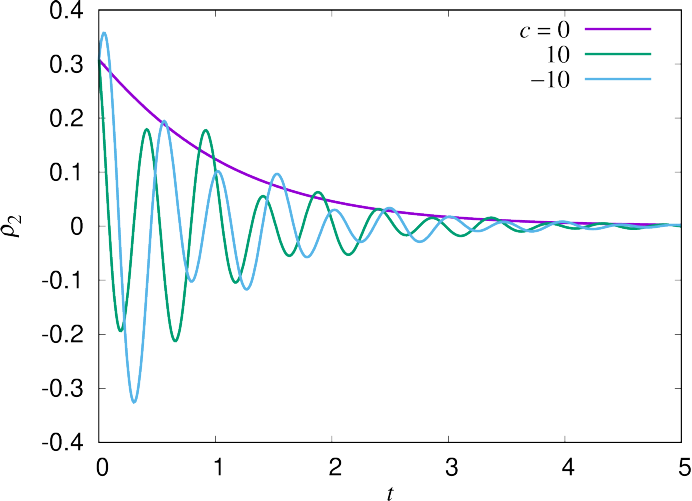}
  \subcaption{$\rho_{2}$.}
  \label{subfig:mimicLBRMHD-t-rho2-c-case1}
 \end{minipage}
\hfill
 \begin{minipage}[t]{0.32\textwidth}
  \centering
  \includegraphics[width=\textwidth]{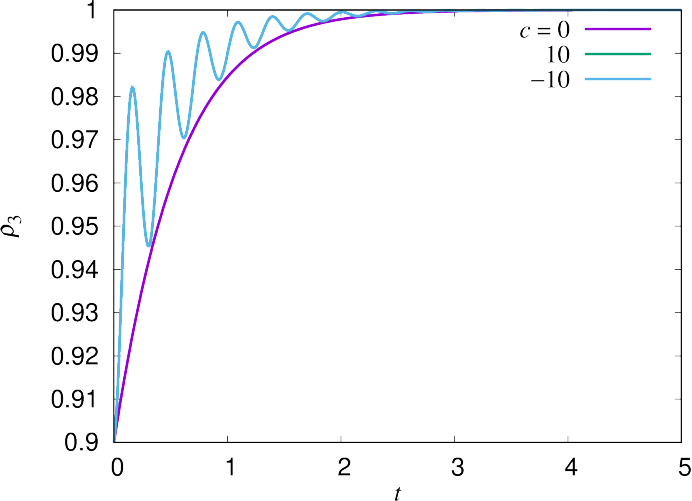}
  \subcaption{$\rho_{3}$.}
  \label{subfig:mimicLBRMHD-t-rho3-c-case1}
 \end{minipage}
\\
 \begin{minipage}[t]{0.32\textwidth}
  \centering
  \includegraphics[width=\textwidth]{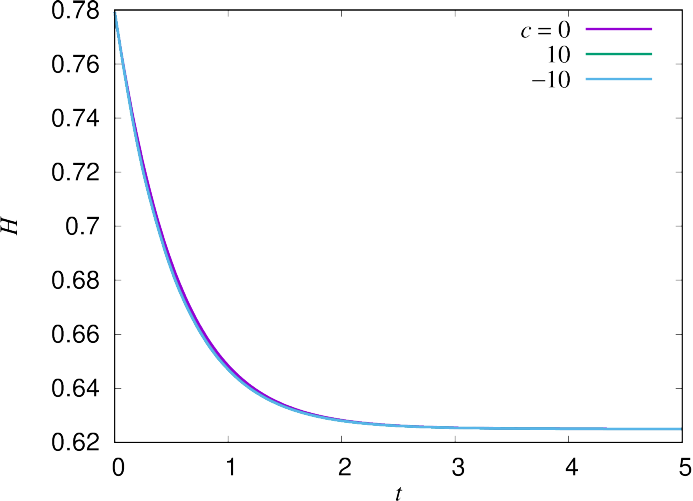}
  \subcaption{$H$.}
  \label{subfig:mimicLBRMHD-t-H-c-case1}
 \end{minipage}
\hfill
 \begin{minipage}[t]{0.32\textwidth}
  \centering
  \includegraphics[width=\textwidth]{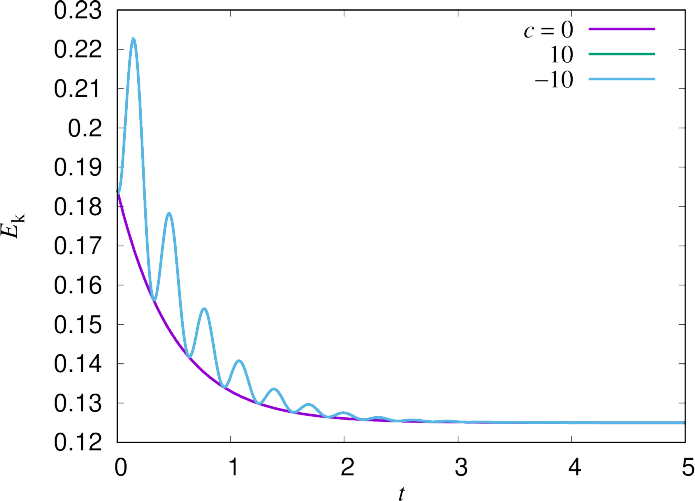}
  \subcaption{$E_{\mathrm{k}}$.}
  \label{subfig:mimicLBRMHD-t-Ek-c-case1}
 \end{minipage}
\hfill
 \begin{minipage}[t]{0.32\textwidth}
  \centering
  \includegraphics[width=\textwidth]{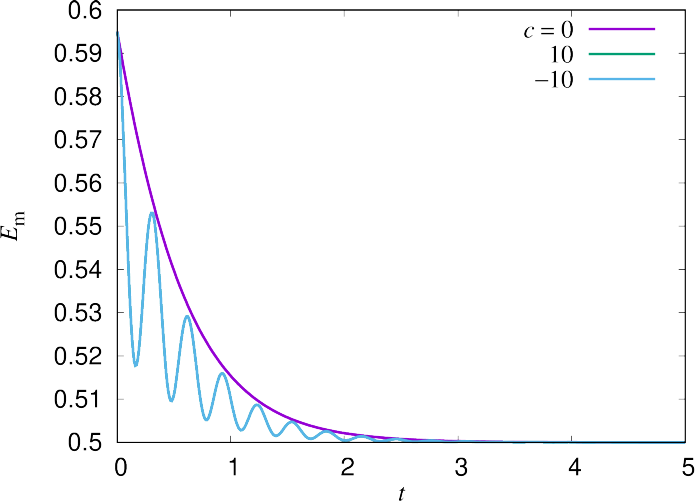}
  \subcaption{$E_{\mathrm{m}}$.}
  \label{subfig:mimicLBRMHD-t-Em-c-case1}
 \end{minipage}
 \caption{
Time evolution of 
Eq.~(\ref{eq:mimicLBMHDEvolutionEqMixHamiltonian})
where Hamiltonian dynamics is added to SA dynamics of the toy model
 mimicking low-beta reduced MHD with 
$I_{1} = I_{2} = I_{3} = 1$, 
$M_{1} = M_{2} = 2$, and $M_{3} = 1$.
The initial condition was a state perturbed from a stable equilibrium
with 
$L_{1} = L_{2} = 0$, $L_{3} = 1/2$, 
$\rho_{1} = \rho_{2} = 0$, and $\rho_{3} = 1$.
}
 \label{fig:mimicLBRMHD-t-L-rho-H-Ek-Em-c-case1}
\end{figure}

The initial condition for SA was chosen  to be a perturbation away from  the equilibrium
of the previous paragraph,  with 
$L_{1} = L_{2} = -0.0649$, $L_{3} = 0.6$, 
$\rho_{1} = \rho_{2} = 0.308$ and $\rho_{3} = 0.9$.
Figure~\ref{fig:mimicLBRMHD-t-L-rho-H-Ek-Em-c-case1} shows comparison of
time evolutions of the variables with $c=0$ and $c=\pm 10$.
A negative $c$ means that the time-reversed Hamiltonian dynamics is
added to the SA dynamics.
Figures~\ref{fig:mimicLBRMHD-t-L-rho-H-Ek-Em-c-case1}\subref{subfig:mimicLBRMHD-t-L1-c-case1},
\subref{subfig:mimicLBRMHD-t-L2-c-case1} 
and \subref{subfig:mimicLBRMHD-t-L3-c-case1}
show time evolution of $L_{1}$, $L_{2}$ and $L_{3}$, respectively.
Similarly, 
Figs~\ref{fig:mimicLBRMHD-t-L-rho-H-Ek-Em-c-case1}\subref{subfig:mimicLBRMHD-t-rho1-c-case1},
\subref{subfig:mimicLBRMHD-t-rho2-c-case1} 
and \subref{subfig:mimicLBRMHD-t-rho3-c-case1}
show time evolution of $\rho_{1}$, $\rho_{2}$ and $\rho_{3}$, respectively.
Figure~\ref{fig:mimicLBRMHD-t-L-rho-H-Ek-Em-c-case1}\subref{subfig:mimicLBRMHD-t-H-c-case1} 
shows the time evolution of the energy.
In
Fig.~\ref{fig:mimicLBRMHD-t-L-rho-H-Ek-Em-c-case1}\subref{subfig:mimicLBRMHD-t-Ek-c-case1},
$E_{\mathrm{k}}$ is the $\vL$ term in the Hamiltonian 
(\ref{eq:mimicLBRMHDHamiltonian}),
while $E_{\mathrm{m}}$ in
Fig.~\ref{fig:mimicLBRMHD-t-L-rho-H-Ek-Em-c-case1}\subref{subfig:mimicLBRMHD-t-Em-c-case1}
is the $\vecrho$ term.
As seen in 
Fig.~\ref{fig:mimicLBRMHD-t-L-rho-H-Ek-Em-c-case1},
the relaxation to a stationary value of $H$ did not differ by much for the different values  of $c$, although each variable showed different time
evolution except for $L_{3}$.
Note that we have also tried $c = \pm 100$, and observed that 
the time evolution of $H$ did not differ much.

\begin{figure}[h]
 \begin{minipage}[t]{0.45\textwidth}
  \centering
  \includegraphics[width=\textwidth]{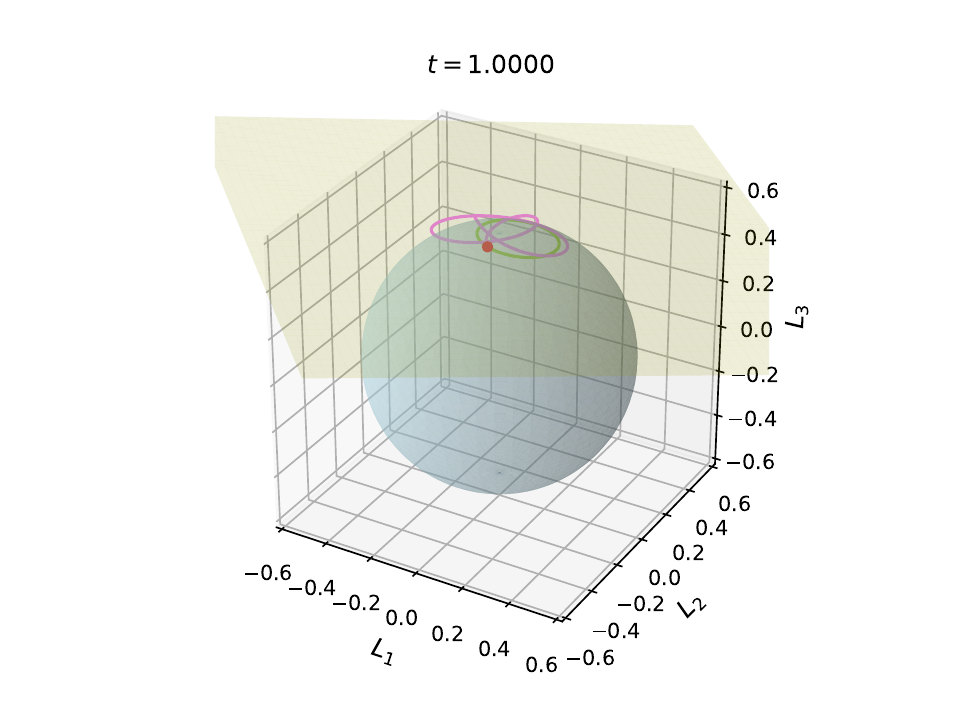}
  \subcaption{$\vL$ space. $c = -10$. $t = 1$.}
  \label{subfig:mimicLBRMHD-L-cm10-t1-case1}
 \end{minipage}
\hfill
 \begin{minipage}[t]{0.45\textwidth}
  \centering
  \includegraphics[width=\textwidth]{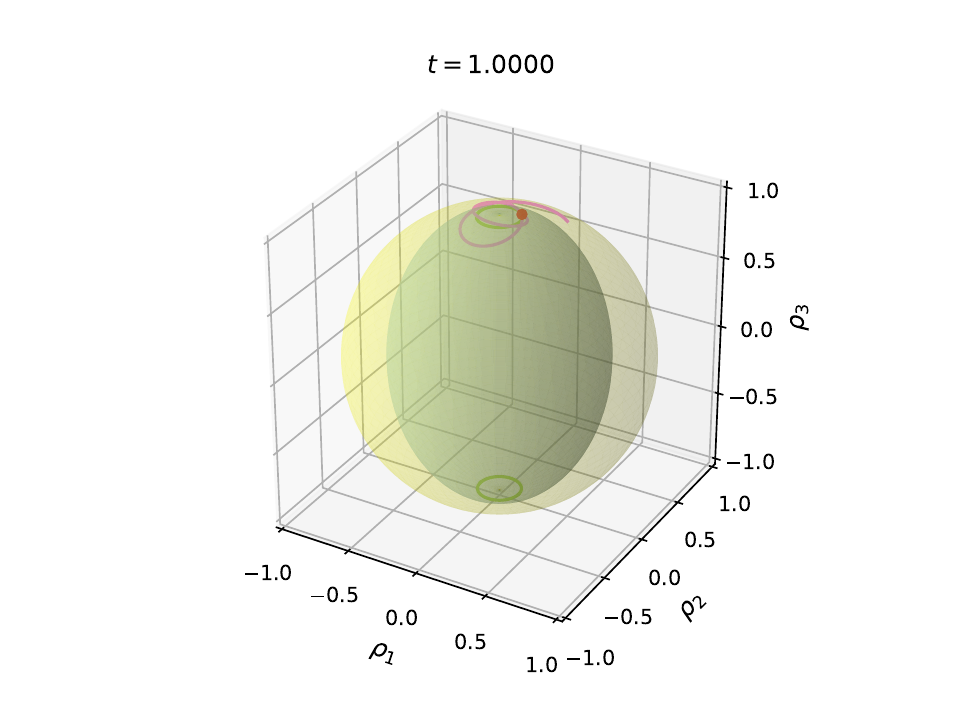}
  \subcaption{$\vecrho$ space. $c = -10$. $t = 1$.}
  \label{subfig:mimicLBRMHD-rho-cm10-t1-case1}
 \end{minipage}
\\
 \begin{minipage}[t]{0.45\textwidth}
  \centering
  \includegraphics[width=\textwidth]{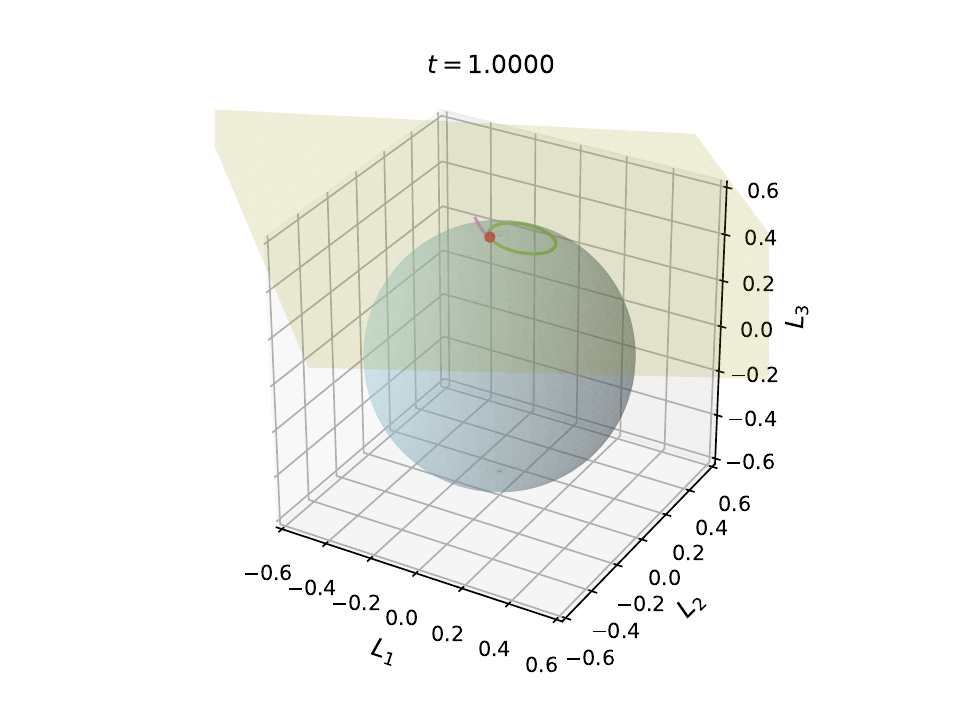}
  \subcaption{$\vL$ space. $c = 0$. $t = 1$.}
  \label{subfig:mimicLBRMHD-L-c0-t1-case1}
 \end{minipage}
\hfill
 \begin{minipage}[t]{0.45\textwidth}
  \centering
  \includegraphics[width=\textwidth]{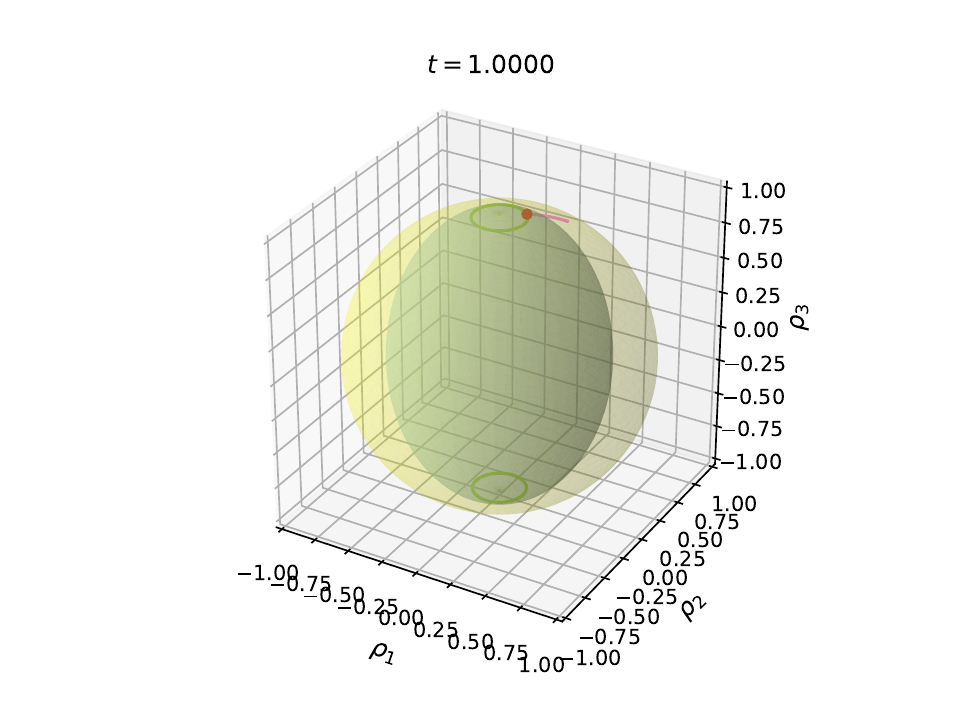}
  \subcaption{$\vecrho$ space. $c = 0$. $t = 1$.}
  \label{subfig:mimicLBRMHD-rho-c0-t1-case1}
 \end{minipage}
\\
 \begin{minipage}[t]{0.45\textwidth}
  \centering
  \includegraphics[width=\textwidth]{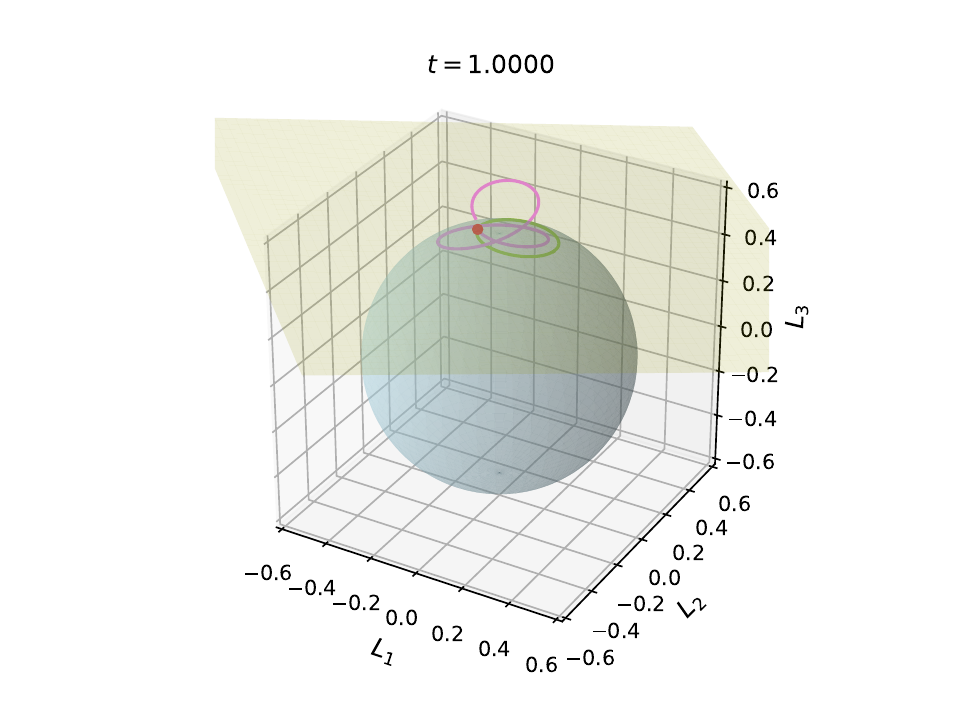}
  \subcaption{$\vL$ space. $c = 10$. $t = 1$.}
  \label{subfig:mimicLBRMHD-L-c10-t1-case1}
 \end{minipage}
\hfill
 \begin{minipage}[t]{0.45\textwidth}
  \centering
  \includegraphics[width=\textwidth]{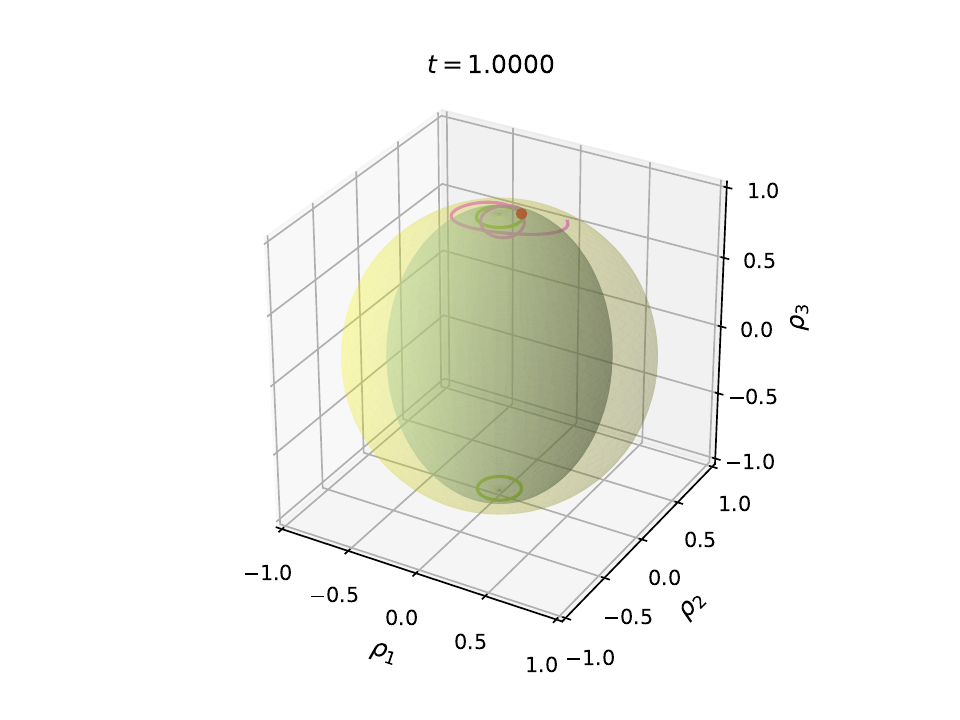}
  \subcaption{$\vecrho$ space. $c = 10$. $t = 1$.}
  \label{subfig:mimicLBRMHD-rho-c10-t1-case1}
 \end{minipage}
 \caption{
Snapshots of phase space at $t = 1$ of dynamics given by 
Eq.~(\ref{eq:mimicLBMHDEvolutionEqMixHamiltonian})
where Hamiltonian dynamics is added to SA dynamics of the toy model
 mimicking low-beta reduced MHD with 
$I_{1} = I_{2} = I_{3} = 1$, 
$M_{1} = M_{2} = 2$, and $M_{3} = 1$.
The initial condition was a state perturbed from a stable equilibrium
with 
$L_{1} = L_{2} = 0$, $L_{3} = 1/2$, 
$\rho_{1} = \rho_{2} = 0$, and $\rho_{3} = 1$.
}
 \label{fig:mimicLBRMHD-L-rho-c-t1-case1}
\end{figure}

Figure~\ref{fig:mimicLBRMHD-L-rho-c-t1-case1} shows 
snapshots of the phase space at $t = 1$.
Figures~\ref{fig:mimicLBRMHD-L-rho-c-t1-case1}\subref{subfig:mimicLBRMHD-L-cm10-t1-case1}
and
\ref{fig:mimicLBRMHD-L-rho-c-t1-case1}\subref{subfig:mimicLBRMHD-rho-cm10-t1-case1}
are for $c = -10$, 
Figs.~\ref{fig:mimicLBRMHD-L-rho-c-t1-case1}\subref{subfig:mimicLBRMHD-L-c0-t1-case1}
and
\ref{fig:mimicLBRMHD-L-rho-c-t1-case1}\subref{subfig:mimicLBRMHD-rho-c0-t1-case1}
are for $c = 0$,
and 
Figs.~\ref{fig:mimicLBRMHD-L-rho-c-t1-case1}\subref{subfig:mimicLBRMHD-L-c10-t1-case1}
and
\ref{fig:mimicLBRMHD-L-rho-c-t1-case1}\subref{subfig:mimicLBRMHD-rho-c10-t1-case1}
are for $c = 10$.
In the $\vL$ space, 
Figs.~\ref{fig:mimicLBRMHD-L-rho-c-t1-case1}\subref{subfig:mimicLBRMHD-L-cm10-t1-case1},
\subref{subfig:mimicLBRMHD-L-c0-t1-case1} and \subref{subfig:mimicLBRMHD-L-c10-t1-case1},
the spherical surface in light blue represents the constant
$E_{\mathrm{k}}$, 
the plane in light yellow represents $\vL \cdot \vecrho = 1/2$, 
the green circle represents the intersection of 
the constant $E_{\mathrm{k}}$ surface and the constant 
$\vL \cdot \vecrho$ surface.
On the other hand, 
in the $\vecrho$ space, 
Figs.~\ref{fig:mimicLBRMHD-L-rho-c-t1-case1}\subref{subfig:mimicLBRMHD-rho-cm10-t1-case1},
\subref{subfig:mimicLBRMHD-rho-c0-t1-case1} and \subref{subfig:mimicLBRMHD-rho-c10-t1-case1},
the ellipsoidal surface in light blue represents the constant
$E_{\mathrm{m}}$, 
the spherical surface in light yellow represents $| \vecrho |^{2} = 1$, 
the green circle represents the intersection of 
the constant $E_{\mathrm{m}}$ surface and the constant 
$| \vecrho |^{2}$ surface.
In each subfigure of Fig.~\ref{fig:mimicLBRMHD-L-rho-c-t1-case1},
the red point and the pink curve 
represent the current position of the system and the trajectory in the
phase space, respectively. 

Without the Hamiltonian dynamics, the trajectories follow  straight
relaxation to the final state as seen in 
Figs.~\ref{fig:mimicLBRMHD-L-rho-c-t1-case1}\subref{subfig:mimicLBRMHD-L-c0-t1-case1}
and \subref{subfig:mimicLBRMHD-rho-c0-t1-case1}.
On the other hand, the trajectories for $c = \pm 10$ turn around.
However, as explained above, the time evolution of energy did not differ
much from that of $c = 0$. 

\section{Using SA for linear stability analysis}
\label{sec:linearStability}

In addition to being useful for equilibrium calculations, SA can be used to assess linear stability, as seen in Sec.~\ref{sec:SAFinDoF}.
 Here we show how this can work for reduced MHD. 
Suppose an equilibrium of an MHD system has been obtained somehow by a
method other than SA.   For example, any cylindrically symmetric state of
the vorticity $U$ and the magnetic flux function $\psi$ is 
an equilibrium of low-beta reduced MHD in cylindrical geometry.
Then, let us perform SA starting from an initial condition that is a small amplitude perturbation away from  the  equilibrium.  
If the perturbation relaxes to  original equilibrium  by SA  dynamics that monotonically decreases the energy of the system,
then the  equilibrium is linearly stable.  On the other hand, 
if the perturbation grows, there are two possibilities:  the equilibrium is linearly unstable or the system linearly stable with  a combination of positive and negative energy modes. To distinguish these two cases a spectral stability analysis would need to be done. The case with all negative energy modes corresponds to the equilibrium located at an energy maximum which is stable.   Upon reversing the direction of time, this case will be detected if relaxation occurs.

Section~\ref{LBRMHDDoubleBracketSAFormulation} introduces the evolution equations of SA used here, while Sec.~\ref{subsec:m2n1perturbation} shows some numerical results of the linear
stability analyses for low-beta reduced MHD in cylindrical geometry.  

\subsection{Formulation}
\label{LBRMHDDoubleBracketSAFormulation}

Consider now the  SA double bracket evolution equations 
for   low-beta reduced MHD in cylindrical geometry.  We begin by defining  the symmetric kernel $\mathcal{K}$ that will be used, i.e., 
\begin{equation}
 \mathcal{K}_{ij} ( \vx^{\pr}, \vx^{\pr\pr}  )
= 
 \alpha_{ij} g( \vx^{\pr} - \vx^{\pr\pr} ),
\quad 
 i, j = 1, 2, 
\label{eq:LBRMHDDoubleBracketSAKernel}
\end{equation}
where the  Green's function is defined by
\begin{equation}
 \bigtriangleup g(\vx) = - \delta^{3} ( \vx ), 
\label{eq:LBRMHDDoubleBracketSA3DGreenFunction}
\end{equation}
with $\bigtriangleup$ being the Laplacian in three dimensions.
We assumed that $\mathcal{K}$ is diagonal with 
positive constants $\alpha_{ii}$ and  
$\alpha_{ij} = 0$ for $i \neq j$.

For simplicity of notation, let us define
the right-hand sides of the original low-beta reduced MHD as 
\begin{equation}
 \{ u^{i} , H \}
\eqqcolon 
f^{i}, \quad i=1,2.
\label{eq:LBRMHDRHS}
\end{equation}
By using the symmetric kernel (\ref{eq:LBRMHDDoubleBracketSAKernel}),
the double bracket can be calculated explicitly, resulting in the following SA equations:
\begin{align}
 \frac{\pd U}{\pd t}
&=
 [ U, \tilde{\vphi} ] + [ \psi, \tilde{J} ]
 - \veps \frac{\pd \tilde{J}}{\pd \zeta},
\label{eq:LBRMHDDoubleBracketSAEvolutionEq1}
\\
 \frac{\pd \psi}{\pd t}
&=
 [ \psi, \tilde{\vphi} ] - \veps \frac{\pd \tilde{\vphi}}{\pd \zeta},
\label{eq:LBRMHDDoubleBracketSAEvolutionEq2}
\end{align}
where the artificial advection fields are defined by
\begin{align}
 \tilde{\vphi}(\vx)
\coloneqq & \,
 \alpha_{11} \int_{\cal D} \td^{3} x^{\pr} \,
  g(\vx, \vx^{\pr}) f^{1}(\vx^{\pr}),
\label{eq:LBRMHDDoubleBracketAdvField1}
\\
\tilde{J}(\vx)
\coloneqq & \,
 \alpha_{22} \int_{\cal D} \td^{3} x^{\pr} \,
 g(\vx, \vx^{\pr}) f^{2}(\vx^{\pr}).
\label{eq:LBRMHDDoubleBracketAdvField2}
\end{align}
As we observe, 
the advection fields of Eqs.~(\ref{eq:LBRMHDDoubleBracketSAEvolutionEq1})
and (\ref{eq:LBRMHDDoubleBracketSAEvolutionEq2})
are replaced by the artificial ones $\tilde{\vphi}$ and $\tilde{J}$
from the advection fields of the low-beta reduced MHD
$\vphi$ and $J$. 
Because of the property of the Poisson tensor,
the Casimir invariants are automatically preserved.

Note that the formulation becomes simpler if we choose Dirac's delta
function instead of the Green's function in the kernel $\mathcal{K}$.
However, in our experience, it is less stable numerically.
The kernel with the Green's function can suppress growth of
fine-scale structure.

\subsection{$m/n = 2/1$ perturbation}
\label{subsec:m2n1perturbation}

We take as a given equilibrium a cylindrically symmetric state
with a safety factor profile 
$q(r) = q_{0} / ( 1 - r^{2}/2 )$ with $q_{0} = 1.75$
\citep[see Fig.~1 of][]{Furukawa-2022}. 
This equilibrium has a  $q=2$ surface  at $r=1/2$ and no  plasma rotation.
This equilibrium is known to be linearly stable against $m=2$ and $n=1$
ideal MHD modes.

A series of $m = 2$ and $n = 1$ perturbations was generated in a
dynamically accessible (Casimir preserving) manner \citep[see][for definition and discussion]{Morrison-1998}. 
Even if we substitute arbitrarily chosen advection fields 
into Eqs.~(\ref{eq:LBRMHDDoubleBracketSAEvolutionEq1})
and (\ref{eq:LBRMHDDoubleBracketSAEvolutionEq2}), ones that 
 are different from $\tilde{\vphi}$ and $\tilde{J}$ 
defined in Eqs.~(\ref{eq:LBRMHDDoubleBracketAdvField1})
and (\ref{eq:LBRMHDDoubleBracketAdvField2}), 
the Casimir invariants are still preserved because of the property of the
Poisson tensor.
Therefore, we use the following advection fields to generate the dynamically
accessible perturbations:
\begin{align}
 \tilde{\vphi}(r, \theta, \zeta)
&=
A_{\vphi} r (1-r) 
\rme^{-\left( \frac{r-r_{0}}{L} \right)^{2}}
 \sin(m \theta - n \zeta),
 \label{eq:LBRMHDDoubleBracketSAAdvFieldDAV1}
 \\
 \tilde{J}(r, \theta, \zeta)
&=
A_{J} r (1-r) 
\rme^{-\left( \frac{r-r_{0}}{L} \right)^{2}}
\cos (m \theta - n \zeta),
 \label{eq:LBRMHDDoubleBracketSAAdvFieldDAV2}
\end{align}
where $A_{\vphi}$, $A_{J}$, $r_{0}$ and $L$ are constants.
The poloidal and toroidal mode numbers are $m = 2$ and $n = 1$, respectively.
A case with $A_{\vphi} = A_{J} = 10^{-3}$, $r_{0} = 0.5$ and $L = 0.1$
was shown in Fig.~2 of \citet{Furukawa-2022}.

The initial condition chosen for generating the series of dynamically accessible
perturbations was the cylindrically symmetric equilibrium.
The time evolution generates the series of helically perturbed states that
are  on the same Casimir leaf as the equilibrium.

As noted, the equilibrium is linearly stable.  Therefore, we expected
that SA would 
recover the cylindrically symmetric equilibrium at least if the given
perturbation is small enough.  
In fact, we observed that the perturbation amplitude became smaller as
the total energy of the system was decreased by SA \citep[see Figs.~10 and 11
of][]{Furukawa-2022}.
We tried some initial perturbations with different ratios of kinetic to
magnetic energies, and we observed that the perturbation tended to disappear in
all cases, i.e., the dynamics relaxes to the equilibrium.    However, the disappearance of the velocity perturbation took
long simulation time, even if we applied an acceleration method which
will be explained in Sec.~\ref{subsec:timeDependentKernel}.

For an equilibrium without the $q = 2$ surface, when $q_{0} = 2.5$ for
example, the perturbation also tended to disappear.
However, since the simulation was performed without the to be explained acceleration technique,
the damping of the velocity part was very slow. 
The magnetic part, on the other hand, disappeared quickly.

We have also tried SA for an unstable equilibrium.
The safety factor was the same as the equilibrium introduced 
above, but  equilibrium poloidal rotation was introduced, according to   
\begin{equation}
 v_{\theta}(r)
  =
   \frac{v_{\theta {\rm max}} (\alpha + 1)^{\alpha + 1}}{\alpha^{\alpha}}
  r ( 1 - r )^{\alpha},
\end{equation}
where $\alpha$ is a positive parameter.  
A radial profile with $v_{\theta {\rm max}} = 0.01$ and $\alpha = 3$
were shown in Fig.~12 of \citet{Furukawa-2022}.
This  equilibrium is linearly unstable against centrifugal instability.

We performed SA, which monotonically decreased the total energy of the
system.  In the course of this evolution, the amplitude of the perturbation grew as  time
proceeded.  The time evolution of the total energy 
and the radial profiles of the perturbation were shown in Figs.~15 and
16 of \citet{Furukawa-2022}, respectively.

\section{Toroidal equilibria}
\label{sec:toroidal}

We have applied SA for high-beta reduced MHD in axisymmetric
toroidal geometry \citep{Furukawa-2018}. 
Section~\ref{HBRMHDDoubleBracketSAFormulation} introduces the evolution
equations of SA, while  Sec.~\ref{subsec:ShafranovEquilibrium} 
describes some numerical results.

\subsection{Formulation}
\label{HBRMHDDoubleBracketSAFormulation}

The symmetric kernel for the double bracket was assumed to be same
as in Sec.~\ref{sec:linearStability};
it was diagonal with positive coefficients.
The advection fields were 
$\tilde{\vphi}$ in Eq.~(\ref{eq:LBRMHDDoubleBracketAdvField1}),
$\tilde{J}$ in Eq.~(\ref{eq:LBRMHDDoubleBracketAdvField2}),
and 
\begin{equation}
 \tilde{h}(\vx)
\coloneqq
 \alpha_{33} \int_{\cal D} \td^{3} x^{\pr} \,
 g(\vx, \vx^{\pr}) f^{3}(\vx^{\pr}),
\label{eq:HBRMHDDoubleBracketAdvField3}
\end{equation}
where 
$f^{3} \coloneqq \{ P, H \}$.

Then the evolution equations of SA read
\begin{align}
 \frac{\pd U}{\pd t}
&=
 [ U , \tilde{\vphi} ]
 + [ \psi , \tilde{J} ] - \veps \frac{\pd \tilde{J}}{\pd \zeta}
 + [ P, \tilde{h} ],
\label{eq:HBRMHDSAVorticityEq}
\\
 \frac{\pd \psi}{\pd t}
&=
 [ \psi , \tilde{\vphi} ] - \veps \frac{\pd \tilde{\vphi}}{\pd \zeta},
\label{eq:HBRMHDSAOhmLaw}
\\
 \frac{\pd P}{\pd t}
&=
 [ P, \tilde{\vphi} ].
\label{eq:HBRMHDSAPressureEq}
\end{align}
Again, the form of the equations are  the same as 
the original high-beta reduced MHD 
Eqs.~(\ref{eq:HBRMHDVorticityEq})--(\ref{eq:HBRMHDPressureEq}), but with the 
 advection fields   replaced by the artificial ones.
And, the  Casimir invariants are preserved, while the energy of the
system monotonically decreases by the time evolution.

\subsection{Large-aspect-ratio, circular-cross-section tokamak equilibrium}
\label{subsec:ShafranovEquilibrium}

For calculating axisymmetric equilibria, 
Fourier components with the toroidal mode number $n = 0$ only were
retained in the simulation.

The initial condition had concentric magnetic surfaces.
The safety factor profile was
$q(r) = q_{0} / ( 1 - r^{2}/2 )$ with $q_{0} = 1.75$.
The pressure profile was assumed to be $P(r) = \beta_{0} ( 1 - r^{2}) $.
Here, the central beta was defined by 
$\beta_{0} := 2 \mu_{0} p_{0} / B_{0}^{2}$, 
where $\mu_{0}$ is the vacuum permeability, $p_{0}$ is the pressure at the
magnetic axis, and $B_{0}$ is the typical magnitude of the magnetic field.
These profiles were plotted in Fig.~1 of \cite{Furukawa-2018}.
The central beta was taken to be $\beta_{0} = 0.1$\%, $0.5$\%,
and $1$\%. Zero poloidal velocity  was assumed.   As the time proceeded, the total energy of the system successfully
decreased, and the stationary states were obtained.
The time evolution of the energy was shown in Fig.~2 of
\cite{Furukawa-2018}.

The flux surfaces of the obtained equilibria showed 
the Shafranov shift as seen in Figs.~3 and 4 of
\cite{Furukawa-2018}.
The distance of the magnetic axis shift was compared with the analytic
theory based on the large-aspect-ratio expansion.
Since the analytic theory includes the toroidicity even if   beta is
zero, the finite Shafranov shift remains even at zero beta.
On the other hand, since the toroidicity drops out completely in  
high-beta reduced MHD, the Shafranov shift was smaller than that
of the analytic theory for all three beta values examined.
However, the results showed reasonable agreement in the increment of the
shift as the beta was increased.

In \cite{Furukawa-2018}, 
some equilibria with poloidal plasma rotation were also calculated by
SA.  
This is an advantage of SA; we just need to solve an initial-value
problem for a given initial condition.  
The resultant stationary states can have plasma rotation.

The initial poloidal velocity was assumed to have a profile 
$v_{\theta}(r) = 4 v_{\theta \mathrm{max}} r ( 1 - r )$ with 
a constant $v_{\theta \mathrm{max}}$.
The Shafranov shift was shown to increase quadratically in the rotation
velocity \citep[see Fig.~7 of][]{Furukawa-2018}.  The quadratic dependence was
explained by a mapping 
between an equilibrium without plasma rotation and poloidally rotating
equilibrium.  

\subsection{Toroidally-averaged stellerator equilibrium}
\label{subsec:ToroidalAverageStelleratorEquilibrium}

Dynamics of toroidally-averaged stellerator plasmas are governed by
equations of the same form as the high-beta reduced MHD.
Numerical results of the obtained equilibria were compared with the
results of a previous study on Heliotron E \citep{Nakamura-1993}.
We   obtained reasonable agreement, although our results did not
completely overlap the previous results.
The difference has  several reasons, e.g.,  our SA calculation could not impose the net 
toroidal current free condition on each magnetic surface, which was
imposed in the previous study.   This may be overcome by using DSA.

\section{Helically deformed equilibria}
\label{sec:helical}

In the present section we show some numerical results where SA leads 
to helically deformed equilibria in cylindrical geometry.
Section~\ref{subsec:m1n1deformation} shows 
a case of internal kink mode like deformation with $m = 1$ and $n = 1$,
and Sec.~\ref{subsec:m2n1deformation} shows 
a case with $m = 2$ and $n = 1$,  where a sheared poloidal rotation was
assumed in the equilibrium. 

\subsection{$m/n = 1/1$ deformation}
\label{subsec:m1n1deformation}

For this case we  performed SA with a safety factor profile
$q(r) = q_{0} / ( 1 - r^{2}/2 )$ with $q_{0} = 0.75$. 
A $q=1$ surface exists at $r=1/2$ in this case. 
The equilibrium plasma rotation was assumed to be zero.
This equilibrium is neutrally stable against ideal internal kink modes.

Dynamically accessible perturbations were generated 
as in Sec.~\ref{subsec:m2n1perturbation}.
The advection fields were given by 
Eqs.~(\ref{eq:LBRMHDDoubleBracketSAAdvFieldDAV1})
and 
(\ref{eq:LBRMHDDoubleBracketSAAdvFieldDAV2})
with $m = 1$ and $n = 1$.
In the numerical results shown in 
the present  section,
$r_{0} = 0.5$ and $L = 0.1$ were used.
The other parameters $A_{\vphi}$ and $A_{J}$ 
were given so as to control the ratio of the perturbed kinetic and
magnetic energies.
The initial condition for generating the dynamically accessible
perturbation was the cylindrically symmetric equilibrium
introduced in the previous paragraph. 

\begin{figure}[h]
 \begin{minipage}[t]{0.45\textwidth}
  \centering
  \includegraphics[width=\textwidth]{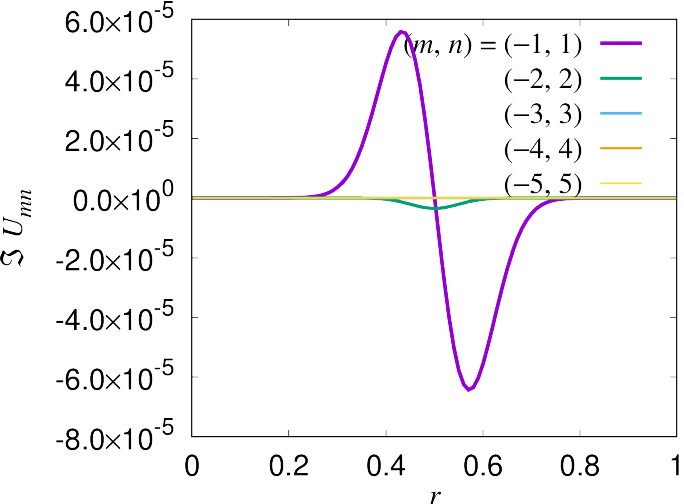}
 \subcaption{$\Im \, U_{mn}$.}
 \label{subfig:LBRMHDSAm1n1deform-r-U_i-t000000_0000}
 \end{minipage}
\hfill
 \begin{minipage}[t]{0.45\textwidth}
  \centering
  \includegraphics[width=\textwidth]{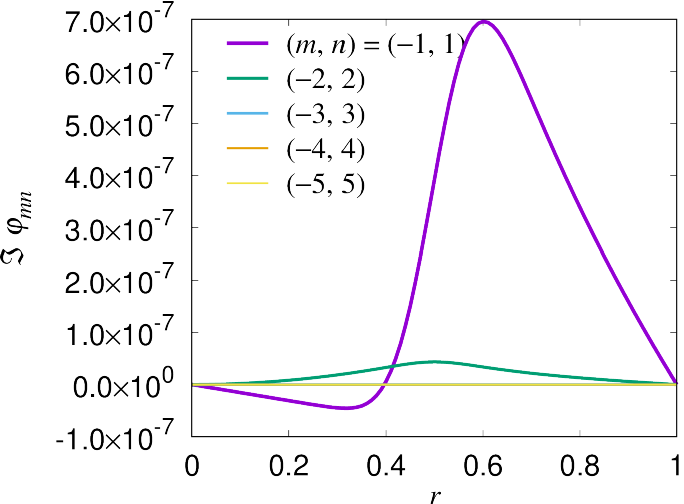}
 \subcaption{$\Im \, \vphi_{mn}$.}
 \label{subfig:LBRMHDSAm1n1deform-r-phi_i-t000000_0000}
 \end{minipage}
\\
 \begin{minipage}[t]{0.45\textwidth}
  \centering
  \includegraphics[width=\textwidth]{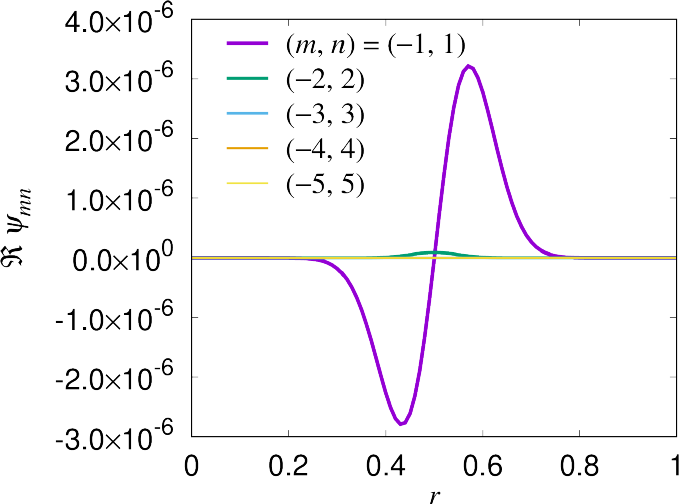}
 \subcaption{$\Re \, \psi_{mn}$.}
 \label{subfig:LBRMHDSAm1n1deform-r-psi_r-t000000_0000}
 \end{minipage}
\hfill
 \begin{minipage}[t]{0.45\textwidth}
  \centering
  \includegraphics[width=\textwidth]{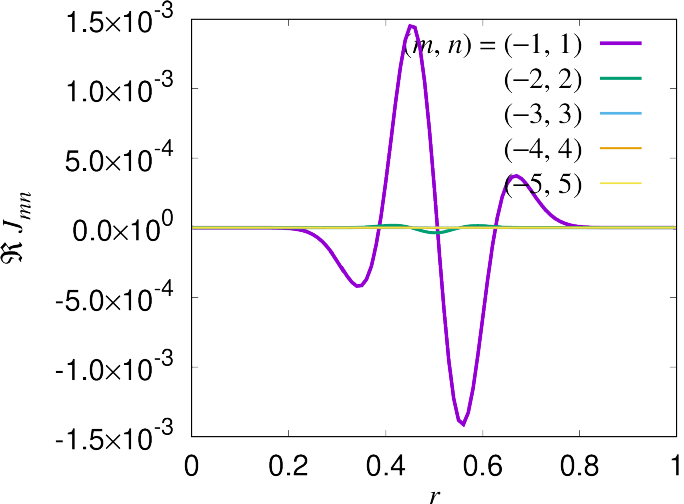}
 \subcaption{$\Re \, J_{mn}$.}
 \label{subfig:LBRMHDSAm1n1deform-r-J_r-t000000_0000}
 \end{minipage}
 \caption{
 Radial profiles of initial condition for SA are plotted: 
(a) $\Im \, U_{mn}$, (b) $\Im \, \vphi_{mn}$, 
(c) $\Re \, \psi_{mn}$, (d) $\Re \, J_{mn}$.
The other components
$\Re \, U_{mn}$, $\Re \, \vphi_{mn}$, $\Im \, \psi_{mn}$ and 
$\Im \, J_{mn}$ were  
zero.
 }
 \label{fig:LBRMHDSAm1n1deform-r-u-t000000_0000}
\end{figure}

\begin{figure}[h]
 \begin{minipage}[t]{0.45\textwidth}
  \centering
  \includegraphics[width=\textwidth]{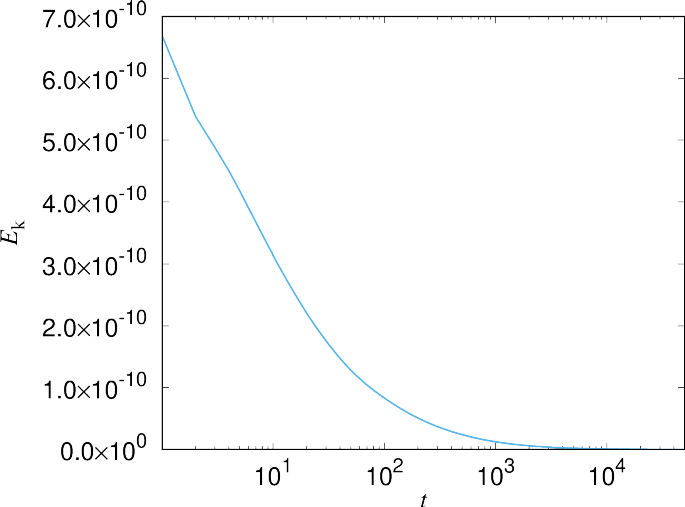}
  \subcaption{Kinetic energy.} 
  \label{subfig:LBRMHDSAm1n1deform-t-Ek}
 \end{minipage}
\hfill
 \begin{minipage}[t]{0.45\textwidth}
  \centering
  \includegraphics[width=\textwidth]{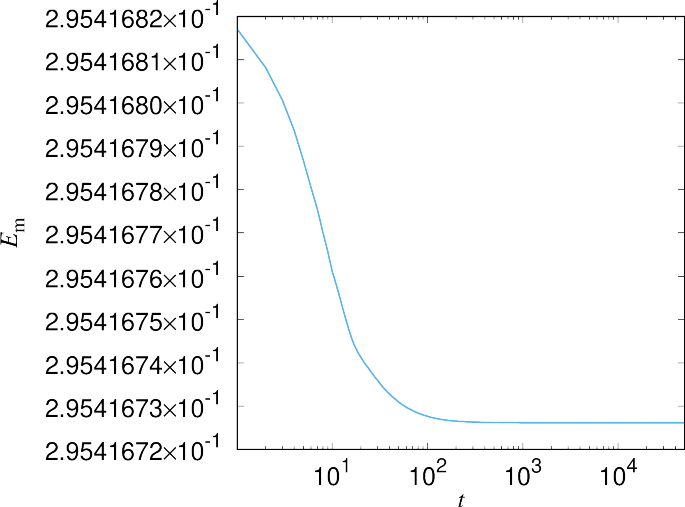}
  \subcaption{Magnetic energy.} 
  \label{subfig:LBRMHDSAm1n1deform-t-Em}
 \end{minipage}
 \caption{Time evolutions of (a) kinetic energy and (b) magnetic energy
 are plotted.  Both decreased monotonically and reached stationary
 values.  Note that the horizontal axis is the log scale in each figure.}
 \label{fig:LBRMHDSAm1n1deform-t-E}
\end{figure}

A numerical example is presented  below, where the  initial condition for SA
is shown in 
Fig.~\ref{fig:LBRMHDSAm1n1deform-r-u-t000000_0000}.
This initial condition corresponds to $A_{\vphi} = 10^{-3}$  
and $A_{J} = 2 \times 10^{-2}$.  
The perturbed kinetic energy is about 0.01 times the perturbed magnetic
energy at $t = 0$.
Time evolution of the energy by SA is shown in 
Fig.~\ref{fig:LBRMHDSAm1n1deform-t-E}.
Kinetic and magnetic energies decrease monotonically and reach  their
stationary values.  Note that the horizontal axis is a  log scale in
each figure.  

\begin{figure}[h]
 \begin{minipage}[t]{0.45\textwidth}
  \centering
  \includegraphics[width=\textwidth]{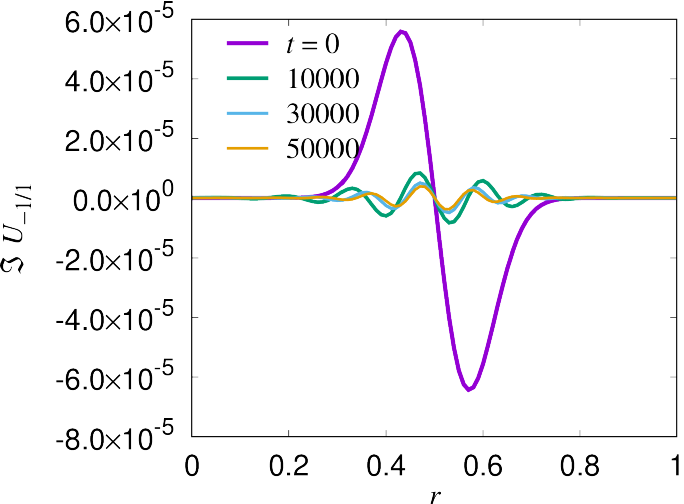}
 \subcaption{$\Im \, U_{-1/1}$.}
 \label{subfig:LBRMHDSAm1n1deform-r-U_i_mm1n1-t}
 \end{minipage}
\hfill
 \begin{minipage}[t]{0.45\textwidth}
  \centering
  \includegraphics[width=\textwidth]{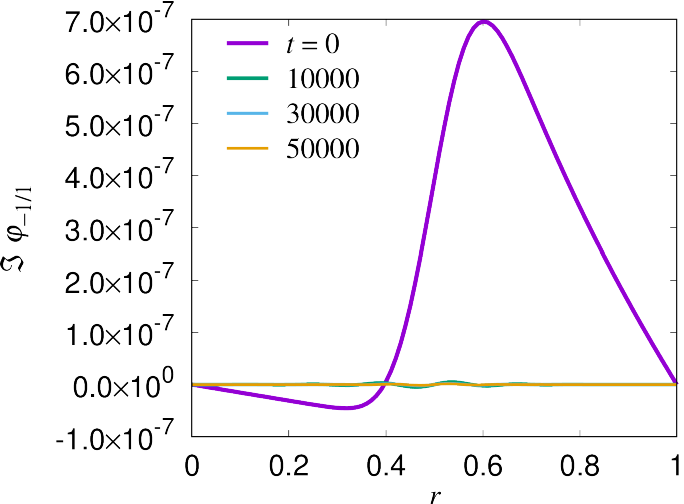}
 \subcaption{$\Im \, \vphi_{-1/1}$.}
 \label{subfig:LBRMHDSAm1n1deform-r-phi_i_mm1n1-t}
 \end{minipage}
\\
 \begin{minipage}[t]{0.45\textwidth}
  \centering
  \includegraphics[width=\textwidth]{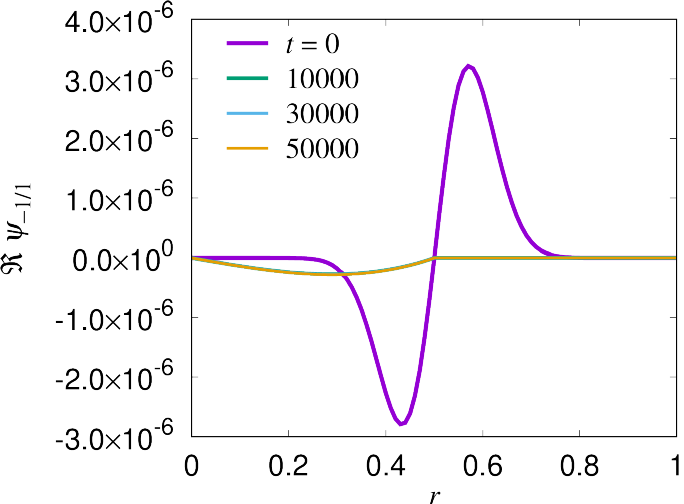}
 \subcaption{$\Re \, \psi_{-1/1}$.}
 \label{subfig:LBRMHDSAm1n1deform-r-psi_r_mm1n1-t}
 \end{minipage}
\hfill
 \begin{minipage}[t]{0.45\textwidth}
  \centering
  \includegraphics[width=\textwidth]{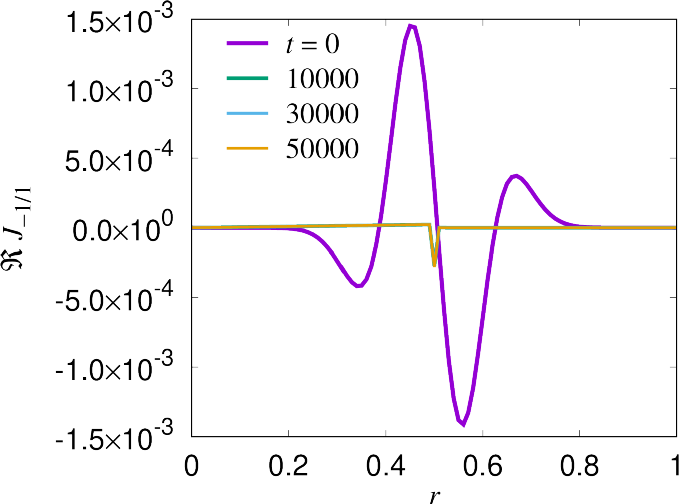}
 \subcaption{$\Re \, J_{-1/1}$.}
 \label{subfig:LBRMHDSAm1n1deform-r-J_r_mm1n1-t}
 \end{minipage}
 \caption{
 Radial profiles of 
(a) $\Im \, U_{-1/1}$, (b) $\Im \, \vphi_{-1/1}$, (c) $\Re \, \psi_{-1/1}$, 
(d) $\Re \, J_{-1/1}$
at $t = 0$, $10000$, $30000$ and $50000$.
The other components
$\Re \, U_{-1/1}$, $\Re \, \vphi_{-1/1}$, $\Im \, \psi_{-1/1}$ and 
$\Im \, J_{-1/1}$ were 
almost zero.
 }
 \label{fig:LBRMHDSAm1n1deform-r-u_mm1n1-t}
\end{figure}

Figure~\ref{fig:LBRMHDSAm1n1deform-r-u_mm1n1-t}
shows radial profiles of 
$\Im \, U_{-1/1}$, $\Im \, \vphi_{-1/1}$, $\Re \, \psi_{-1/1}$ and 
$\Re \, J_{-1/1}$
at $t = 0$, $10000$, $30000$ and $50000$.
The other components
$\Re \, U_{-1/1}$, $\Re \, \vphi_{-1/1}$, $\Im \, \psi_{-1/1}$ and 
$\Im \, J_{-1/1}$ as well as higher $(m,n)$ modes were 
almost zero.
The damping of the velocity part was slow,  as was  the case of 
Sec.~\ref{subsec:m2n1perturbation},
even though  the acceleration method, to be explained in 
Sec.~\ref{subsec:timeDependentKernel},
was used.  Although the vorticity $\Im \, U_{-1/1}$ still remains finite,
the stream function $\Im \, \vphi_{-1/1}$ almost disappears. 

The magnetic part remains almost unchanged after $t > 10^{4}$, and 
 is  finite at the stationary state.
The final state has a  structure similar to that of  an internal kink mode, 
although it may be difficult to observe since the amplitudes at the
 stationary state are much smaller than the initial amplitudes.
The magnetic flux function $\Re \, \psi_{-1/1}$ has a finite amplitude
 at $r < 1/2$, and zero at $r > 1/2$.  
Also, the current density $\Re \, J_{-1/1}$ has a spiky structure around
 $r = 1/2$.  This is typical of the internal kink mode.

We have performed SA with different initial conditions where 
(i) the perturbed kinetic energy is 100 times the perturbed
magnetic energy, and 
(ii) the perturbed kinetic and magnetic energies are almost the same.  
In all cases examined, we obtained helically deformed equilibria.
The spatial structures were similar to the internal kink mode.

We have also performed SA with a different equilibrium with $q_{0} = 1.1$, which has 
no $q = 1$ surface inside the plasma.
The initial conditions for SA were generated by using the 
advection fields
 Eqs.~(\ref{eq:LBRMHDDoubleBracketSAAdvFieldDAV1})
and 
(\ref{eq:LBRMHDDoubleBracketSAAdvFieldDAV2})
with $m = 1$ and $n = 1$.
We   generated three initial conditions for SA:
(i) the perturbed kinetic energy was 100 times the perturbed
magnetic energy,
(ii) they were almost same,
(iii) perturbed kinetic energy was 0.01 times the perturbed
magnetic energy.
In all these cases, the perturbation went away as the total energy of
the system decreased monotonically by SA.


\subsection{$m/n = 2/1$ deformation}
\label{subsec:m2n1deformation}

\begin{figure}[h]
 \centering
 \includegraphics[width=0.5\textwidth]{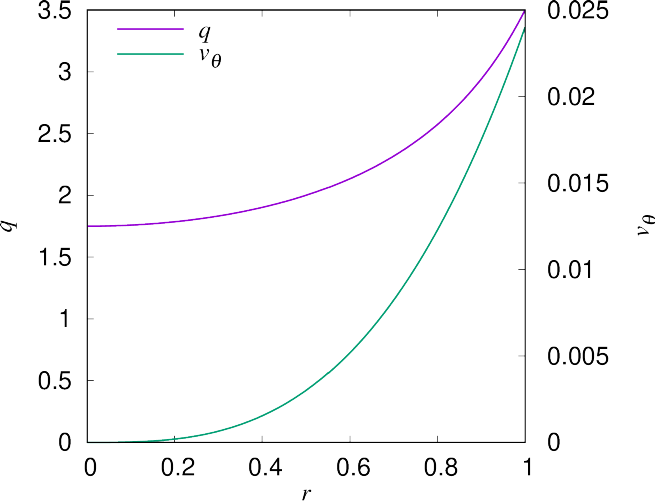}
 \caption{A poloidal velocity profile with
 $v_{\theta} (r) = 8 v_{\theta \mathrm{s}} r^{3}$ and
 $v_{\theta \mathrm{s}} = 0.003$ is shown, 
together with the $q$ profile.}
\label{fig:LBRMHDSAmm2n1deform-r-q-vth}
\end{figure}

Another example of helically deformed equilibrium 
with $m = 2$ and $n = 1$ structure is shown in this 
section. 
Here, we consider the cylindrically symmetric equilibrium with the same
$q$ profile as in  Sec.~\ref{subsec:m2n1perturbation}, where the   $q = 2$ resonant surface exists at $r = 1/2$.  
We assumed a sheared poloidal rotation velocity 
$v_{\theta} (r) = 8 v_{\theta \mathrm{s}} r^{3}$.
The poloidal rotation velocity at the resonant surface 
is $v_{\theta \mathrm{s}}$.  
A poloidal rotation velocity profile with $v_{\theta \mathrm{s}} = 0.003$ is
shown in Fig.~\ref{fig:LBRMHDSAmm2n1deform-r-q-vth},
together with the $q$ profile.

\begin{figure}[h]
 \centering
 \includegraphics[width=0.5\textwidth]{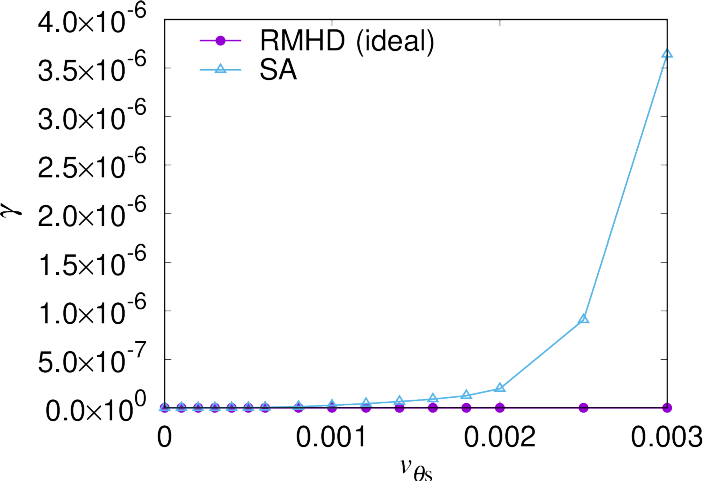}
 \caption{
Linear growth rates $\gamma$ are plotted versus the poloidal rotation
 velocity $v_{\theta \mathrm{s}}$ at $r = 1/2$.
In the figure, 
``RMHD (ideal)`` denotes the linearized low-beta reduced MHD,
while ``SA'' denotes the linearized SA equations.
The equilibria are spectrally stable against ideal MHD modes,
while unstable in SA at finite rotation velocity.  
}
\label{fig:LBRMHDSAmm2n1deform-vths-gam-parabolic-shear}
\end{figure}

Figure~\ref{fig:LBRMHDSAmm2n1deform-vths-gam-parabolic-shear}
shows the spectral stability of the equilibria with the sheared poloidal
rotation.
The horizontal axis is $v_{\theta \mathrm{s}}$, and the vertical axis is
the linear growth rate.
In the figure, ``RMHD(ideal)'' denotes the linear growth rates obtained by
the spectral analyses of the linearized ideal low-beta reduced MHD.  
We observed that the equilibria are stable even with a finite rotation
velocity with $0 \leq v_{\theta \mathrm{s}} \leq 0.003$.

Also in Fig~\ref{fig:LBRMHDSAmm2n1deform-vths-gam-parabolic-shear},
``SA'' denotes the linear growth rates by the spectral analyses of 
the linearized SA equations
(\ref{eq:LBRMHDDoubleBracketSAEvolutionEq1})
and (\ref{eq:LBRMHDDoubleBracketSAEvolutionEq2}),
although the symmetric kernel was taken to be diagonal and 
\begin{equation}
 \mathcal{K}_{ii} ( \vx^{\pr}, \vx^{\pr\pr}  )
= 
 \delta^{3}( \vx^{\pr} - \vx^{\pr\pr} ),
\quad 
 i = 1, 2
\label{eq:LBRMHDSAmm2n1deformKernelLinear}
\end{equation}
for simplicity.
The linearized SA equation shows instability at finite 
$v_{\theta \mathrm{s}}$.
This indicates that the equilibria with the sheared poloidal rotation
are not energy minima.

We have performed SA with an initial condition that is a summation of
the cylindrically symmetric equilibrium and a dynamically accessible
perturbation with $m = 2$ and $n = 1$.
For generating the dynamically accessible perturbation,
we used the advection fields of 
Eqs.~(\ref{eq:LBRMHDDoubleBracketSAAdvFieldDAV1})
and 
(\ref{eq:LBRMHDDoubleBracketSAAdvFieldDAV2})
with $m = 2$, $n = 1$, $r_{0} = 0.8$ and $L = 0.1$.
Then the initial perturbation has larger amplitudes around 
$r = 0.8$.  This was because   the eigenmode structure of the
linearized SA equation has  larger amplitudes at larger radii.
Then the relaxation by SA to a stationary state can occur in a shorter 
simulation time.

\begin{figure}[h]
 \begin{minipage}[t]{0.45\textwidth}
  \centering
  \includegraphics[width=\textwidth]{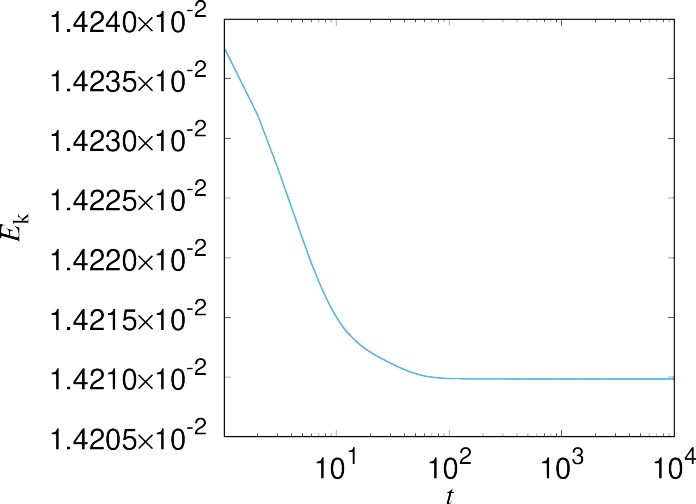}
  \subcaption{Kinetic energy.} 
  \label{subfig:LBRMHDSAm2n1deform-t-Ek}
 \end{minipage}
\hfill
 \begin{minipage}[t]{0.45\textwidth}
  \centering
  \includegraphics[width=\textwidth]{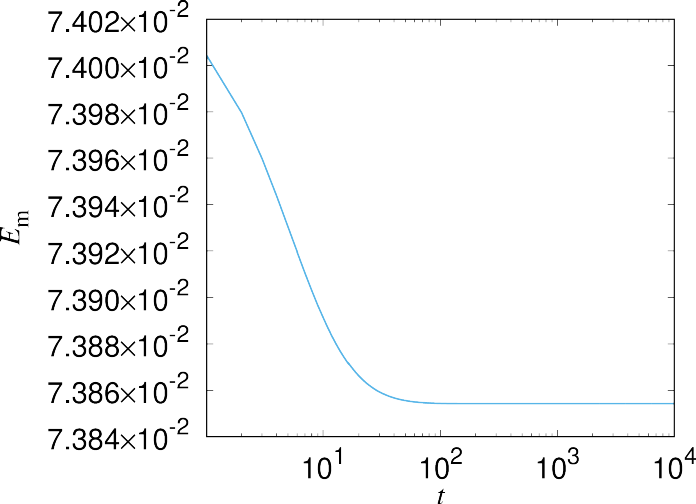}
  \subcaption{Kinetic energy.} 
  \label{subfig:LBRMHDSAm2n1deform-t-Em}
 \end{minipage}
 \caption{Time evolutions of (a) kinetic energy and (b) magnetic energy
 are plotted.  Both decreased monotonically and reached stationary
 values.  Note that the horizontal axis is the log scale in each figure.}
 \label{fig:LBRMHDSAm2n1deform-t-E}
\end{figure}

Figure~\ref{fig:LBRMHDSAm2n1deform-t-E} shows time evolution of energy.
Both kinetic and magnetic energies decreased monotonically and reached their
stationary values.  Note that the horizontal axis is a  log scale in
each figure.  

\begin{figure}[h]
 \begin{minipage}[t]{0.45\textwidth}
  \centering
  \includegraphics[width=\textwidth]{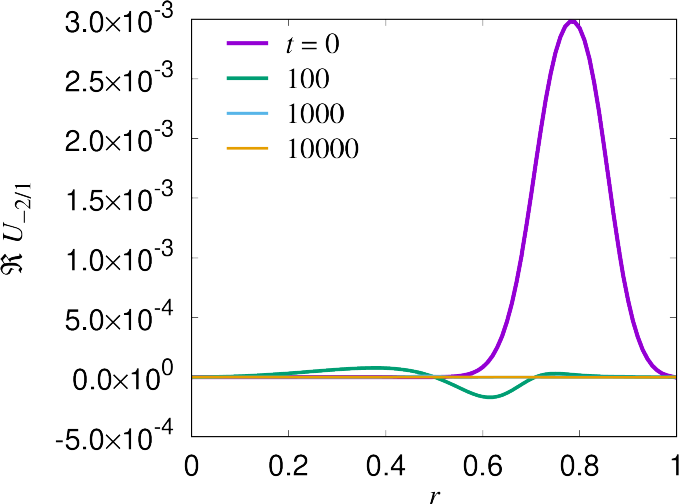}
 \subcaption{$\Re \, U_{-2/1}$.}
 \label{subfig:LBRMHDSAm2n1deform-r-U_r_mm2n1-t}
 \end{minipage}
\hfill
 \begin{minipage}[t]{0.45\textwidth}
  \centering
  \includegraphics[width=\textwidth]{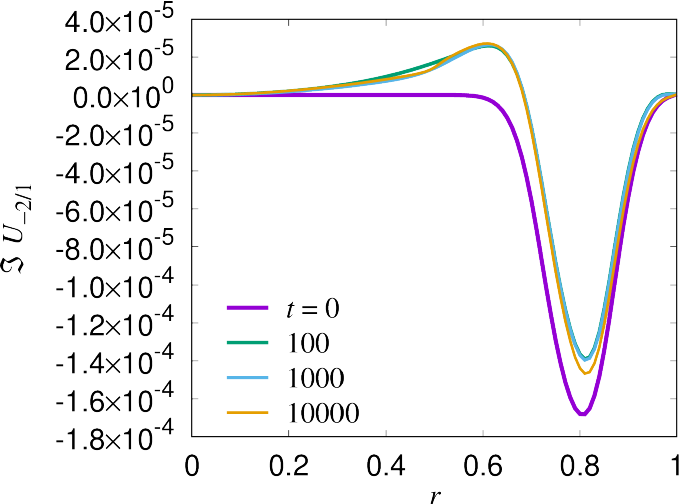}
 \subcaption{$\Im \, U_{-2/1}$.}
 \label{subfig:LBRMHDSAm2n1deform-r-U_i_mm2n1-t}
 \end{minipage}
\\
 \begin{minipage}[t]{0.45\textwidth}
  \centering
  \includegraphics[width=\textwidth]{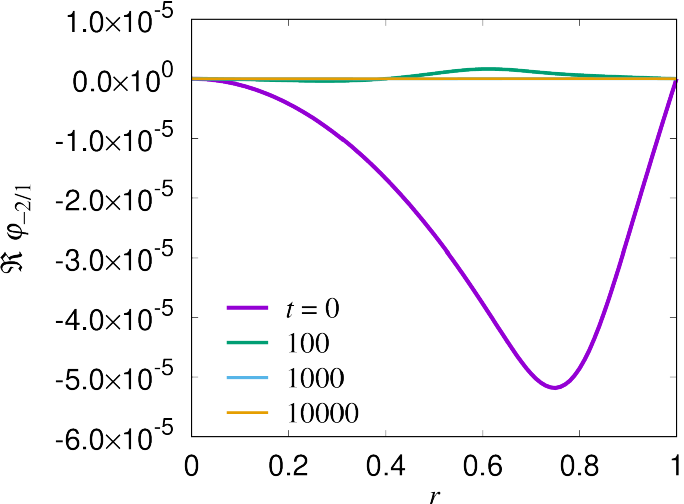}
 \subcaption{$\Re \, \vphi_{-2/1}$.}
 \label{subfig:LBRMHDSAm2n1deform-r-phi_r_mm2n1-t}
 \end{minipage}
\hfill
 \begin{minipage}[t]{0.45\textwidth}
  \centering
  \includegraphics[width=\textwidth]{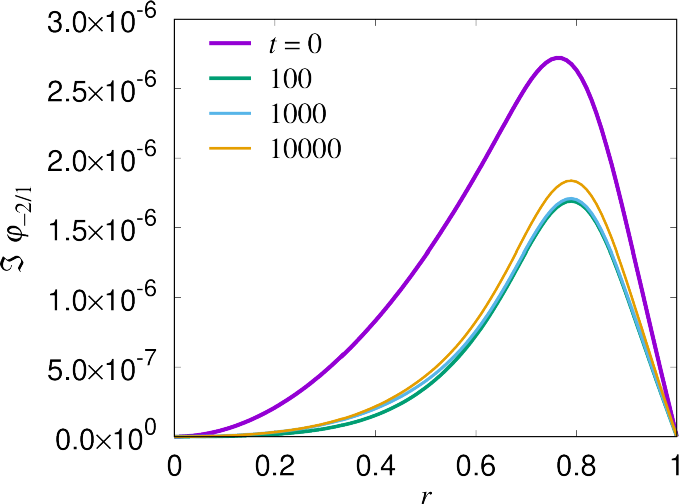}
 \subcaption{$\Im \, U_{-2/1}$.}
 \label{subfig:LBRMHDSAm2n1deform-r-U_i_mm2n1-t}
 \end{minipage}
\\
 \begin{minipage}[t]{0.45\textwidth}
  \centering
  \includegraphics[width=\textwidth]{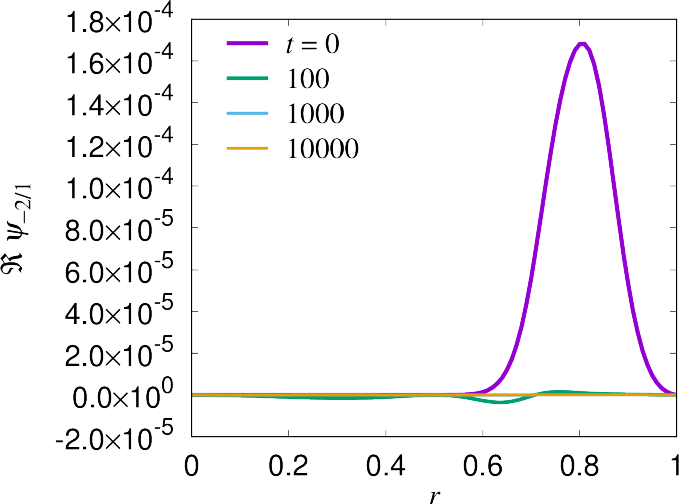}
 \subcaption{$\Re \, \psi_{-2/1}$.}
 \label{subfig:LBRMHDSAm2n1deform-r-psi_r_mm2n1-t}
 \end{minipage}
\hfill
 \begin{minipage}[t]{0.45\textwidth}
  \centering
  \includegraphics[width=\textwidth]{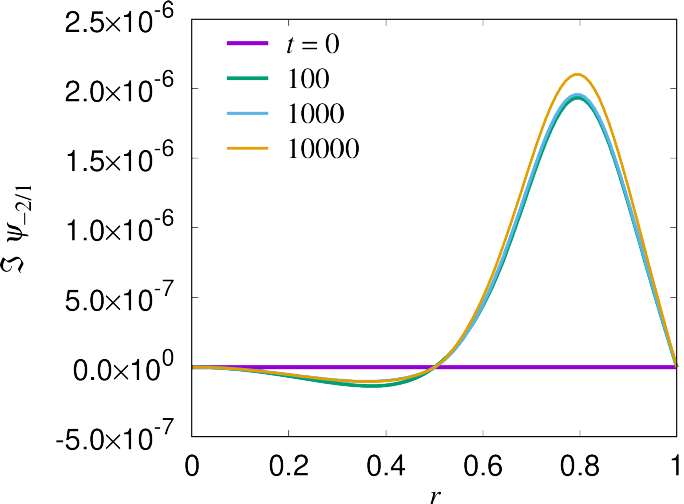}
 \subcaption{$\Im \, \psi_{-2/1}$.}
 \label{subfig:LBRMHDSAm2n1deform-r-psi_i_mm2n1-t}
 \end{minipage}
\\
 \begin{minipage}[t]{0.45\textwidth}
  \centering
  \includegraphics[width=\textwidth]{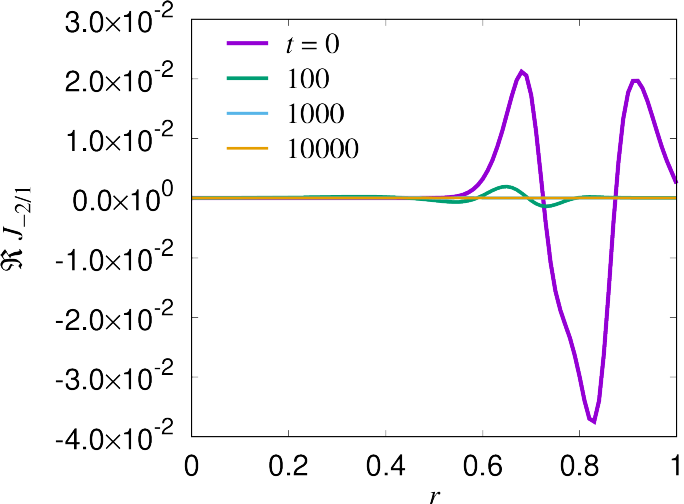}
 \subcaption{$\Re \, J_{-2/1}$.}
 \label{subfig:LBRMHDSAm2n1deform-r-J_r_mm2n1-t}
 \end{minipage}
\hfill
 \begin{minipage}[t]{0.45\textwidth}
  \centering
  \includegraphics[width=\textwidth]{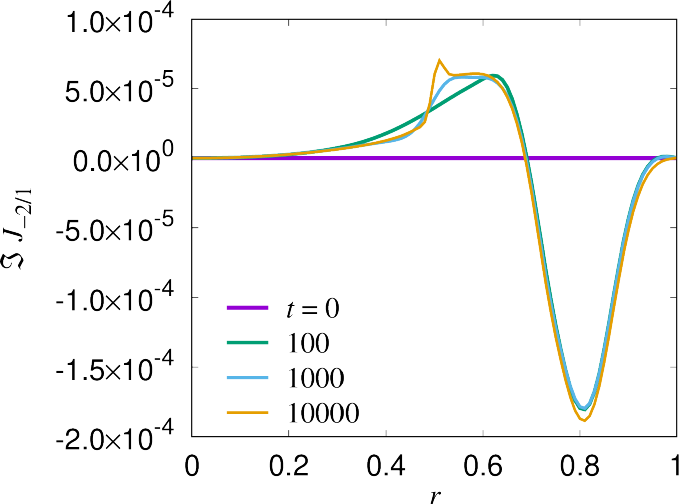}
 \subcaption{$\Im \, J_{-2/1}$.}
 \label{subfig:LBRMHDSAm2n1deform-r-J_i_mm2n1-t}
 \end{minipage}
 \caption{
 Radial profiles of 
(a) $\Re \, U_{-2/1}$, (b) $\Im \, U_{-2/1}$, 
(c) $\Re \, \vphi_{-2/1}$, (d) $\Im \, \vphi_{-2/1}$, 
(e) $\Re \, \psi_{-2/1}$, (f) $\Im \, \psi_{-2/1}$, 
(g) $\Re \, J_{-2/1}$, (h) $\Im \, J_{-2/1}$
at $t = 0$, $100$, $1000$ and $10000$.
 }
 \label{fig:LBRMHDSAm2n1deform-r-u_mm2n1-t}
\end{figure}

Figure~\ref{fig:LBRMHDSAm2n1deform-r-u_mm2n1-t}
shows time evolution of the radial profiles of 
$U_{-2/1}$,
$\vphi_{-2/1}$,
$\psi_{-2/1}$,
$J_{-2/1}$,
at $t = 0$, $100$, $1000$ and $10000$.
The real parts of both the velocity and the magnetic parts were
initially finite.
However, they almost disappeared.
On the other hand, the imaginary parts of the velocity and the magnetic
parts appear to be  generating some structure with finite amplitudes.
The structure did not change significantly after $t > 100$, 
although the amplitudes were still getting larger slowly on  the long
time scale.

\section{Super-Alfv\'enic equilibria}
\label{sec:superAlfvenic}

When applying  SA to  low- or high-beta reduced MHD,
the total energy of the system is minimized in order to  reach a stationary state
with a smooth spatial structure.
On the other hand, when applying  SA to  two-dimensional Euler flow,
the total energy of the system is maximized to reach a stationary 
state with a smooth spatial structure \citep{Vallis-1989, Carnevale-1990,
Flierl-Morrison-2011}.
If the energy is minimized in the two-dimensional Euler flow by SA,
the system approaches a state called  Kelvin's sponge
\citep[see][]{Flierl-Morrison-2011}.

The numerical results shown in 
Secs.~\ref{sec:linearStability}, \ref{sec:toroidal}, and \ref{sec:helical}
were obtained by minimizing the total energy of the system 
by SA for reduced MHD.
In most of these numerical cases,  plasma flow was  absent.
Even in the case with  finite plasma flow, 
the flow velocity was small compared to the Alfv\'en velocity, so that the 
 system was dominated by magnetic energy.  Thus, the question arises of  what will happen if we perform SA to minimize
the total energy when the kinetic energy is 
comparable to or even larger than the magnetic energy.
Some examples of this case were  shown in \citet{Chikasue-2015-PoP},
where SA was performed for   low-beta reduced MHD in a doubly-periodic
rectangular domain.

Figure~5 of \citet{Chikasue-2015-PoP} shows  the time evolution 
of $U$, $\vphi$, $\psi$ and $J$ for a case with comparable kinetic and
magnetic energies.  We observed that fine spatial structures in $U$ and
$J$ remained,  although the system seemed to be trying to generate a smooth and symmetric
circular spatial structure, such as  that reached in the sub-Alfv\'enic case
of  Fig.~8 of \citet{Chikasue-2015-PoP}.   Figures~11 and 14 of \cite{Chikasue-2015-PoP} show  the time evolution
of $U$, $\vphi$, $\psi$ and $J$ for super-Alfv\'enic cases.
The case of Fig.~14 had larger ratio of the kinetic energy to the
magnetic energy.  In both cases, we observed fine spatial structures.
This indicates that the system behaved more like a two-dimensional
neutral fluid.  

\section{Equilibrium with magnetic islands}
\label{sec:island}

As explained in Sec.~\ref{subsec:SAdoubleBracket},
the double bracket dynamics of SA preserves all the Casimir invariants.
Therefore, the magnetic field topology is also preserved.
If there is no magnetic island in the initial condition for SA,
then theoretically magnetic islands should  never appear. 

We  tried an initial condition with magnetic islands, and  
obtained an equilibrium with  magnetic islands by SA of low-beta
reduced MHD in cylindrical geometry \citep{Furukawa-2017}.
The initial condition was a sum  of a cylindrically symmetric
equilibrium and a small-amplitude helical perturbation.
The safety factor $q$ profile of the equilibrium was the
same as the one in Sec.~\ref{subsec:m2n1perturbation};
the $q$ profile is monotonic and 
there exists a $q=2$ surface at $r = 1/2$.
Equilibrium  flow was absent and the  helical perturbation had Fourier mode numbers $m=2$ and $n=1$.

\begin{figure}[h]
 \begin{minipage}[t]{0.5\textwidth}
  \centering
  \includegraphics[width=\textwidth]{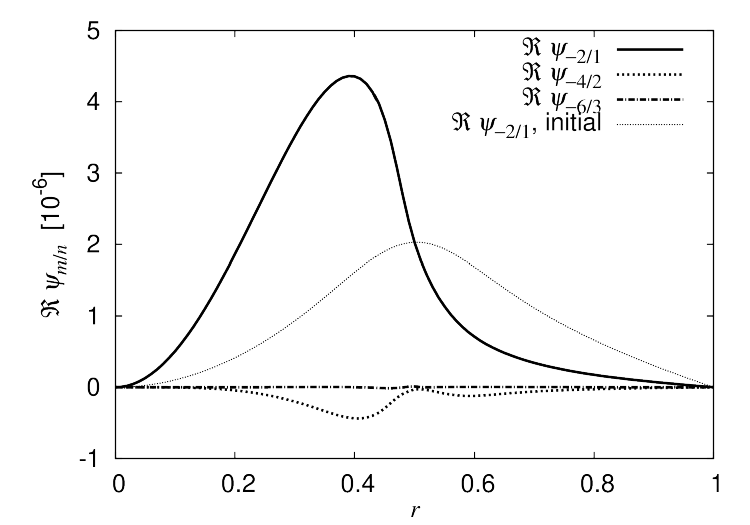}
  \subcaption{Radial profiles of $\Re \, \psi_{mn}$.}
  \label{subfig:LBRMHDSA-island-r-psi}
 \end{minipage}
\hfill
 \begin{minipage}[t]{0.4\textwidth}
  \centering
  \includegraphics[width=\textwidth]{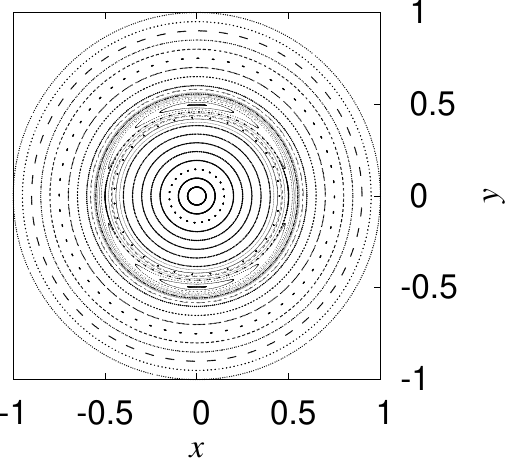}
  \subcaption{Poincar\`e plots of magnetic field lines on a poloidal cross section.}
  \label{subfig:LBRMHDSA-island-PoincarePlot}
 \end{minipage}
 \caption{
(Reprinted from M. Furukawa, J. Plasma Fusion Res. \textbf{94}, 341--384
 (2018), Fig.~4 (in Japanese).)
An equilibrium with magnetic islands were calculated by SA of
 low-beta reduced MHD in cylindrical geometry.
 The $\Re \, \psi_{mn}$ at the $q = 2$ resonant
  surface at $r = 1/2$ did not change during the time evolution of SA
 because of preservation 
  of Casimir invariants.  The island
  width did not change from the initial condition.
}
 \label{fig:LBRMHDSA-island}
\end{figure}

The radial profiles of $\Re \, \psi_{mn}$ and the Poincar\`e plots of
magnetic field lines on a poloidal cross section 
at the stationary state are shown in Fig.~\ref{fig:LBRMHDSA-island}.
The initial $\Re \, \psi_{-2/1}$ is also plotted in 
Fig.~\ref{fig:LBRMHDSA-island}\subref{subfig:LBRMHDSA-island-r-psi}.
The value of $\Re \, \psi_{mn}$ at the $q = 2$ resonant surface at $r = 1/2$ did
not change during the time evolution of SA because of preservation of
Casimir invariants.  Therefore, the island width did not change from the
initial condition. 

\section{Accelerated relaxation}
\label{sec:acceleration}

For both equilibrium and stability calculations, SA solves an initial-value problem. 
Generally, the computations are time consuming, especially as one gets near the energy
minimum.  Therefore, accelerated relaxation to the stationary state is
indispensable for SA to be practically useful.  We have examined two methods for acceleration, which are explained in
this section.  Section~\ref{subsec:timeDependentKernel} explains 
the first method,  where time dependence was introduced in the kernel 
when defining the double bracket in Eq.~(\ref{eq:doubleBracketInfDoF}).
This certainly had an acceleration effect.
The other method  is explained in 
Sec.~\ref{subsec:additionHamiltonianDynamics},
where the original Hamiltonian dynamics was added to the SA dynamics.
We observed both acceleration and deceleration of relaxation using this method.
Further examination of methods for acceleration are  under investigation.

\subsection{Time dependent kernel}
\label{subsec:timeDependentKernel}

In \citet{Furukawa-2022},
we found that the magnetic energy decreases quickly, while the kinetic
energy  changes over a significantly  longer time scale.  This is the reason why the
system requires a long time to approach a stationary state.
Therefore, it seemed better to find a relaxation path such that 
the kinetic and magnetic energies decrease at  comparable rates.

The idea was to introduce time dependence in the symmetric kernel.
Explicitly, we controlled the magnitudes of 
$\alpha_{ii}$ in Eqs.~(\ref{eq:LBRMHDDoubleBracketAdvField1})
and (\ref{eq:LBRMHDDoubleBracketAdvField2})
so that $\tilde{\vphi}$ and $\tilde{J}$ become comparable.
If $f^{i}$, the right-hand side of the original low-beta
reduced MHD equations, is small (large), then
$\alpha_{ii}$ is changed to a larger (smaller) value
at each time step.  This method successfully accelerated the relaxation to the stationary
state, although it may still be possible to improve how the time
dependence is implemented.

\subsection{Addition of Hamiltonian dynamics}
\label{subsec:additionHamiltonianDynamics}

We have also examined whether the relaxation can be accelerated if we add the
original Hamiltonian dynamics to the SA dynamics, which uses  the double
bracket.    In Sec.~\ref{subsec:SAmimicLBRMHD},
we observed that the time required to approach the stationary state did not
differ significantly with  the inclusion of the Hamiltonian dynamics in the
case of the toy model mimicking  low-beta reduced MHD.
In the low-beta reduced MHD case, on the other hand, 
we found that the relaxation could be either accelerated or decelerated 
\citep{Furukawa-2023-CCP, Furukawa-2023-AAPPS-DPP}.

Recall 
$f^{i} \coloneqq  \{ u^{i} , H \}$
in
Eq.~(\ref{eq:LBRMHDRHS}),
and define
\begin{equation}
 \tilde{f}^{i} := (( u^{i} , H )), 
\label{eq:LBRMHDSARHS}
\end{equation}
which gives  the right-hand sides of the evolution equations of SA with the
double bracket.
In the simulation results shown here,
the symmetric kernel was chosen to be diagonal, and the coefficients
$\alpha_{ii}$ were taken to be constant during the simulations.
Then the  mixed dynamics was  generated by
\begin{equation}
 \frac{\pd u^{i}}{\pd t}
= 
 \tilde{f}^{i} + c f^{i}.
\label{eq:additionHamiltonianDynamicsEvolutionEq}
\end{equation}
The parameter $c$ is a constant
representing the ratio of the Hamilton dynamics to the SA dynamics.
When $c < 0$, the time-reversed Hamiltonian dynamics is added to the SA
dynamics.  Pure SA dynamics corresponds to $c = 0$.

\begin{figure}[h]
 \begin{minipage}[t]{0.45\textwidth}
  \centering
  \includegraphics[width=\textwidth]{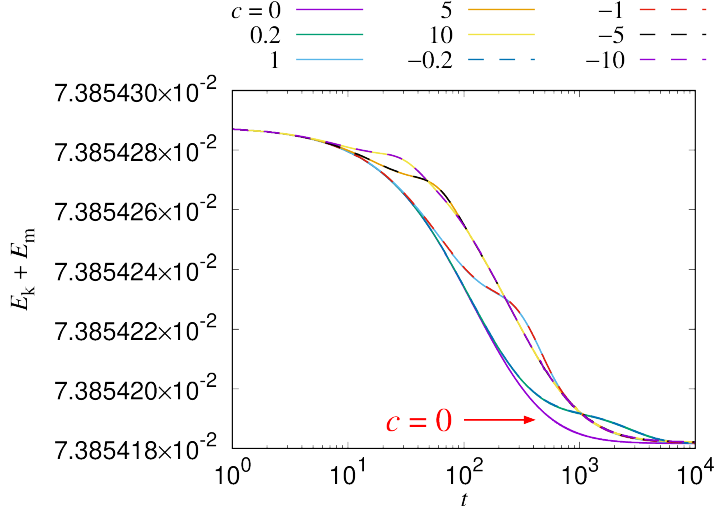}
  \subcaption{Time evolution of the total energy is shown.  The
  perturbed kinetic energy was much smaller than the perturbed magnetic
  energy in the initial condition.  The relaxation was decelerated by
  adding the Hamiltonian dynamics in either sign of $c$.}
  \label{subfig:LBRMHDSA-additionHamiltonianDynamics-t-E-1}
 \end{minipage}
\hfill
 \begin{minipage}[t]{0.45\textwidth}
  \centering
  \includegraphics[width=\textwidth]{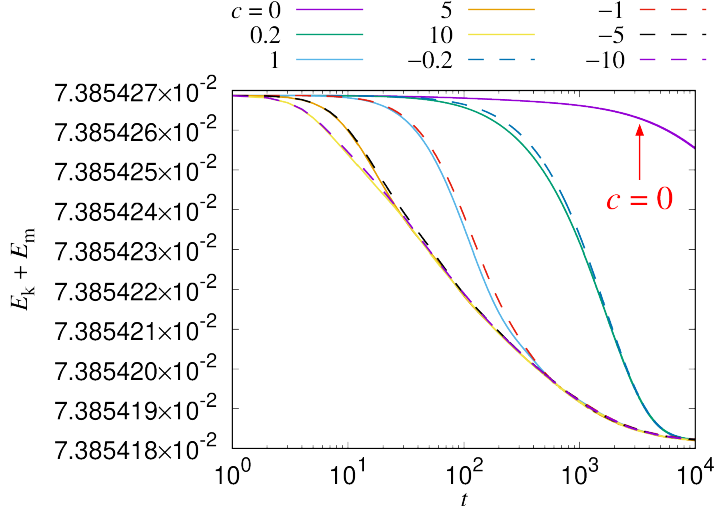}
  \subcaption{Time evolution of the total energy is shown.  The
  perturbed kinetic energy was much larger than the perturbed magnetic
  energy in the initial condition.  The relaxation was accelerated by
  adding the Hamiltonian dynamics in either sign of $c$.}
  \label{subfig:LBRMHDSA-additionHamiltonianDynamics-t-E-2}
 \end{minipage}
 \caption{
The relaxation to the stationary state was either decelerated or
 accelerated by adding the Hamiltonian dynamics.  
The time direction of the Hamiltonian dynamics,
 expressed by the sign of $c$, did not generate significant differences
 in the relaxation. 
}
 \label{fig:LBRMHDSA-additionHamiltonianDynamics-t-E}
\end{figure}

Figure~\ref{fig:LBRMHDSA-additionHamiltonianDynamics-t-E} shows the time
evolution of SA for the same equilibrium presented in 
Sec.~\ref{subsec:m2n1perturbation};
the equilibrium with the monotonic $q$ profile with the $q = 2$ surface at
$r = 0.5$ and without plasma rotation.
The initial perturbations were   dynamically
accessible; i.e.,  they were generated by the advection fields 
(\ref{eq:LBRMHDDoubleBracketSAAdvFieldDAV1}) 
and 
(\ref{eq:LBRMHDDoubleBracketSAAdvFieldDAV2}),
where $m = 2$, $n = 1$, $r_{0} = 0.5$, and $L = 0.1$.
In Fig.~\ref{fig:LBRMHDSA-additionHamiltonianDynamics-t-E}\subref{subfig:LBRMHDSA-additionHamiltonianDynamics-t-E-1},
the initial condition for SA were generated with 
$A_{\vphi} = A_{J} = 10^{-3}$, 
for which the perturbed kinetic energy was much smaller than the perturbed
magnetic energy at the initial time.
On the other hand, 
in Fig.~\ref{fig:LBRMHDSA-additionHamiltonianDynamics-t-E}\subref{subfig:LBRMHDSA-additionHamiltonianDynamics-t-E-2},
the initial condition for SA were generated with 
$A_{\vphi} = 10^{-4}$ and $A_{J} = 2 \times 10^{-1}$,
for which the perturbed kinetic energy was much larger than the perturbed
magnetic energy at the initial time.

In Fig.~\ref{fig:LBRMHDSA-additionHamiltonianDynamics-t-E}\subref{subfig:LBRMHDSA-additionHamiltonianDynamics-t-E-1},
we observe that the relaxation was fastest when $c = 0$.
The addition of the Hamiltonian dynamics decelerated the relaxation. 
Moreover, the sign of $c$ did not generate a visible difference in the
time evolution of the total energy. On the other hand, 
in Fig.~\ref{fig:LBRMHDSA-additionHamiltonianDynamics-t-E}\subref{subfig:LBRMHDSA-additionHamiltonianDynamics-t-E-2},
we observe that the relaxation was slowest when $c = 0$.
As explained in Sec.~\ref{subsec:m2n1perturbation}, the relaxation to a
stationary state becomes considerably slow when the initial perturbation
has a large kinetic energy.  
Addition of the Hamiltonian dynamics significantly accelerated the
relaxation.  
The time evolutions of the total energy were slightly different
depending on the sign of $c$ with a same magnitude, however, the
difference was not significant.

We observed that the time evolution of the kinetic energy 
may be a key to understanding  what causes  the relaxation to be  accelerated or
decelerated.  This issue is still under investigation.

\section{Discussion}
\label{sec:discussion}

An issue to be clarified was raised in Sec.~\ref{sec:superAlfvenic} 
regarding the equilibria with large plasma flow velocities.
In the context of magnetically confined fusion plasmas,
it may be unusual to have a super-Alfv\'enic flow velocity.
However, if we do need to calculate an equilibrium with a super-Alfv\'enic
flow velocity, we may perform SA maximizing the total energy of the
system to obtain an equilibrium with smooth spatial structure, 
according to the results in \citet{Chikasue-2015-PoP}.
On the contrary, when the kinetic and magnetic energies are
comparable, we do not know whether the total energy should be minimized
or maximized to obtain a stationary state by SA.  

As explained in Sec.~\ref{sec:acceleration},
accelerated relaxation is especially important if SA is utilized to
obtain a stationary state of a Hamiltonian system with infinite degrees
of freedom because SA requires solving an initial value problem.
In Sec.~\ref{subsec:timeDependentKernel}, 
we explained that the relaxation can be accelerated by introducing 
 time dependence in the kernel of the double bracket.
The key was to control the advection fields to have comparable
magnitudes.  However, we have to determine what magnitudes are appropriate.
If the magnitudes are too large, the time evolution likely becomes
numerically unstable.  If we have a numerically more stable algorithm
for the time evolution, the magnitudes can be larger.
Normally, such algorithms use implicit methods, which  require  iteration
to solve nonlinear equations.  For these,   an efficient preconditioning is required to  realize a 
large time step,  which is an advantage of implicit methods.

SA can be applied to any Hamiltonian system; hence, a natural future step might be  to apply it  to the full MHD system in toroidal
geometry.  Then we may be able to calculate an MHD equilibrium with
magnetic islands and/or even magnetic chaos. 
In such a case, it may be important to recognize on which Casimir leaf the
equilibrium exists.  Since the Casimir invariants do not change during
the time evolution of SA, we need to adjust the values of the Casimir
invariants of the initial condition for SA.  
It was demonstrated that we could adjust the values of the Casimir
invariants of the initial condition for two-dimensional Euler flow
and the low-beta reduced MHD  in \citet{Chikasue-2015-JFM}.
This adjustment method will be useful when applying SA to full MHD, or 
even kinetic models that are Hamiltonian.

Regarding   numerical stability, 
spatial discretization methods should be also important in addition to
the time integration methods. 
The numerical results introduced in this paper on the reduced MHD systems
in cylindrical and toroidal geometry 
used second-order central differences in the radial direction
and   Fourier decomposition in the poloidal and the toroidal directions
for all variables equally.  It should be advantageous to implement the
discretization based on finite element exterior calculus 
\citep{Arnold-2006, Kraus-2017} for improving numerical stability.

Such improved numerical stability may enable us to obtain another
equilibrium by SA when an equilibrium is unstable. 
As explained in Sec.~\ref{sec:linearStability},
SA succeeded in identifying a linearly unstable equilibrium. 
However, after the initial growth of the helical perturbation, 
a spiky behavior appeared in the radial profile of the variables.
Therefore, the time evolution of SA was stopped.  
Although it is unclear whether such spiky behavior is because of
physics or is a  numerical artifact, it is anyway better to adopt numerically
stable algorithms. 

Another future possibility is  to explore the calculation of free boundary equilibria.  The numerical results explained in the paper were all
obtained under fixed boundary conditions,  except for the doubly-periodic
boundary condition in the two-dimensional rectangular domain in 
Sec.~\ref{sec:superAlfvenic}.   This is certainly possible theoretically.

\section{Summary and conclusions}
\label{sec:summaryConclusion}

Simulated annealing (SA) is a method for obtaining equilibria and analyzing stability 
of Hamiltonian systems.  
Starting from any  Hamiltonian system, 
an artificial dynamics is derived that monotonically changes the total 
energy of the system, while preserving all  the Casimir invariants.  
These are accomplished by using the double bracket obtained from  the Poisson bracket.  
By solving an initial-value problem of the artificial dynamics, the
system may reach a state with a stationary energy that is an equilibrium.
If the energy is minimized or maximized, the equilibrium is stable from an  energy
standpoint.

This paper reviewed Hamiltonian structure, formulation of SA, and
 described numerical demonstrations of SA for 
some Hamiltonian systems of  both finite and infinite degrees of freedom. 
The numerical  results for  reduced MHD systems,  obtained by double bracket SA,  included  cylindrical as well as axisymmetric
toroidal equilibria, linear stability, helically deformed equilibria,
flowing equilibria, and equilibria with magnetic islands.
We also explained the importance of  accelerated relaxation,
and introduced two methods for doing so, although one of the methods  is
still under investigation. Some issues for future work  were also discussed.

We hope that this paper succeeded in sharing interesting aspects of SA,
and revealing how  SA can be applied to many other Hamiltonian systems.  Then,  the theoretical and practical use of SA might be further developed in the future.

\bmhead{Acknowledgements}

M.F. was supported by JSPS KAKENHI (Grant No. JP15K06647, JP21K03507 and
 No. JP24K06993), while P.J.M. supported by the United States Department of Energy (Grant No.  DE-FG02- 04ER54742).
Both authors  would like to acknowledge the JIFT program for the
support of M.F. to visit IFS in the Spring of 2019 when a portion of this
work was carried out.
M.F. is grateful for discussions with Y. Chikasue, Takahiro Watanabe,
K. Goto, and K. Ichiguchi.  Lastly, we  sincerely appreciate the late Emeritus Professor Robert
L. Dewar for encouraging us to write this RMPP paper.

\bmhead{Data availability}
The data that support the findings of this study are available from the
corresponding author upon reasonable request.

\section*{Declarations}

\textbf{Conflict of interest}
Authors state that there is no conflict of interest.

\begin{appendices}

\section{Detailed explanation of Eq.~(\ref{eq:2DEulerEvolutionEqFunctionalPoissonBracket})}
\label{sec:detail2DEulerEvolutionEqFunctionalPoissonBracket}

First, we  recognize that the arguments of the right-hand side of 
Eq.~(\ref{eq:2DEulerEvolutionEqFunctionalPoissonBracket})
are functionals, mappings from functions to real numbers.
The Hamiltonian functional  defined by Eq.~(\ref{eq:2DEulerHamiltonian}), which is
a number as a result of spatial integration.
The vorticity $U$ can  also be interpreted  as  a functional given by
\begin{equation}
 U( \vx_{0} , t )
=
 \int_{\mathcal{D}} \td^{2} x  \, 
 \delta^{2} ( \vx - \vx_{0} ) U ( \vx , t ),
\label{eq:vorticityFunctional}
\end{equation}
where $\delta^{2}(\vx)$ is a two-dimensional Dirac  delta function.
The spatial integration gives us a value of $U$ at $\vx = \vx_{0}$
and time $t$.

In order to evaluate  
Eq.~(\ref{eq:2DEulerFunctionalPoissonBracket})
or 
Eq.~(\ref{eq:2DEulerFunctionalPoissonBracket-2}) when Eq.~(\ref{eq:vorticityFunctional})  is inserted,
we need $\delta U (\vx_{0} , t) / \delta U ( \vx , t )$.
This is obtained through 
\begin{align}
 \delta U ( \vx_{0} , t )
&=
 \lim_{\veps \rightarrow 0}
 \frac{1}{\veps}
 \int_{\mathcal{D}} \td^{2} x  \,
   \delta^{2} ( \vx - \vx_{0} )
 \left( 
   \left(
     U ( \vx , t ) + \veps \delta U ( \vx , t )
   \right)
  - U ( \vx , t )
 \right)
\\
&=
 \int_{\mathcal{D}} \td^{2} x  \,
 \delta U ( \vx , t ) \delta^{2} ( \vx - \vx_{0} )
\end{align}
which implies 
\begin{equation}
 \frac{\delta U ( \vx_{0} , t )}
      {\delta U ( \vx , t )}
 =
 \delta^{2} ( \vx - \vx_{0} ).
\end{equation}
On the other hand,
\begin{align}
 \delta H [ U ]
&=
 \int_{\mathcal{D}} \td^{2} x  \,
   \nab \vphi ( \vx , t )
   \cdot 
   \nab \delta \vphi ( \vx , t ) 
\\
&=
 \int_{\mathcal{D}} \td^{2} x  \,
   \delta U ( \vx , t ) ( -\vphi ( \vx, t ) ),
\end{align}
where $\delta U ( \vx , t ) = \bigtriangleup_{\perp} \vphi ( \vx , t )$
and an integration by parts were used.
Therefore
\begin{equation}
 \frac{ \delta H [ U ] }{ \delta U ( \vx , t ) }
=
 - \vphi( \vx , t  ).
\end{equation}
Then, Eq.~(\ref{eq:2DEulerFunctionalPoissonBracket-2}) reads
\begin{align}
 \{ U ( \vx_{0} , t ) , H \}
&=
 \int_{\mathcal{D}} \td^{2} x  \,
    \delta^{2} ( \vx - \vx_{0} )
    \left[
      - \vphi( \vx , t  ) , U ( \vx , t )
    \right]
\nonumber
\\
&=
 \left.
 \left[
   U ( \vx , t ) , \vphi( \vx , t  )
 \right]
 \right\vert_{\vx_{0}} 
\end{align}
and  Eq.~(\ref{eq:2DEulerEvolutionEqFunctionalPoissonBracket})
gives 
\begin{equation}
 \frac{\pd U ( \vx_{0} , t )}{\pd t} 
= 
 \{ U ( \vx_{0} , t ) , H [ U ] \}
=
 \left.
 \left[
   U ( \vx , t ) , \vphi( \vx , t  )
 \right]
 \right\vert_{\vx_{0}}.
\label{eq:2DEulerEvolutionEqX0}
\end{equation}

The evolution equations of   low-beta RMHD in two dimensions 
(\ref{eq:rectLBRMHDEvolutionEqFunctionalPoissonBracket}), 
those in cylindrical geometry
(\ref{eq:cylLBRMHDEvolutionEqFunctionalPoissonBracket})
as well as 
(\ref{eq:cylLBRMHDHelFlxEvolutionEqFunctionalPoissonBracket})
for single helicity dynamics,
and those in toroidal geometry 
(\ref{eq:torHBRMHDEvolutionEqFunctionalPoissonBracket})
are understood similarly.

Equation~(\ref{eq:evolutionEqSAInfDoF}) may also needs some
explanation. Writing  the arguments of the variables and functionals as
\begin{equation}
 \frac{\pd u^{i} ( \vx_{0} , t )}{\pd t}
= (( u^{i} ( \vx_{0} , t ) , H [ \vu ]  )),
 \label{eq:evolutionEqSAInfDoFX0}
\end{equation}
we see its evaluation, leading to the  evolution equations  \eqref{eq:evolutionEqSAInfDoF}, is similar  to that for the Poisson bracket above. 
The same applies for the 
the Dirac bracket on functionals  
(\ref{eq:DiracDoubleBracketInfDoF}), leading to   (\ref{eq:evolutionEqDSAInfDoF}),  and  metriplectic dynamics as given by 
(\ref{eq:metriplecticEvolutionEqInfDoF}).

\end{appendices}



\end{document}